Roadmap

# Machine Learning on Heterogeneous, Edge, and Quantum Hardware for Particle Physics

Julia Gonski,[1,*] Jennifer Ott,[2,*] Shiva Abbaszadeh,[3] Sagar Addepalli,[1] Matteo Cremonesi,[4] Jennet Dickinson,[5] Giuseppe Di Guglielmo,[6,7] Yigit Erdem Ertorer,[4] Lindsey Gray,[6] Ryan Herbst,[1] Christian Herwig,[8] Tae Min Hong,[9] Benedikt Maier,[10] Maryam Bayat Makou,[11] David W. Miller,[12] Mark S. Neubauer,[13] Cristián Peña,[6] Dylan Rankin,[14] Seon-Hee Seo,[6] Giordon Stark,[15] Alexander Tapper,[10] Audrey Corbeil Therrien,[16] Ioannis Xiotidis,[17] Keisuke Yoshihara,[18] Nural Akchurin,[20] Carlos Argüelles,[21] Saptaparna Bhattacharya,[11] Lorenzo Borella,[22,23] Christian Boutan,[24] Tom Braine,[24] James Brau,[25] Marco Carminati,[29] Antonio Chahine,[10] Yuan-Tang Chou,[27] Talal Ahmed Chowdhury,[26] Seokju Chung,[28] Ryan Coffee,[1] Alberto Coppi,[22,23] Mariarosaria D'Alfonso,[30] Abhilasha Dave,[1] Chance Desmet,[24] Angela Di Fulvio,[31] Karri DiPetrillo,[12] Javier Duarte,[19] Auralee Edelen,[1] Jan Eysermans,[30] Yongbin Feng,[20] Emmett Forrestel,[32] Julián García Pardiñas,[30] Loredana Gastaldo,[33] Lino Gerlach,[34] Abarajithan Gnaneswaran,[19] Loukas Gouskos,[32] Katya Govorkova,[30] Carl Grace,[35] Christopher Grant,[36] Philip Harris,[30] Ciaran Hasnip,[17] Timon Heim,[35] Abraham Holtermann,[30] Gian Michele Innocenti,[30] Koji Ishidoshiro,[37] Miaochen Jin,[21] Jyothisraj Johnson,[35] Stephen Jones,[24] Andreas Jung,[38] Nicholas Kamp,[21] Georgia Karagiorgi,[28] Ryan Kastner,[19] Doojin Kim,[39] Kyoungchul Kong,[26] Katie Kudela,[30] Bo-Cheng Lai,[40] Yun-Tsung Lai,[41] Jelena Lalic,[30] Tommy Lam,[42] Jeffrey Lazar,[43] Aobo Li,[19] Zepeng Li,[18] Haoyun Liu,[10] Qibin Liu,[1] Vladimir Lončar,[30] Zhenghua Ma,[19] Luca Macchiarulo,[44] Christopher Madrid,[20] Prashansa Mukim,[45] Victoria Nguyen,[30] Sungbin Oh,[6] Isobel Ojalvo,[34] Hideyoshi Ozaki,[46] Simone Pagan Griso,[49] Santosh Parajuli,[13] Myeonghun Park,[47] Benjamin Parpillon,[6] Christoph Paus,[30] Sara Pozzi,[52] Ema Puljak,[17] Benjamin Ramhorst,[48] Amy Roberts,[50] Larry Ruckman,[1] Sebastian Schmitt,[30] Kate Scholberg,[51] Noah Singer,[38] Eluned Anne Smith,[30] Alexandre Sousa,[53] Michael Spannowsky,[54] Kaito Sugizaki,[14] Sioni Summers,[17] Yanwen Sun,[1] Daniel Tapia Takaki,[26] Antonino Tumeo,[24] Caterina Vernieri,[1] Belina von Krosigk,[55] Yash Vora,[32] Linyan Wan,[6] Michael H. L. S. Wang,[6] Amanda Weinstein,[56] Andy White,[57] Simon Williams,[54] Liangyu Wu,[1] and Felix Yu[21]

[1]*SLAC National Accelerator Laboratory, Menlo Park, California, USA*
[2]*Department of Electrical and Computer Engineering, University of Hawai'i Mānoa, Honolulu, Hawaii, USA*
[3]*Santa Cruz Institute for Particle Physics, UC Santa Cruz, Santa Cruz, USA*
[4]*Department of Physics, Carnegie Mellon University, Pittsburgh, Pennsylvania, USA*
[5]*Department of Physics (also CLASSE) Cornell University, Ithaca, New York, USA*
[6]*Fermi National Accelerator Laboratory, Batavia, Illinois, USA*
[7]*Electrical and Computer Engineering, Northwestern University, Evanston, Illinois, USA*
[8]*Department of Physics, University of Michigan, Ann Arbor, Michigan, USA*
[9]*Department of Physics and Astronomy, University of Pittsburgh, Pittsburgh, Pennsylvania, USA*
[10]*Imperial College London, London, United Kingdom*
[11]*Department of Physics, Southern Methodist University, University Park, Texas, USA*
[12]*Department of Physics, Enrico Fermi Institute, Kavli Institute for Cosmological Physics, University of Chicago, Chicago, Illinois, USA*
[13]*Department of Physics, University of Illinois, Urbana-Champaign, USA*
[14]*Department of Physics and Astronomy, University of Pennsylvania, Philadelphia, Pennsylvania, USA*
[15]*Santa Cruz Institute for Particle Physics (SCIPP), University of California Santa Cruz, Santa Cruz, California, USA*
[16]*Département de Génie Électrique et de Génie Informatique, Université de Sherbrooke, Sherbrooke, Québec, Canada*
[17]*CERN, Geneva, Switzerland*
[18]*Department of Physics and Astronomy, University of Hawai'i Mānoa, Honolulu, Hawaii, USA*
[19]*University of California San Diego, La Jolla, California, USA*
[20]*Department of Physics and Astronomy, Texas Tech University, Lubbock, Texas, USA*
[21]*Department of Physics, Harvard University, Cambridge, Massachusetts, USA*
[22]*Department of Physics and Astronomy, Istituto Nazionale di Fisica Nucleare (INFN), Padova, Italy*
[23]*University of Padua, Padova, Italy*
[24]*Nuclear and Particle Physics, Pacific Northwest National Laboratory, Richland, Washington, USA*
[25]*University of Oregon, Eugene, Oregon, USA*
[26]*Department of Physics and Astronomy, University of Kansas, Lawrence, Kansas, USA*
[27]*Department of Physics, University of Washington, Seattle, Washington, USA*

[*]Contact author: jgonski@slac.stanford.edu

[28]*Department of Physics, Columbia University, New York, New York, USA*
[29]*Department of Electronics, Information and Bioengineering (DEIB), Politecnico di Milano, Milano, Italy*
[30]*Massachusetts Institute of Technology, Cambridge, Massachusetts, USA*
[31]*Department of Nuclear, Plasma, and Radiological Engineering, University of Illinois, Urbana-Champaign, USA*
[32]*Department of Physics, Brown University, Providence, Rhode Island, USA*
[33]*Kirchhoff Institute for Physics, Heidelberg University, Heidelberg 69120, Germany*
[34]*Department of Physics, Princeton University, Princeton, New Jersey 08544, USA*
[35]*Lawrence Berkeley National Laboratory, Berkeley, California, USA*
[36]*Department of Physics, Boston University, Boston, Massachusetts, USA*
[37]*Research Center for Neutrino Science (RCNS), Tohoku University, Sendai, Miyagi, Japan*
[38]*Department of Physics and Astronomy, Purdue University, West Lafayette, Indiana, USA*
[39]*Department of Physics, University of South Dakota, Vermillion, South Dakota, USA*
[40]*Department of Electronics Engineering, National Yang Ming Chiao Tung University, Taiwan*
[41]*High Energy Accelerator Research Organization, Japan*
[42]*Virginia Tech, Blacksburg, Virginia, USA*
[43]*Centre for Cosmology, Particle Physics and Phenomenology (CP3), Université Catholique de Louvain, Louvain-la-Neuve, Belgium*
[44]*Nalu Scientific LLC, USA*
[45]*Brookhaven National Laboratory, Upton, New York, USA*
[46]*Department of Physics and Astronomy, Tokyo University of Science, Shinjuku City, Tokyo 162-8601, Japan*
[47]*School of Liberal Arts, Seoul National University of Science and Technology, South Korea*
[48]*ETH Zurich, Switzerland*
[49]*Physics Division, Lawrence Berkeley National Laboratory, Berkeley, California, USA*
[50]*Department of Physics, University of Colorado Denver, Denver, Colorado, USA*
[51]*Department of Physics, Duke University, Durham, North Carolina, USA*
[52]*Dept. of Nuclear Engineering & Radiological Sciences (joint w/ Physics), University of Michigan, Ann Arbor, Michigan, USA*
[53]*Department of Physics University of Cincinnati, Cincinnati, Ohio, USA*
[54]*Institute for Particle Physics Phenomenology (IPPP), Dept. of Physics, Durham University, Durham, United Kingdom*
[55]*Kirchhoff Institute for Physics, Heidelberg University, Heidelberg, Germany*
[56]*Department of Physics and Astronomy, Iowa State University, Ames, Iowa, USA*
[57]*Department of Physics, University of Texas Arlington, Arlington, Texas, USA*

The next generation of particle physics experiments will face an era of challenges in data acquisition, due to unprecedented data rates and volumes along with extreme environments and operational constraints. Harnessing these data for scientific discovery demands real-time inference and decision-making, intelligent data reduction, and efficient processing architectures beyond current capabilities. Crucial to the success of this experimental paradigm are several emerging technologies, such as artificial intelligence and machine learning, silicon microelectronics, and the advent of quantum algorithms and processing. Their intersection includes areas of research such as low-power and low-latency devices for edge computing, heterogeneous accelerator systems, reconfigurable hardware, codesign and synthesis strategies, readout for cryogenic or high-radiation environments, and analog computing. This white paper presents a community-driven vision to identify and prioritize research and development opportunities in hardware-based ML systems and corresponding physics applications, contributing toward a successful transition to the data frontier of fundamental science.

**CONTENTS**

## I. INTRODUCTION

The scientific ambitions of particle physics over the coming decades, covering precision measurements and discovery-driven experiments across a huge range of energy scales, are tightly coupled to advances in detector instrumentation and computing. The potential impacts of basic research at the most fundamental level are vast and transformational, with the promise of a more detailed understanding of the evolution and fate of the universe. The technological advancements required for this research also lead to breakthroughs such as magnetic resonance imaging and the World Wide Web, transforming society and improving lives. What is more, research at the most fundamental level of physics represents a pursuit of inherent human curiosity, arguably responsible for the most broad and enduring advances in scientific understanding and technological innovation.

In recent years, the United States particle physics community has sought to articulate these scientific goals and potential enabling technologies through a variety of strategic planning processes, such as the 2021 Snowmass process [1], the 2023 Particle Physics Project Prioritization Panel (P5) report [2], the Department of Energy (DOE) Basic Research Needs (BRN) reports for detector R&D [3], and the AI-native high-energy physics (HEP) ecosystem vision [4]. These studies collectively point to an era of experiments that face unprecedented data rates, growing detector complexity, and real-time decision-making requirements, all under increasingly stringent constraints on power, cost, and facility resources. Enabling continued discovery in this environment demands computing paradigms that maximize physics reach while making efficient use of limited resources.

Advancing to the next stage of fundamental physics requires the integration of several critical emerging technologies. Artificial intelligence (AI) and machine learning (ML) are essential, due to their ability to extract complex patterns from high-dimensional data, enabling the scaling of human insight to regimes of data volume and complexity that are otherwise inaccessible. In parallel, silicon microelectronics technologies can improve modern scientific data acquisition systems, determining achievable bandwidth, latency, power efficiency, and reliability from the detector front end through downstream processing. The integration of AI/ML directly into advanced microelectronics enables real-time inference at the “edge” (i.e., at the source of data), transforming data acquisition (DAQ) from a passive readout into an active, intelligent component of the experiment. The concept of AI/ML to enhance real-time data-taking systems goes back as far as the 1990s [5,6], making it a long-standing research direction that is now becoming broadly practical thanks to modern high-density, low-power complementary metal-oxide semiconductor (CMOS), fast on-chip memory, and reconfigurable/heterogeneous hardware accelerators.

Relatedly, quantum sensors, networks, and processors have the potential to fundamentally reshape precision measurement and information processing, enabling classes of data acquisition and inference beyond the reach of classical systems, while simultaneously serving as platforms for fundamental physics measurements in their own right. Optimal use of these technologies in combination with the cutting-edge advances offered by AI/ML-enabled electronics motivates the study of data acquisition architectures that optimize dataflow across heterogeneous computing platforms, ensuring efficient movement, transformation, and processing of information across classical, accelerated, and quantum resources. The dedicated R&D of all such critical technologies under consideration of particle physics drivers, and the study of their benefits and implementation opportunities into particle physics experiments, is a unique area of study defined here as Machine Learning on Heterogenous, Edge, and QUantum hardware for Particle Physics (ML-HEQUPP) (Fig. 1).

Achieving the full potential of ML-HEQUPP requires a codesign philosophy that tightly integrates algorithms, hardware, firmware, and systems, as well as close collaboration between physicists and engineers. Rather than treating machine learning models and computing platforms as independent layers, a holistic approach to system design is necessary to optimize performance, latency, power consumption, and reliability.

Investment in ML-HEQUPP technologies for particle physics aligns with U.S. mission priorities in AI/ML and microelectronics, supporting technology transfer, cross-sector partnerships, and international competitiveness.

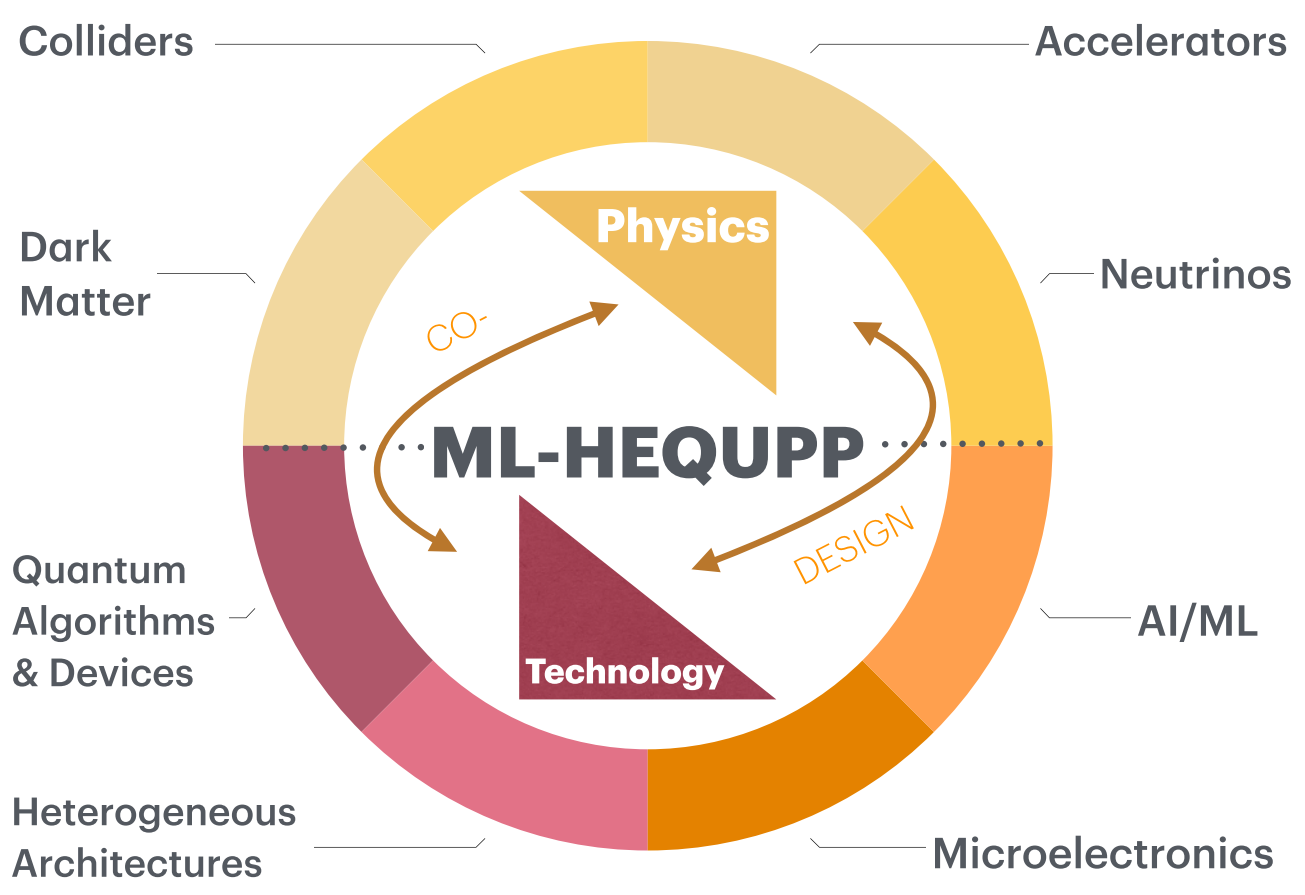

FIG. 1. Diagram indicating the physics and hardware areas under the ML-HEQUPP purview.

It also strengthens workforce development at the intersection of particle physics, machine learning, and hardware engineering, building specialized talent in high demand sectors across government, academia, national labs, and industry. Furthermore, this program offers a risk-balanced portfolio spanning varying investment timescales: near-term deployment of edge ML capabilities in operating experiments; midterm design and optimization of future experiments with heterogeneous platforms; and longer-term exploratory work in electronics and quantum/hybrid workflows for unprecedented physics reach. The 2025 U.S. government announcement of the Genesis mission [7] further underscores the timeliness of these investments by emphasizing the national importance of building secure, scalable, and energy-efficient AI/ML capabilities that can be translated from frontier research into broad economic and societal benefit.

In this spirit, the machine learning for front-end (ML4FE) workshop was launched in 2025 to connect physicists and engineers interested in front-end ML developments and applications to share their latest results, brainstorm future directions, build collaborations, pursue funding, and share infrastructure. The inaugural workshop was hosted by the University of Hawai'i at Manoa in Honolulu, Hawaii [8] and was held in a hybrid format with a total of 77 registrants (approximately 25 of whom participated in person) spanning physicists and engineers across a variety of specialties. Invitations for speakers and abstract submissions were advertised broadly in related communities; namely, technology collaborations, i.e., FastML [9] and the Coordinating Panel on Advanced Detectors (CPAD) [10]; AI interest groups within existing experiments; and future collider study efforts such as the U.S. Higgs Factory Circular Collider Ref. [11] and international Future Circular Collider (FCC) Detector Concepts communities. This document captures the content and discussions that took place at the workshop, where it was realized that the "front-end" indication was both ambiguous and limiting. Therefore, the white paper effort was initiated following the workshop with significant expansion to cover work and interests across the ML-HEQUPP community. While it does not match the reach of exercises like a BRN or Snowmass (particularly considering its origin and thus relative focus in collider physics), it encapsulates a growing community while providing a means for collaboration and sharing of ideas.

This white paper is organized as follows. The workshop was divided into two core but interrelated tracks, namely, the core technology required for ML-HEQUPP and the related physics applications, each covered in Secs. II and III, respectively. Community context, including interdisciplinary overlaps, and analysis facilities are discussed more in Sec. IV. Finally, the covered material is synthesized into a set of key R&D topics listed and described in Sec. V.

## II. TECHNOLOGY

The main challenge driving the ML-HEQUPP initiative is the dramatic increase in data rates associated with next-generation scientific facilities. There is a growing need in many experiments for flexible, high-throughput, low-latency processing solutions that can operate under the extreme bandwidth and timing constraints characteristic of modern instruments. The improvement in detector resolution and scale is leading to an increase of data volumes from gigabytes per second to hundreds of terabytes per second. The "record everything" approach that has traditionally used by many experiments has become technically and economically infeasible. In response, scientific data pipelines are now incorporating intelligent real-time filtering steps near the data generation source itself, which preserve salient scientific information while reducing data streams to manageable rates. Generally, addressing this challenge can require a distributed and heterogeneous computing ecosystem that spans detectors, front-end electronics (FEE), and offline computing infrastructure. Details of the implementation can make use of application-specific integrated circuits (ASICs), field programmable gate arrays (FPGAs), embedded processors, graphics processing units (GPUs), other specialized compute platforms, and/or a combination of many, depending on the constraints specific to a given application.

Several current science-driven examples demonstrate the potential and benefits of this distributed approach, from the edge to the data center. Attosecond streaking detectors requires megahertz-rate reconstruction of ultrafast x-ray pulses at the SLAC Linac Coherent Light Source (LCLS); fusion devices like DIII-D benefit from real-time prediction of disruptive events; and high-energy physics applications at the Large Hadron Collider (LHC) rely on intelligent triggers to reduce overwhelming event rates. In these examples and more, this community emphasizes that real-time and edge/distributed machine learning is required to manage data volumes and support responsive experimental control in today's detectors.

Furthermore, the technological approaches described here offer a path to meet the physics demands of future experiments considering full system-level design. The detector edge varies from experiment to experiment and has different requirements depending on many factors such as environment and data rate. In this way, an ML/hardware codesign approach enables the reimagining of sensor arrays as adaptive, hierarchical networks capable of localized processing, real-time feature extraction, intertile communication, and energy-optimized data reduction. This view drives heterogeneous and modu-lar architectures, where future systems can mix FPGAs for submicrosecond decisions, GPUs for flexible low-latency inference, ASICs for ultra-low-power always-on tasks, quantum hardware upon availability, and yet-to-be-imagined hardware continuously tested and integrated. It furthermore opens the door to embedded calibration, monitoring, fault detection, and validation, setting a path toward the autonomous laboratories that will comprise the next era of scientific progress.

### A. ASICs

ASICs are a key enabler to push machine learning into the detector front end and enable real-time "smart" decisions at the sensor boundary. As data rates climb from GB/s to TB/s, custom silicon offers the most direct path to low-latency, low-power, and tightly integrated processing that can operate close to the signal source and reduce downstream data movement. ASIC implementations also include substantial nonrecurring engineering costs, which must typically be amortized over high-volume production to be economically viable. In addition, ASIC development cycles are comparatively long; the end-to-end timeline from specification through tape-out, fabrication, and validation can extend significantly, making ASICs less suitable for rapidly evolving requirements or short time-to-market constraints. In light of these considerations, successful ML deployment is rarely a simple "port" of a software model; it requires codesign, where model structure, numerical precision, and hardware architecture are optimized together under stringent area and power budgets.

This section highlights complementary ASIC approaches to edge intelligence, spanning various ML architectures and objectives (filtering, data compression, feature extraction, etc.), and covers both current and future experimental implementations. Emerging open-source electronic design automation (EDA) and automation ecosystems are noted as ways to broaden access to ASIC prototyping and accelerate iteration on ML-centric front-end designs.

Codesign of artificial neural networks (NNs) for real-time feature extraction in AIML65P1 ASIC AIML65P1 [12] is a smart readout ASIC, developed at BNL to address the growing data challenges in modern physics experiments. AIML65P1 is fabricated in a 65-nm CMOS process and integrates 23 independent channels, each containing a complete analog and digital processing chain. The chip layout and architecture are shown in Fig. 2. At the front end, a charge-sensitive amplifier followed by a third-order analog antialiasing filter shapes the detector signal [13]. A 12-bit hybrid analog-to-digital converter (ADC) [14] then digitizes the waveform, feeding a digital back end that includes a programmable finite impulse response shaping filter, waveform alignment, and baseline subtraction circuitry. The digital signal processing blocks are followed by an embedded artificial neural network (ANN), which is a two-layer multilayer perceptron network with programmable weights and biases that can be configured for regression (e.g., pulse amplitude and decay-time estimation) or classification (e.g., pulse shape discrimination or signal/noise separation) [15]. The ANN enables per-channel intelligent processing, effectively functioning as an edge AI engine within each readout channel. Further, the deployment of trainable ANNs partially mitigates the rigidity of ASIC electronics compared to FPGAs.

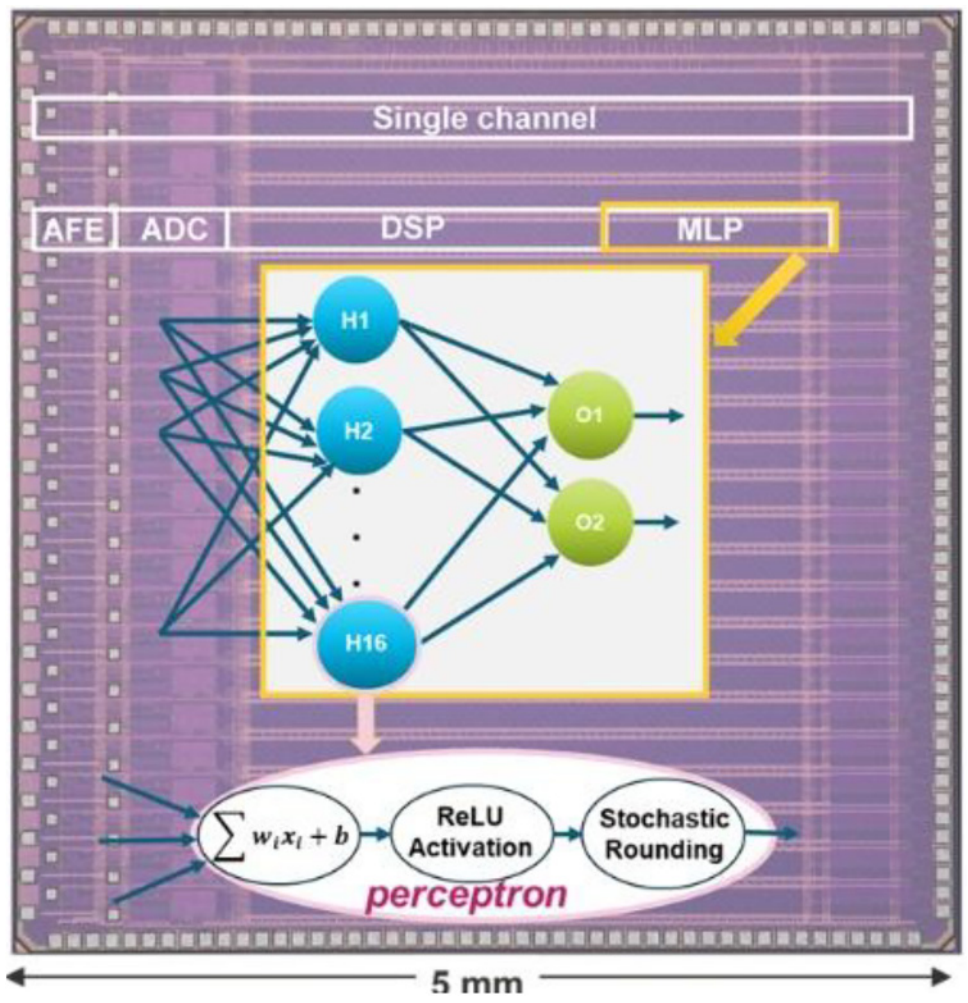

FIG. 2. AIML65P1 ASIC layout and architecture.

To support flexible on-chip learning and deployment, the ANN parameters were quantized to 8 bits through an optimization pipeline. The ASIC additionally implements stochastic rounding [16] in hardware, enabling *in situ* learning and ensuring unbiased gradient behavior even under quantized arithmetic—a crucial requirement for effective training with low-precision weights, activations, and partial sums. Each channel contains nearly 2000 programmable bits, accessible via an $I^2C$ interface. The chip has been fabricated and is currently undergoing testing and performance characterization.

A major component of this work is the codesign strategy that guided the ANN architecture [17]. Because area and power budgets in front-end ASICs are extremely constrained, neural networks cannot be overdesigned; the ML architecture must be co-optimized with the hardware implementation. The team developed an automated, hardware-aware codesign framework that uses multiobjective Bayesian optimization to explore the joint space of ANN parameters (depth, width, bit precision) and synthesis strategies. Importantly, the optimization loop incorporated actual synthesis results rather than analytical proxies, enabling accurate evaluation of area, power, and timing. Over 22 000 networks were trained and synthesized to map the Pareto frontier across accuracy, area, and latency. From this exploration, key design insights emerged: For in-pixel or in-channel implementations, two-layer networks provide the best trade-off; quantization must be carefully allocated across layers; and hardware-aware training is vital for high-precision regression tasks.

In this way, the AIML65P1 ASIC provides a key example of next-generation AI-enabled front ends by integrating low-noise analog front ends, early digitization, nonlinear digital signal processing, embedded machine learning, and a rigorous codesign methodology. The design philosophy centers on shifting signal processing into the digital domain at the earliest feasible stage, enabling real-time intelligent data reduction, adaptive system behavior, and ultimately improved scientific discovery potential.

*End cap concentrator trigger ASIC for CMS high-granularity calorimeter.* At the high-luminosity Large Hadron Collider (HL-LHC), to cope with the increasing radiation in the forward region and with the larger number of particles from pileup, the CMS Collaboration will install a high-granularity calorimeter (HGCAL) in the end caps of the detector [18]. It represents one of the most data-intensive calorimeters ever built, estimated to produce raw rates of ∼5 Pb/s (per end cap), comparable to the global internet traffic rate [19]. To handle this unprecedented data volume, the front-end electronics must perform substantial on-detector data reduction and feature extraction before transmission off detector.

Data from the HGCAL will be used in the CMS level-1 trigger (L1T). Together with the information on tracks, available for the first time at the HL-LHC, the high granularity will facilitate the usage of particle flow [20] in L1T, increasing the ability to deal with particles from pileup.

As the first stage of HGCAL's trigger system, the goal of the end cap concentrator trigger (ECON-T) chip is to reduce data volume before transmission off detector. It is an on-detector radiation-tolerant digital ASIC that receives information on trigger cells (TCs), an aggregation of raw signals from multiple detector cells, computed at an earlier stage of the on-detector data processing. The ECON-T compresses and transmits TC information off detector under stringent latency ($< 250$ ns) and power ($< 2.5$ mW/channel) constraints. To do that, the ECON-T runs different rule-based data reduction algorithms. It also supports a hardware implementation of an NN encoder for data compression [21].

ECON-T is the first radiation-tolerant, digital ASIC to provide ML inference directly at the detector edge for particle physics applications. The NN encoder itself operates with latency less than 50 ns and power dissipation less than 50 mW, which corresponds to 1.2 nJ per inference.

Integrating ML into such a device requires careful co-optimization between algorithmic design and ASIC implementation, balancing physics performance against area, power, and timing requirements. Quantization, fixed-point arithmetic, and model pruning are essential to meeting these constraints while maintaining low-latency operation.

Ongoing R&D efforts explore the use of a conditional autoencoder (CAE) [22] architecture within the ECON-T ASIC. The CAE extends previous autoencoder-based compression strategies [23] by conditioning the latent representation on sensor-specific metadata—such as sensor type and geometric location in the detector—and by training models for varying bandwidth constraints. This approach aims to achieve a physics-aware compression that adapts to the heterogeneous detector configuration while respecting the tight throughput requirements of the trigger system. Although results are forthcoming, this work demonstrates a broader paradigm shift toward intelligent, on-detector processing, where codesigned ML algorithms and custom ASIC architectures jointly enable real-time data reduction for next-generation particle-physics experiments.

*SmartPix: in-pixel integration of signal processing and ML data filtering for HGCAL.* The data volume challenge of modern collider experiments is exemplified in the HGCAL [18] (described in detail in Sec. III A 3). The CMS L1T correlator uses information from CMS HGCAL clusters to form high-level physics quantities and reduce data bandwidth. However, pixel information is not used at level-1 due to latency and throughput limits, reducing sensitivity to displaced signatures and low-momentum ($p_{\mathrm{T}}$) clusters, which should be rejected by the trigger as they do not carry relevant physics information. Here, a readout ASIC prototype is presented that integrates charge processing and machine-learning data filtering directly in-pixel, enabling on-detector data reduction at the bunch-crossing scale.

*Architecture.* The ASIC, fabricated in 28 nm bulk CMOS, implements two $32 \times 8$ superpixel matrices at 25 μm pitch. The overall architecture is illustrated in Fig. 3 and described in detail in Refs. [24–26]. Each pixel comprises a charge-sensitive preamplifier followed by a 2-bit flash ADC operating at 40 MHz. Comparator thresholds $V_{\mathrm{th}[0,1,2]}$ define transition levels and are adjustable off chip. Pixel outputs are latched, encoded, and projected into 16 compressed row buses, preserving spatial information while reducing bit width by an order of magnitude. A fully parallel two-layer neural network, implemented purely in combinational logic, classifies local clusters without cycle-to-cycle latency. The network is derived from QKeras quantization and synthesized via `hls4ml` [27,28] (see Sec. II B) and Siemens Catapult high-level synthesis (HLS) [29].

*Analog performance.* Charge is injected via programmable in-matrix capacitors to emulate realistic clusters. Repeated $S$-curve scans across injected charge extract both threshold dispersion and equivalent noise charge (ENC). The prototype exhibits ENC ∼58 $e^-$ with an average conversion gain of 58.5 μV/$e^-$. These values confirm preamplifier stability and comparator resolution and align with expectations from simulation.

*On-chip ML filtering.* Cluster profiles derived from CMS simulation were injected into the pixel matrix and propagated through the digital chain. Across 104 test patterns, ASIC outputs match register transfer level (RTL) at 99.86%, demonstrating robust inference, correct timing, and a fully functioning DAQ readout.

The results demonstrate that in-pixel charge processing combined with embedded machine learning can be realized in physical silicon. By performing data reduction at the earliest stage of readout, this approach offers a promising path to overcome bandwidth and latency limitations that currently restrict the use of tracker information in real-time decisions. These developments open the door to data-driven readout strategies for future high-energy physics experiments.

While final performance studies are ongoing, this prototype already exemplifies a broader shift in detector design: Intelligence is being pushed directly into the front end. By co-optimizing machine-learning models and custom ASIC architectures, pixel detectors can perform real-time feature extraction and data suppression at the point of capture, easing bandwidth and latency bottlenecks that have historically limited physics sensitivity. Equally important, the development of this chip has strengthened a model of industry-laboratory codesign—most notably through our collaboration with Siemens EDA—which has been essential for advancing ML-centric hardware design methodologies. Together, these efforts chart a path toward scalable,

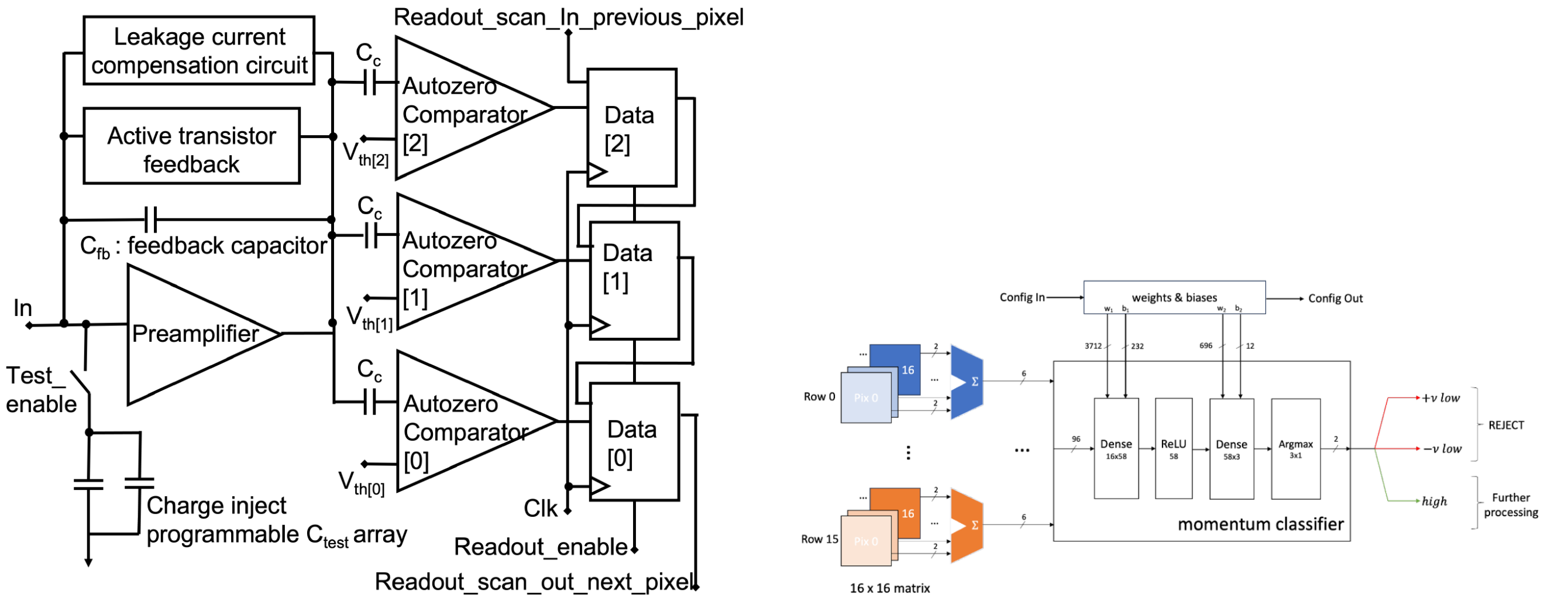

FIG. 3. Overview of the SmartPix architecture: Analog pixel front end (left) and digital projection with embedded in-pixel classifier (right).

data-driven readout systems for next-generation collider experiments.

*Open-source ASIC design: tools, automation, and test bed infrastructure.* Open-source EDA tools eliminate commercial licensing costs (often exceeding $1M annually per institution), democratizing ASIC development for academic collaborations. The OpenROAD project [30,31] provides complete RTL-to-graphic database system II (GDSII) flows with multiple open process design kits (PDKs): predictive nodes (FreePDK 45 nm, ASAP 7 nm) for early-stage research and production-ready nodes (SkyWater 130 nm, GlobalFoundries 180 nm) for fabrication. The SODA Synthesizer exemplifies this fully open approach, providing Python-to-GDSII compilation through multilevel intermediate representation (MLIR)-based front end, Bambu HLS back end, and OpenROAD physical design [32], enabling automated synthesis of arbitrary Python algorithms including data preprocessing, control logic, ML inference, and data formatting stages.

Building on these open-source tools, the end-to-end codesign for performance, energy fficiency, and security in AI-enabled computational science (ENCODE) initiative addresses the next challenge: reducing chip design complexity and turnaround time [33]. ENCODE's intelligent hardware design thrust researches ML-driven optimization algorithms and scalable co-optimization methodologies to drastically reduce design time for advanced packaging technologies including chiplet-based systems and three-dimensional (3D) integration. By leveraging tools like SODA and OpenROAD within automated design flows, ENCODE enables rapid exploration of design spaces that would be prohibitively time-consuming with manual approaches. The advanced architecture test bed thrust establishes verification/validation methodologies and physical test bed capabilities for presilicon characterization and postsilicon validation. The expectation is that this will soon enable access to low-cost fabrication for prototyping research ideas.

For HEP applications, this integrated ecosystem democratizes custom ASIC development: from detector-specific readout ASICs and ML inference accelerators to radiation-hardened designs requiring rapid design iteration and validation. Open-source flows facilitate integration of radiation-hardening techniques (triple modular redundancy, error correction, specialized standard cell libraries) without proprietary tool restrictions, while automated design methodologies reduce expertise barriers and development cycles. Combined test bed access addresses the fabrication cost and complexity challenges that have historically limited academic ASIC research, enabling HEP collaborations to prototype custom designs for future detector systems considering total ionizing dose (TID), single-event effects (SEE), and thermal management from early design stages.

### B. FPGAs

Software frameworks such as Keras, TensorFlow, and PyTorch have revolutionized model development by enabling rapid prototyping, expressive architectures, and efficient training on GPUs, enabling highly parametrized models optimized primarily for accuracy with abundant resources. In contrast, deployment platforms such as FPGAs operate under strict constraints on logic utilization, on-chip memory, bandwidth, latency, and energy consumption. This mismatch frequently exposes a profound disconnect between software-first model design and hardware feasibility. Simply converting a trained model into a hardware-compatible representation is rarely sufficient; instead, successful deployment requires a fundamental shift in how models are conceived, optimized, and validated.

Addressing this challenge requires a hardware-aware design philosophy in which model capacity, parameter efficiency, and inference cost are considered alongside accuracy from the earliest stages of development. Excessively large models often contain significant redundancy, and careful analysis of model capacity can reveal opportunities to enhance parameter efficiency (e.g., pruning) while mitigating overfitting. Quantization can also dramatically reduce memory footprint and computational complexity with minimal impact on accuracy.

FPGA chips are the technology of choice in HEP experiments for real-time trigger systems as well as situations where electronic controls need to be reconfigurable. The configurable integrated circuit allows for customized parallel operations on real-time data. They allow for more parallelism in algorithm execution compared to CPU and GPU at the expense of added complexity from the circuitry involved in implementation. They also allow for more flexibility as reprogrammable devices compared to ASIC chips as the block resources, such as look-up tables (LUTs), flip-flops (FFs), and digital signal processors (DSPs), can be dynamically "field programmed" through the interconnects. The number of logic gates has risen exponentially, from thousands in the 1980s to tens of millions today, as well as specialized compute units, such as Advanced Micro Devices, Inc. (AMD)'s AI engine tiles with specific setups to optimize DSP usage. As discussed later in this document, there are approaches that expand and create hybrid systems where FPGA mingles with technologies, such as embedded systems. Nevertheless, FPGAs are and will continue to be a staple of high-energy physics.

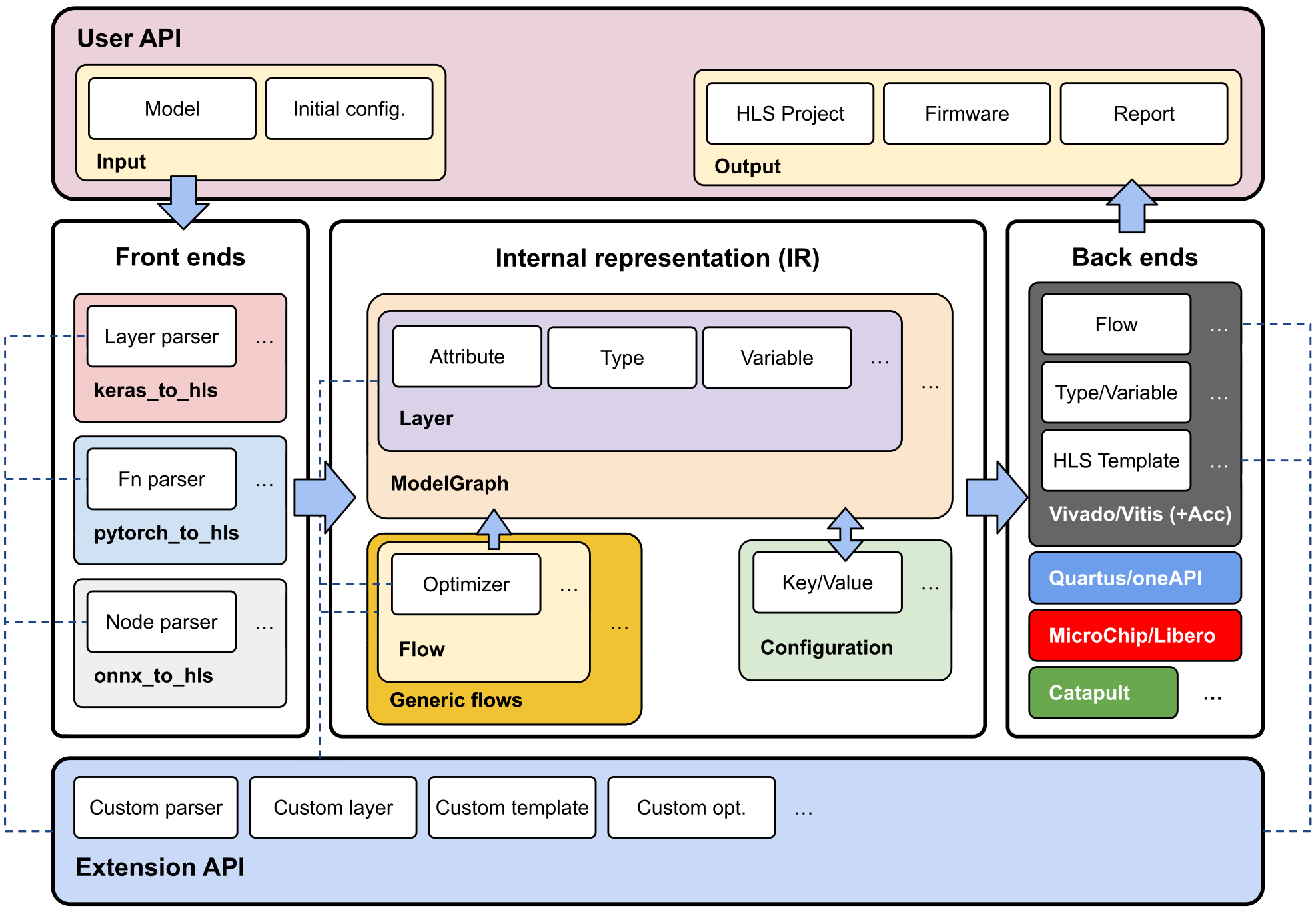

FIG. 4. `hls4ml` model conversion flow; reproduced from Ref. [34].

### *1.* `hls4ml`

Machine learning has become ubiquitous in edge and real-time systems, creating a growing demand for efficient, low-latency hardware capable of meeting strict latency, resource, and power constraints. While FPGAs and ASICs deliver high performance and energy efficiency for such workloads, their development remains complex and requires significant expertise. `hls4ml` [28,34] is a flexible, open-source platform that addresses these hardware development challenges by converting high-level NNs from PyTorch, Keras, or similar frameworks into HLS kernels suitable for deployment on FPGAs and ASICs. More discussion of open-source tools for the ML-HEQUPP scope can be found in Sec. II D 3.

`hls4ml` is structured in a modular fashion, following the design of modern compilers. The overall model conversion and compilation flow is illustrated in Fig. 4. Models from supported frameworks are first translated into a unified intermediate representation (IR), which captures the computation graph and layer configurations independently of the model framework or target hardware. The IR then undergoes a series of hardware-aware optimizations, including precision adjustment, data layout transformation, and operator fusion, producing a computation graph optimized for dataflow inference on the target platform. The final hardware design is obtained by composing templated implementations of the optimized operators, which can be configured to balance between latency and resource consumption. When synthesizing the hardware, `hls4ml` abstracts the interaction with EDA tools such as Vivado and Quartus, automating synthesis, placement, and routing, allowing users to generate a deployable bitstream in a few lines of Python code.

To reduce resource utilization and inference latency, `hls4ml` emphasizes the use of quantization and low-precision arithmetic. `hls4ml` can directly parse models from popular quantization-aware training (QAT) libraries, including QKeras [35] and Brevitas. Complementary tools developed by the same community, such as heterogeneous quantization [36] and DSP-aware pruning [37], provide hardware-aware compression techniques that can further reduce resources when targeting FPGAs. Finally, `hls4ml` allows users to configure precision heterogeneously on a per-layer and per-variable basis, enabling fine-grained post-training quantization that tailors numerical precision to the sensitivity of each operation.

`hls4ml` supports a broad set of neural network architectures, including multilayer perceptrons, convolutional networks, recurrent networks, and more recently, transformers. While most standard layers are fully supported, custom operations can be incorporated through the extension application programming interface (API), which allows users to provide custom HLS implementations of specific operators. `hls4ml` differentiates itself from other hardware acceleration platforms through its support for multiple front ends (i.e., Keras, PyTorch, and Quantized ONNX (QONNX) [38]), multiple FPGA back ends (e.g., Vitis HLS and Intel oneAPI), and an ASIC back end through Siemens HLS. More recently, `hls4ml` has been extended with support for Google XLS and the Libero back end for radiation-hard FPGAs.

Targeting FPGA deployments, `hls4ml` has been used to accelerate ML in high-energy physics experiments [39,40], spacecraft [41] and autonomous vehicles [42], quantum computing [43], and heart signal monitoring [44]. In addition, `hls4ml` has been used to generate ASIC designs for a number of on-detector front-end applications, including data compression using an NN encoder for the CMS HGCAL [21]. Similarly, a "smart pixel" readout integrated circuit (ROIC) was prototyped using `hls4ml` with an implementation of an NN to process 256 pixels to determine the momentum of the particle track [45,46]. These front-end applications underscore the unique capabilities of `hls4ml` and the need for hardware-model codesign for the next generation of experiments.

### 2. *cgra4ml*

CGRA4ML [47] is an open-source, modular hardware/software codesign framework designed to implement deep neural networks (DNNs) for scientific edge computing applications. While existing tools like `hls4ml` [28,34] and FINN [48] excel at small, low-latency models, they struggle to scale to deeper networks due to their layer-by-layer dataflow architecture, which consumes increasing hardware resources as model depth grows. CGRA4ML addresses this by employing a lightweight and highly parametrizable coarse-grained reconfigurable array (CGRA), shown in Fig. 5 that reuses processing elements (PEs) across layers, enabling deployment of larger models such as ResNet-50 and PointNet on FPGAs and ASICs.

The framework provides a holistic, full-stack infrastructure, as shown in Fig. 6. Central to this flow is a modular and extensible IR based on the concept of "bundles." These are deterministic units of execution that encapsulate groups of DNN layers, facilitating clean hardware/software partitioning. Scientists define and train quantized models using a Python API based on QKeras, which supports sub-8-bit quantization-aware training. The front end decomposes these models into these bundles to bridge the gap between high-level models and hardware. Compute-intensive operations, such as convolutions and matrix multiplications, are offloaded to the CGRA, while complex pointwise operations and edge cases are handled by the host CPU.

Architecturally, CGRA4ML generates synthesizable, vendor-agnostic SystemVerilog RTL. The CGRA engine (Fig. 5) features a two-dimensional (2D) array of PEs where columns can dynamically regroup at run-time to process varying convolution kernel widths and output channels. It includes a weight cache with ping-pong buffers for full throughput and a pixel shifter that exploits vertical data locality to reduce required input bandwidth. A key contribution is the automated generation of C run-time firmware that manages the unified dataflow and coordinates software-side tensor operations across diverse processors like advanced RISC machines (ARM), reduced instruction set computer five (RISC-V) and x86. For verification, the framework uses a, Direct Programming Interface for C (DPI-C)-enabled suite that runs the production firmware in-the-loop with the RTL simulation. This allows users to emulate real-world conditions, such as randomized memory congestion, to ensure bit-accurate parity between the software model and physical implementation.

CGRA4ML has been experimentally validated on AMD-Xilinx FPGAs and Cadence ASIC flows. Iso-resource microbenchmarking on a Pynq-Z2 at 150 MHz (Fig. 7) shows that while `hls4ml` excels at small workloads, fitting larger workloads requires an exponentially increasing reuse factor, and a $16 \times 16$ CGRA4ML array maintains stable latency and higher throughput when extreme reuse is required to meet resource limits. Deployment is further streamlined via automated tool command language (TCL) scripts that handle synthesis, floorplanning, and SoC integration, facilitating a rapid transition from research prototypes to silicon realization. CGRA4ML also provides an integration with Ibex SoC [49], a popular RISC-V platform. By filling the gap between small-model dataflow tools and traditional 8-bit AI accelerators, CGRA4ML enables the scientific community to move more sophisticated processing to the extreme edge hardware.

### 3. *SLAC neural network language*

The SLAC neural network library (SNL) [50–54] is a lightweight, high-performance ML inference framework tailored for deployment on FPGAs in low-latency, high-throughput environments. SNL bridges the gap between widely used training platforms such as Keras, TensorFlow, and PyTorch and the hardware constraints of edge devices by providing a seamless workflow from model training to FPGA inference (Fig. 8). Developed using HLS in C++, SNL provides a Keras-inspired API and supports neural networks with tens of thousands of parameters, depending on architectural complexity and available FPGA resources. With end-to-end latencies ranging from microseconds to a few milliseconds, SNL offers a pipelined, streaming dataflow architecture that balances low-latency processing with high frame-rate requirements.

SNL and `hls4ml` are best understood as complementary frameworks, as each targets a different point in the design space. `hls4ml` fuses a model's weights and biases directly into the FPGA fabric, producing a customized implementation tailored to a specific trained network. This enables extremely fast inference for small models (ideal for ultra-low-latency applications such as those in collider physics) and gives the user fine-grained control over optimizations such as pruning, since the implementation is specialized to a fixed set of parameters. The trade-off is that any change to the model, including retraining, requires regenerating and resynthesizing the design, which can introduce hours or days of engineering effort and carries the risk of timing closure failures or the network no longer fitting on the device.

SNL provides a complementary benefit by enabling dynamic reloading of neural network weights and biases without requiring resynthesis or reconfiguration, facilitating rapid model iteration and adaptive inference strategies. This targets models in the microsecond range and somewhat larger architectures, and crucially allows a network to be retrained without resynthesizing the FPGA design. Once a model is mapped to the device, there is no risk of it growing beyond the available resources or reintroducing timing failures on retraining, which provides a high degree of deployment stability and saves significant engineering time. This flexibility comes with its own trade-off: Because the optimal pruning of a model can change with each retraining, SNL does not support pruning whereas hls4ml does, owing to its fixed, network-specific implementation. In summary, hls4ml excels at highly optimized, customized deployment of small and fast networks, while SNL prioritizes deployment stability and the ability to host generally larger models under evolving experimental conditions.

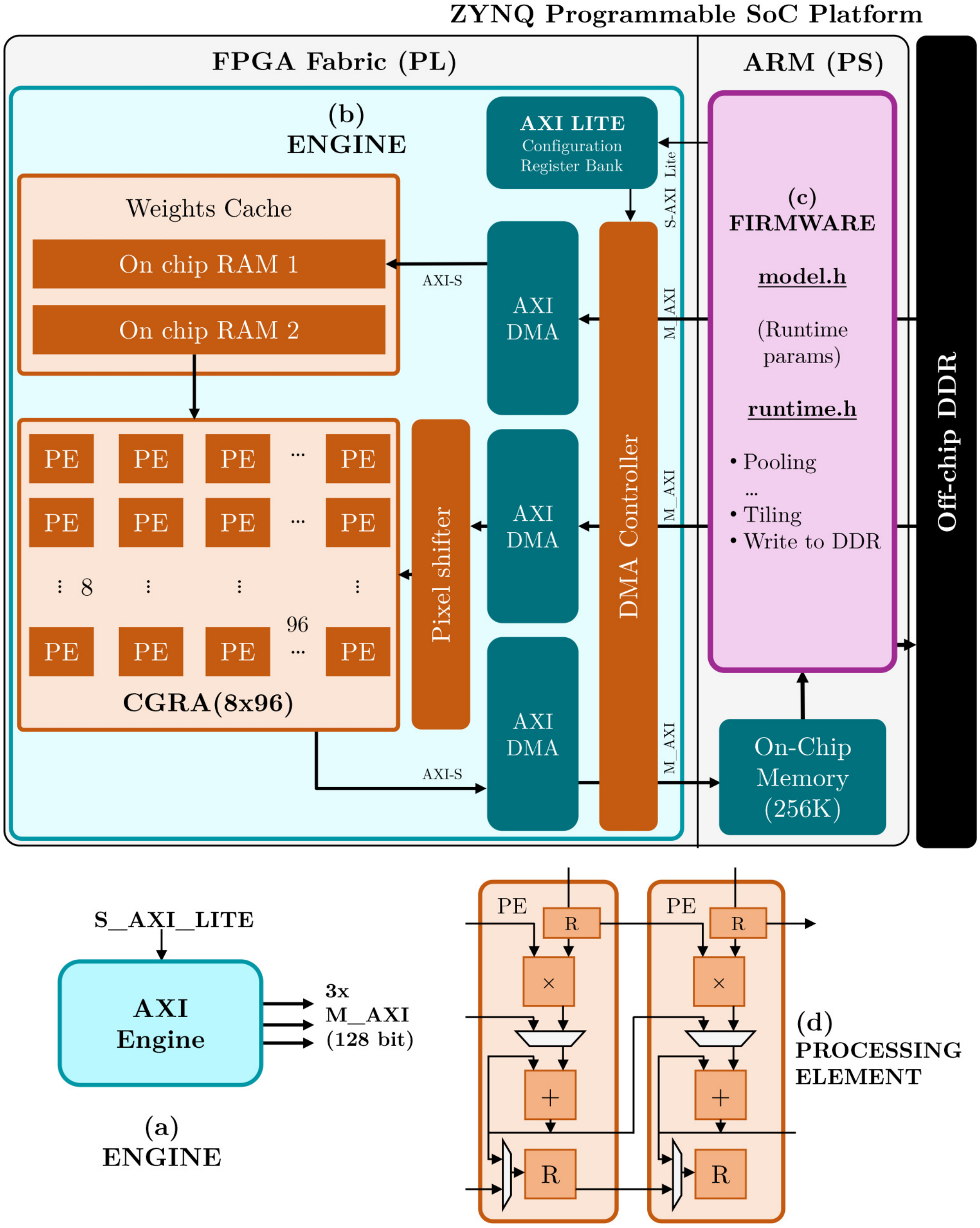

FIG. 5. The CGRA4ML architecture features a CGRA engine with three 128-bit AXI manager ports and parametrizable PEs, bit widths, and weights cache. The PEs are kept small to increase compute density. CGRA4ML offloads compute-heavy tasks to hardware while emitting ARM-based software for pixelwise operations and edge cases.

While pruning is a widely used technique for removing nonessential parameters, it is currently not supported within SNL due to the challenges associated with efficiently implementing sparse representations on FPGA hardware. In practice, comparable benefits can often be achieved through careful architectural simplification and capacity control. Additional gains can be realized through preprocessing at the data source, knowledge distillation, and focusing on inherently lightweight and resource-efficient algorithms, rather than retrofitting for deployment.

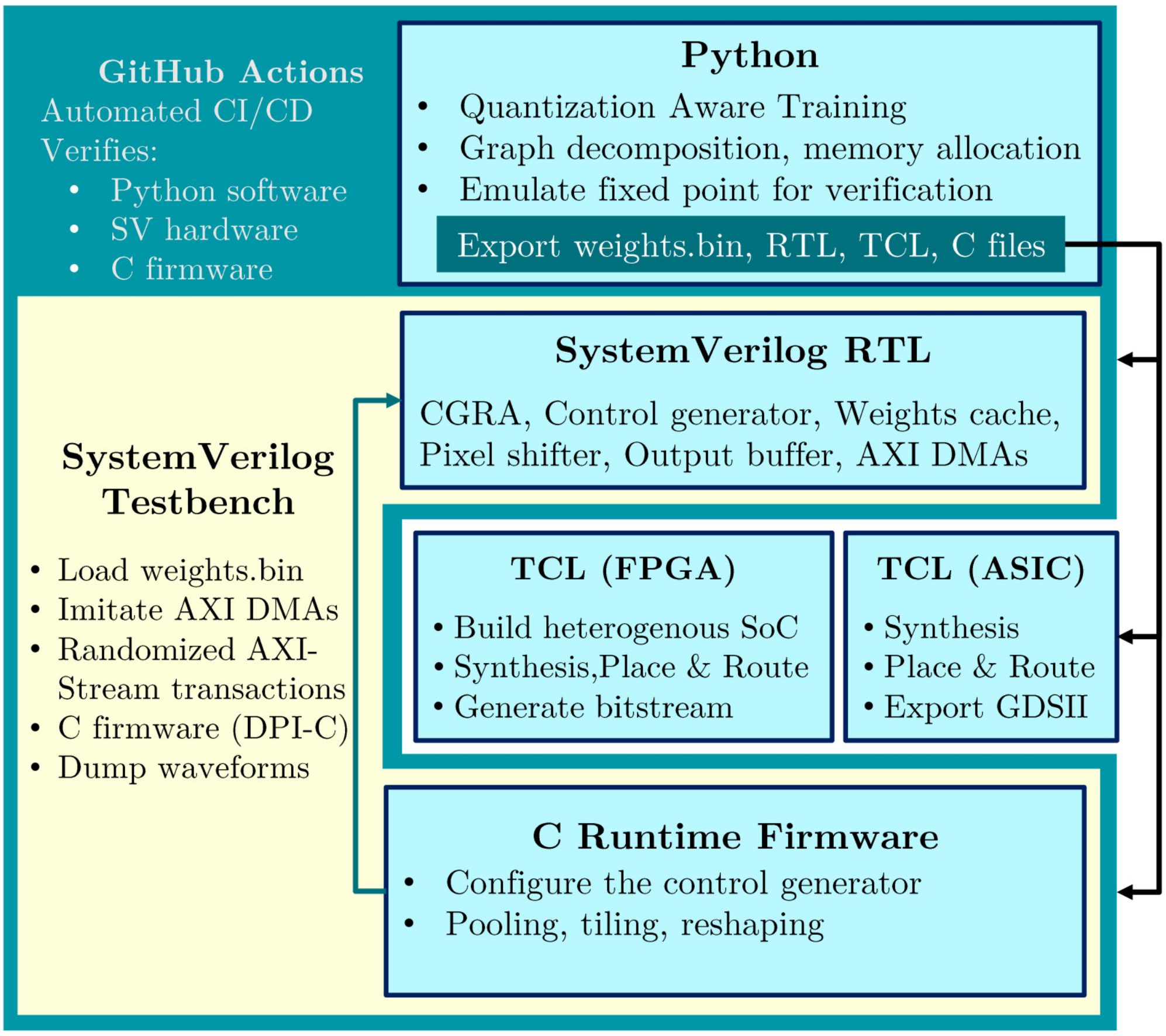

FIG. 6. Python API utilizes QKeras to extract fixed-point tensors, generating binary weights, model specifications, vendor-agnostic SystemVerilog RTL, C firmware, and TCL toolflows. A DPI-C verification suite and continuous integration and continuous delivery/deployment (CI/CD) pipeline validate the integrated hardware/software system, emulating real-world memory congestion.

A compelling use case of SNL is provided by the LCLS-II SpeckleNN [51] application. In this use case, neural networks are employed to analyze diffraction patterns generated by molecular imaging experiments and to decide, in real time, whether individual frames contain scientifically valuable information that should be retained or vetoed. The original GPU-based SpeckleNN model achieved 94% accuracy and 98% data reduction but, with 5.5 million parameters, was too large for FPGA deployment at the detector edge. Using SNL and a hardware-aware optimization process, the model was reengineered to 65 000 parameters, maintaining 91% accuracy while improving compression to 98.8%, and enabling FPGA deployment with microsecond latency. This example illustrates how modest sacrifices in accuracy can yield substantial gains in deployability and overall system performance. Similar efforts are underway in the domain of collider physics, where expensive transformer-based learning for classification of beyond-standard-model (BSM) event signatures can be replaced by lightweight multilayer perceptron (MLP) mixer [55] models, offering similar performance while eliminating expensive operations.

#### *4. Decision forests*

Decision forests (DF) (e.g., ensembles of decision trees) are a class of ML models that maintain a good balance of performance and efficient inference in hardware. While various flavors of NN architecture excel on low-level data, Decision forests have remained state of the art in tasks on tabular data [56,57], only recently being superseded in performance by transformer architectures. At the same time, decision forests can be extremely lightweight ML models, making them suitable for constrained inference for edge applications. Decision forests have been shown to be robust against irrelevant features, making them achieve good out-of-the-box performance, and also able to perform well in anomaly detection (AD) tasks [58]. In the domain of edge ML brought about by tools such as `hls4ml`, CGRA4ML, and SNAC-PAC, decision forests especially shine in extremely constrained applications. In these cases, resource usage of 100 FPGA LUTs, latency of ~10 ns, and power consumption in the milliwatt regime are achievable.

Tools for producing hardware synthesizable products from trained decision forest models include `fwXmachina` and `conifer` [59]. The `conifer` tool is designed as a source-to-source compiler. Converters are provided for a variety of popular DF training libraries, such as scikit-learn, xgboost, yggdrasil decision forests, and others. Across these libraries, support is available for classification, regression, and AD (isolation forest) tasks. From the converted model, hardware projects can be produced using implementations in Vitis HLS or VHSIC hardware description language (VHDL). These implementations target ultralow latency by fully unrolling the DF logic and exploring all decision paths in parallel using Boolean logic. The Vitis HLS implementation targets AMD/Xilinx devices, and the VHDL implementation enables support for other FPGA vendors and even ASIC toolflows. Additional infrastructure is provided for producing out-of-the-box FPGA accelerators with Advanced eXtensible Interface (AXI) interfaces for high-throughput, low-latency-boosted decision tree (BDT) inference offloaded from a host CPU. Figure 9 shows an FPGA floorplan of an accelerated model on a pynq-z2 board, also highlighting the parallel logic of the decision forest. A third, reprogrammable, implementation is provided, termed the "forest processing unit." This design is implemented with Vitis HLS, but implements a generic DF processor, which exposes memories that encode a specific DF model onto an interface such that DF models can be loaded and exchanged at run-time. A C++ and Python implementation are also provided for software emulation and validation.

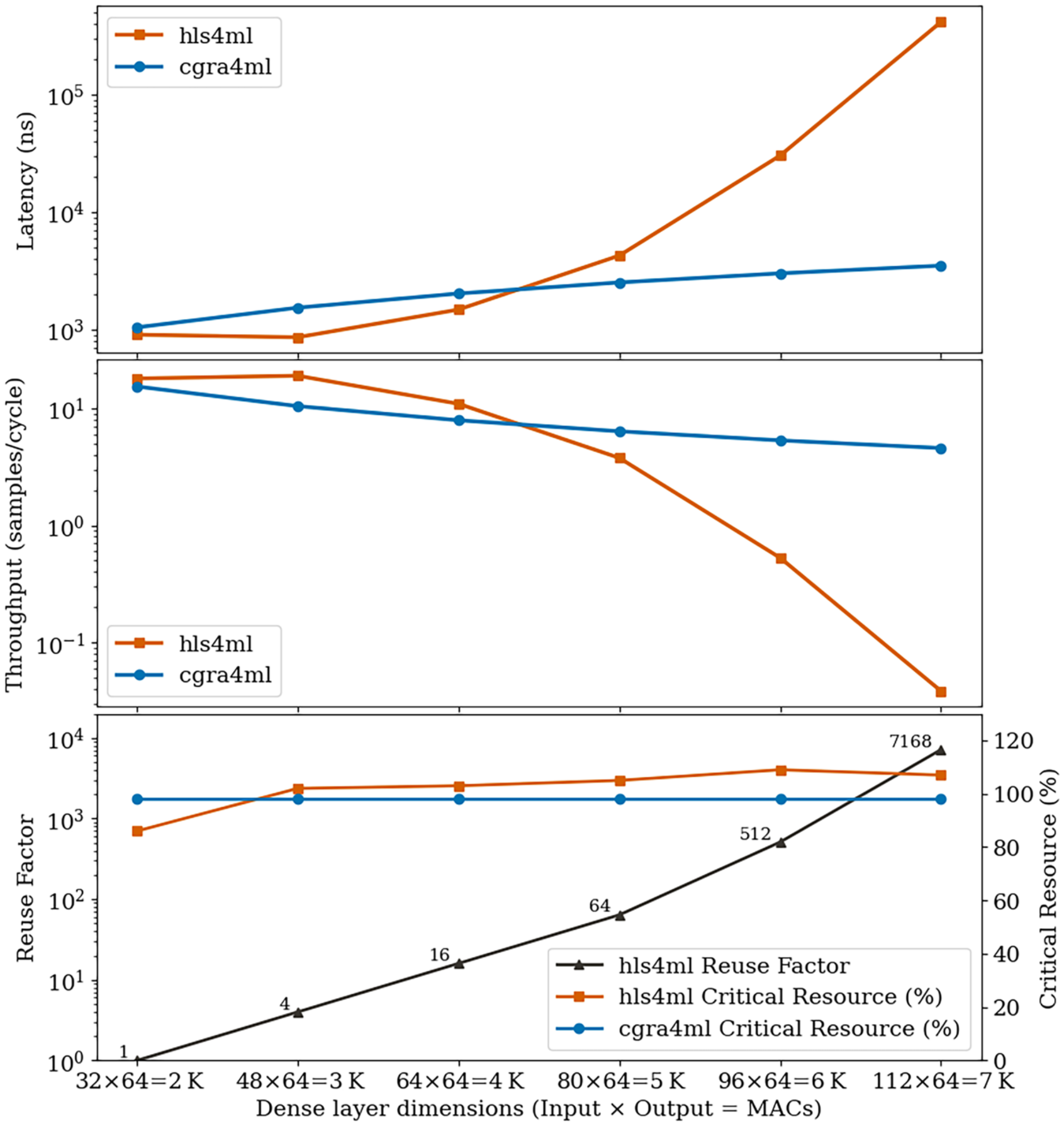

FIG. 7. Iso-resource microbenchmarking on Pynq-Z2 compares a $16 \times 16$ CGRA4ML array against `hls4ml`. While `hls4ml` requires exponential reuse factors to fit larger workloads, causing performance to suffer, CGRA4ML performs better at high-reuse conditions by maintaining stable latency and higher throughput.

The `fwXmachina` framework is an open-source toolchain for translating BDT models into ultra-low-latency FPGA implementations suitable for real-time inference in high-energy physics environments such as hardware trigger systems [60]. Originally developed to support nanosecond-scale classification tasks—achieving O(10 ns) latency with minimal FPGA resource usage by restructuring BDT logic and exploiting a parallel decision paths architecture, `fwXmachina` has since been extended to support deep boosted trees for regression [61] and interpretable decision-tree-based autoencoders for AD [62] in trigger systems. Figure 10 shows the parallel decision path structure used for the FPGA implementation, where all decision paths through each decision tree are evaluated in parallel. The framework integrates with standard ML training workflows, optimizing bit-precision per input and BDT structure for both physics performance and hardware efficiency, then synthesizing the design via HLS into firmware that meets stringent timing constraints typical of level-1 triggers at collider experiments. Benchmarks on physics tasks—from vector boson fusion classification to missing transverse momentum estimation and exotic decay AD—demonstrate competitive physics performance with latency and resource footprints tailored for stringent embedded contexts, providing a complementary tool to other edge ML compilers in the particle physics ecosystem.

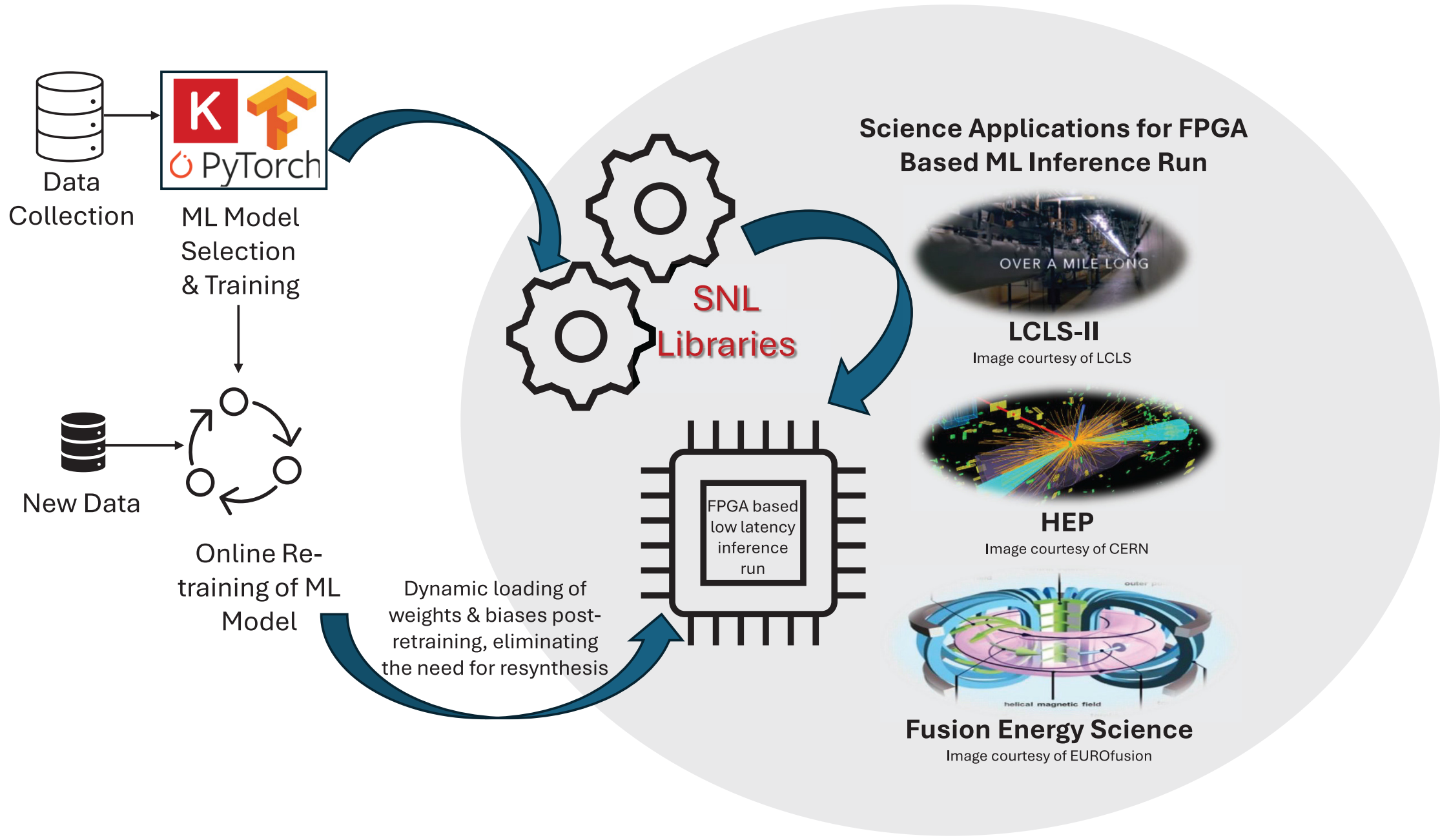

FIG. 8. SNL high-level design flow [51].

Applications of decision forests in trigger and data acquisition systems at modern collider experiments go back to early

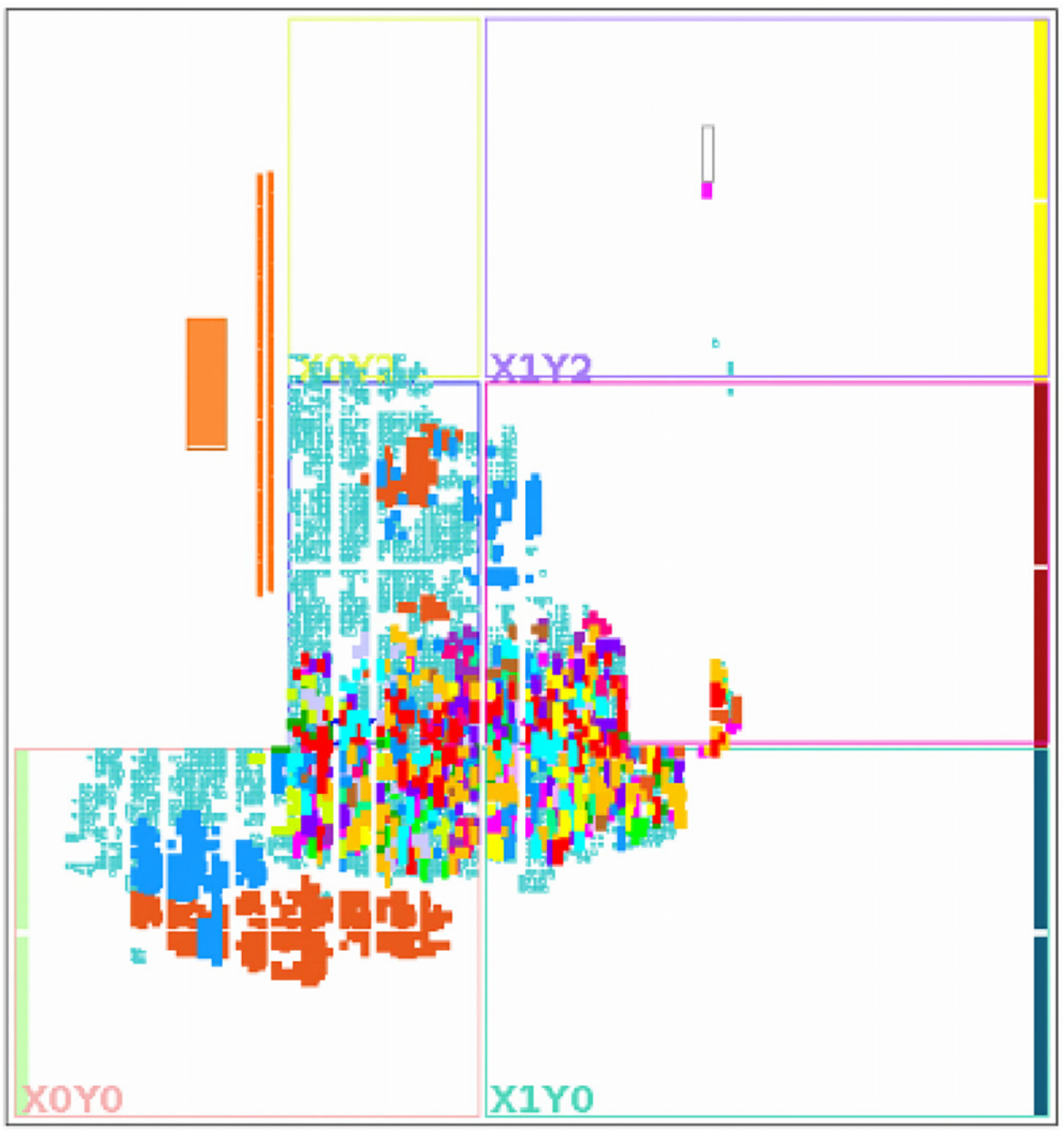

FIG. 9. FPGA floorplan of a boosted decision tree with 50 trees with a maximum depth of 3, implemented with `conifer` for an AMD pynq-z2 device. The small multicolored regions of cells are each of the 50 decision trees, mapped onto independent logic cells for parallel evaluation. The large orange and blue areas are infrastructure for input-output ports—only provided for acceleration workflows.

LHC data taking, where the LHCb experiment high-level trigger made extensive use of so-called "bonsai" boosted decision trees, highlighting their robustness and fast inference [63]. In hardware triggering, BDTs were used in the CMS muon trigger system, using a custom external low-latency look-up table module that enumerated the BDT over address space of the features [64]. This model was able to reduce the trigger background rate by a factor of 3 without compromising the signal efficiency.

More recent applications have deployed DFs directly in FPGA logic, as supported by `conifer` and `fwXmachina`. The ATLAS experiment has used a BDT during LHC Run 3 to select hadronically decaying $\tau$ leptons using information from the calorimeter [65]. This algorithm is more resource efficient than the previous heuristic algorithm and provides a valuable platform to study the real-world deployment of ML in detector data taking.

Both the ATLAS and CMS experiments plan extensive use of ML in their upgraded hardware trigger systems for the HL-LHC. In the HL-LHC data-taking conditions of a significantly increased number of simultaneous proton-proton collisions (pileup), ML will provide a powerful handle to maintain signal efficiency without overwhelming detector readout and data acquisition. The CMS experiment plan to use BDTs to reject fake tracks that are reconstructed from false coincidences of tracker hits in its Level 1 Track Finder development [66]. The BDT algorithm is more performant than basic cuts, while being sufficiently resource, latency, and throughput efficient to fit within the tight system constraints. A BDT will also be used in the reconstruction of electrons in the level-1 trigger, where a calorimeter cluster and charged particle track are linked [67]. After a first loose linking of pairs of calorimeter clusters with tracks, based on kinematic properties, a BDT ultimately identifies the most likely pairings originating from electrons, suppressing background and maintaining high efficiency. In this application, ten instances of a small BDT are required in the densely packed FPGA in order to fit resource, latency, and throughput constraints. Sizeable use of ML is also envisaged at the ATLAS experiment for its HL-LHC upgrade. Several event-selection level tasks have been developed for its Global Event Processor, including classifying events of Higgs bosons produced by vector boson fusion (a priority for HL-LHC data taking) and regressing the missing transverse energy (MET) using BDTs.

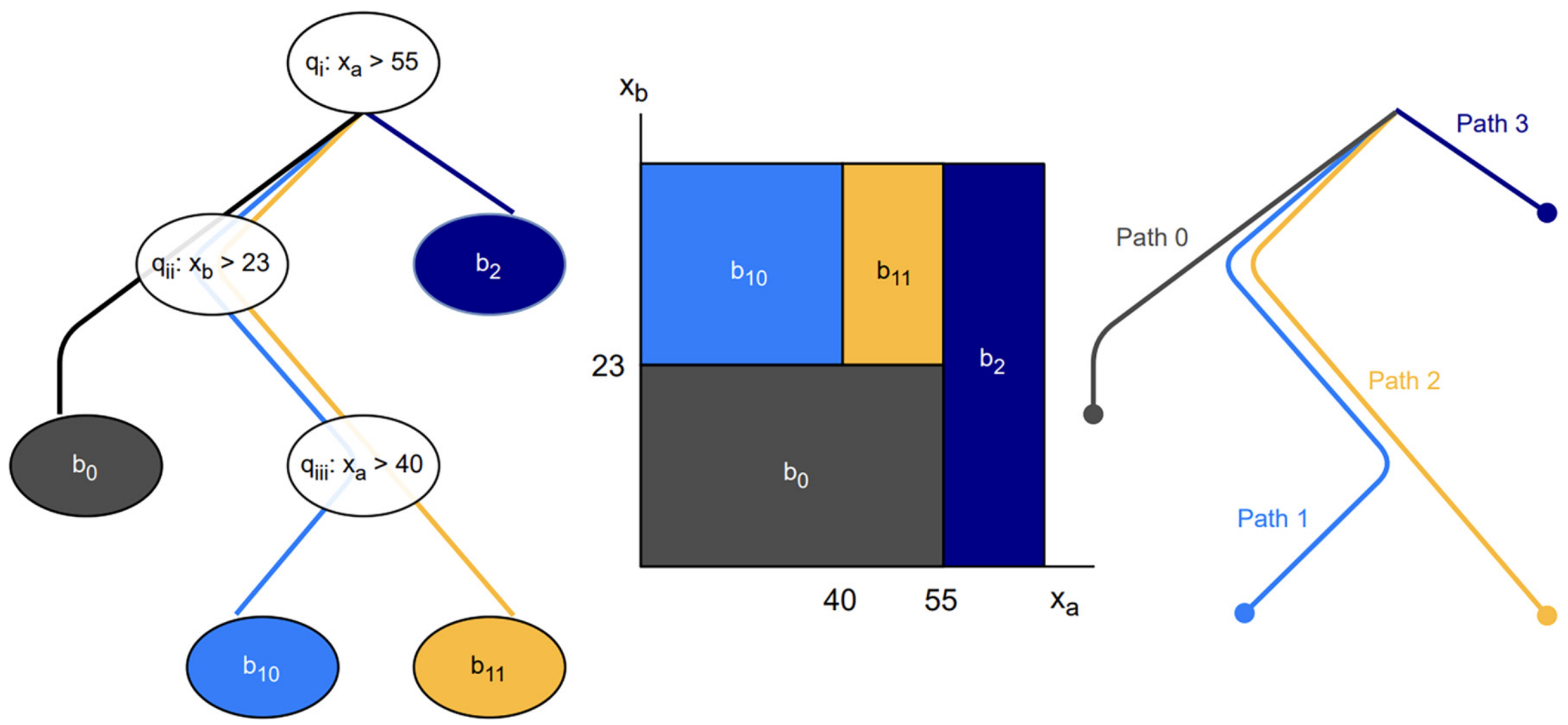

FIG. 10. Visualization of the mapping of decision tree structure onto decision paths in `fwXmachina` [61]. Left: the structure of the decision tree; center: the visualization of the splits in feature space; right: the decision paths of the tree, which are evaluated in parallel in the FPGA implementation.

The task of AD has also become prominent in online data section, with NN-based anomaly detectors in use in data taking at both CMS and ATLAS during Run 3 (see Sec. III A 4). DF-based AD algorithms are also under study, seeking to combine the robust and hardware-efficient inference they offer with the signal-agnostic target of AD. Methods include quantile regression using BDTs [68], tree-based autoencoders [62], and isolation forest [69], proving that AD with latencies of tens of nanoseconds is achievable.

Having become well established in software and now hardware triggering, use of DFs closer to sensor readout is also emerging as a viable approach to on-detector intelligence. At the Belle II muon CDC system, ML in FPGAs is being explored as a means to reduce the high rate of noise resulting in fake tracks being reconstructed in the level-1 trigger system [70]. Even with a very small model, BDTs emerged as the most performant model at rejecting noise hits. In this case, extreme resource efficiency is required as one FPGA processes measurements from tens of detector channels simultaneously, requiring correspondingly many instances of the ML model operating in parallel. A single decision tree was deployed onto an extremely resource-constrained embedded field programmable gate array (eFPGA)—with only 448 LUTs—as part of a pixel readout chip with on-chip data selection rejecting hits from particles with low transverse momentum [71]. Using a hardware-aware codesign methodology, the eFPGA resource requirement for a task of on-chip classification between neutron/gamma incidences has been explored [72]. Both NN and BDT models were used to explore the trade-off between classification performance, hardware cost, and throughput. BDTs were found to be resource efficient and capable of low-latency and high-throughput inference [72].

In summary, decision forests are robust, efficient ML models highly suited to edge inference in the resource-constrained, low-power, low-latency, and high-throughput regimes. This makes these simple ML models especially relevant for in-hardware inference, such as in hardware trigger systems and in front-end readout chips. The capability to efficiently reject background signals at high throughput will ensure future experiments can read out their desired signal data while maintaining reasonable data rates in data acquisition.

### C. Quantum technology

Quantum technologies comprise engineered physical systems that harness controllable quantum degrees of freedom, such as qubits, collective quantum states, or single photons, for computation, communication, and sensing, together with high-fidelity control and measurement. Over the past decade, multiple quantum computing platforms have matured from laboratory prototypes into programmable devices accessible via cloud interfaces, as well as deployable quantum sensors and readout components integrated into experimental systems. Each type of system reflects distinct trade-offs between coherence and noise, operation speed, scalability, fabrication complexity, and integration with classical control and cryogenic infrastructure. No single approach has yet emerged as dominant, and it is increasingly clear that multiple quantum modalities may coexist, each optimized for particular application domains and scaling strategies [73].

In particle physics, quantum processors are being explored as computational tools to push beyond the limits of classical simulation and inference in both theory and experiment [74]. On the theory side, quantum computers lend themselves most naturally to the simulation of quantum systems [75], making quantum algorithms a unique tool for tackling complex system simulations [76,77]. For collider experiments, quantum computers can be used to supplement traditional Monte Carlo based simulations [78,79]. All existing quantum processor technologies (superconducting, trapped ion, photonic, etc.) remain within the noisy intermediate-scale quantum (NISQ) regime, where error correction is costly and computational advantage is limited to specialized tasks [73]. Across platforms, the central challenge is the realization of scalable, fault-tolerant logical qubits with acceptable physical-qubit overhead. Progress is increasingly driven by system-level considerations, including control electronics, packaging, cryogenics, and software-hardware codesign, rather than qubit performance alone. Rather than converging on a single hardware solution, the field is likely to evolve toward a heterogeneous quantum computing ecosystem in which different processor technologies serve complementary roles in computation, communication, and sensing.

Additionally, quantum and quantum-inspired ML may offer a more performant approach than traditional ML for tasks involving complex and long-range correlated datasets, such as event classification, AD, track and jet reconstruction, and combinatorial problems in triggering and offline reconstruction [80,81]. Critically, establishing a credible quantum advantage, whether arising from the algorithm, the hardware, or their codesign, relative to best-in-class classical approaches is an essential prerequisite for incorporating these technologies into modern scientific computing infrastructure.

In light of the current challenges for quantum technology, within the ML-HEQUPP scope, there are two primary categorizations of present-day efforts:

(1) Quantum, quantum-inspired, and hybrid classical-quantum algorithms accelerated for classical hardware edge deployments (e.g., for FPGAs).

(2) Use of quantum devices for ultrasensitive detector design [e.g., low-mass dark matter (DM), high-energy charged particles].

This section covers the underlying technologies that support such research in particle physics, while Sec. III includes the applications of these technologies to existing and future experiments.

While near-term quantum approaches are not necessarily expected to outperform state-of-the-art classical pipelines broadly, these application-driven efforts are helping identify niches where quantum advantage could plausibly emerge as hardware matures. Furthermore, building out the computational infrastructure needed to pursue such research ensures "quantum readiness" of particle physics experiments and workflows, such that the field is prepared to take advantage of at-scale quantum processors at the moment they become available (see Sec. II E). Reaching this point would introduce computational efficiency and speed that will transform our ability to probe the big data frontier of particle physics.

### *1. Quantum algorithms*

*Circuits and qubits*. Most near-term quantum algorithms are formulated in the quantum circuit model, where computation is performed by applying sequences of parameterized quantum gates to an initial register of qubits, followed by measurement. In this framework, algorithms are typically expressed as variational quantum circuits (VQC), which combine fixed entangling gate patterns with tunable single- and multiqubit rotations. The expressivity of a circuit is controlled by its depth, connectivity, and gate set. This circuit-based paradigm is largely hardware agnostic and maps naturally onto most leading quantum processor technologies, making it the dominant abstraction for quantum computing in the NISQ era.

Different physical realizations of qubits give rise to distinct circuit primitives and computational trade-offs. Discrete-variable qubit systems, such as superconducting circuits, trapped ions, neutral atoms, and semiconductor spins, implement gates that directly correspond to unitary operations on two-level systems [82–85]. In contrast, photonic and bosonic architectures often operate in higher-dimensional Hilbert spaces, where information is encoded in continuous degrees of freedom such as photon number, quadratures, or temporal modes [86,87]. These bosonic systems enable compact representations and naturally support operations like squeezing and displacement, but require careful control of loss, noise, and non-Gaussian resources. From an algorithmic perspective, many circuit constructions can be translated between qubit and bosonic encodings, albeit with different resource requirements and error sensitivities.

The long-term viability of circuit-based quantum algorithms hinges on fault tolerance and error correction. While large-scale surface-code implementations remain out of reach, recent experiments have demonstrated small logical qubits and early forms of error suppression [88]. As quantum hardware matures, circuit-based quantum models are expected to evolve alongside advances in compilation, control, and hardware-software codesign, forming a foundational layer for quantum-enhanced AI applications in particle physics.

*Quantum annealing*. In contrast to the gate-based circuit model, quantum annealing implements computation through the controlled adiabatic evolution of a many-qubit system governed by a time-dependent Hamiltonian. The system is initialized in the ground state of a simple transverse-field Hamiltonian and is slowly evolved toward a problem Hamiltonian whose ground state encodes the solution to a combinatorial optimization problem, typically expressed in Ising or quadratic unconstrained binary optimization (QUBO) form [89,90]. Quantum annealers natively realize this continuous-time dynamics using fixed qubit connectivity and analog control of couplings, rather than discrete quantum gates. While quantum annealing is not universal for quantum computation and cannot directly implement general quantum circuits, it occupies a complementary niche to circuit-based processors, offering a physically scalable approach to solving structured optimization problems.

In high-energy physics and related domains, quantum annealing and quantum-inspired annealing algorithms have been explored for applications such as track reconstruction [91], clustering [92,93], and pattern recognition [94]. A quantifiable and generalizable quantum advantage over state-of-the-art classical CPU/GPU methods in such experimental challenges remains to be seen. The advantage is primarily anticipated where the problem structure maps naturally to annealing, but this comes with nontrivial costs in problem reformulation/embedding overheads, limited

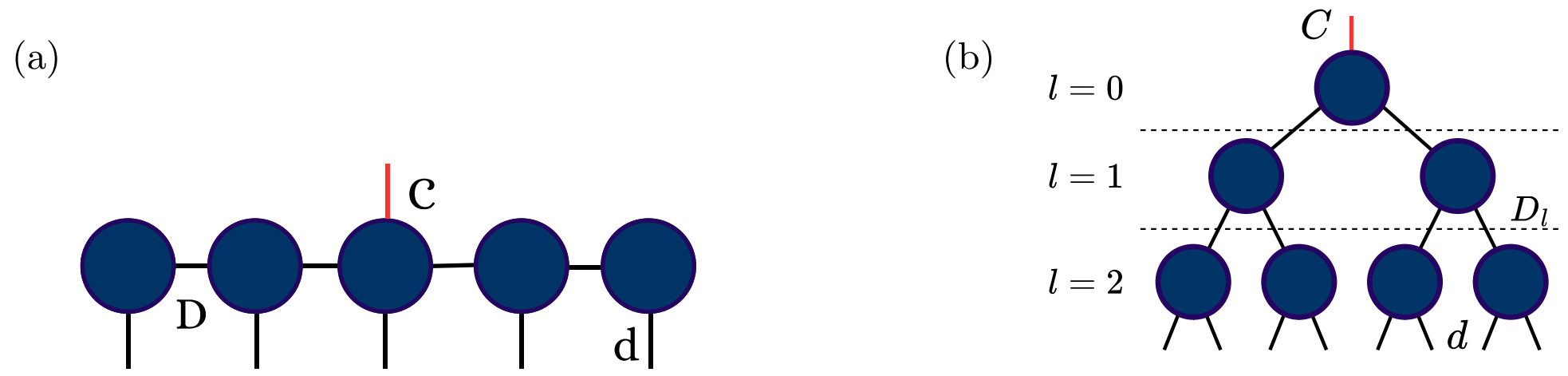

FIG. 11. Tensor network architectures used for classification: (left) MPS and (right) TTN. (a) MPS structure designed for a classification task with five tensors connected via bond indices of dimension $D$, where each tensor has a physical (input) dimension $d$. The central tensor has an additional output leg corresponding to the class index $C$. (b) TTN designed for a classification task, approximating an eight-order tensor with three layers $l$, bond dimension $D_l$ for each layer, and an output index $C$ corresponding to the number of classes.

connectivity and precision, hardware access/latency, and integration complexity.

*Tensor networks.* Tensor network (TN) models provide an algorithmically hardware-efficient alternative to conventional deep learning architectures [95–97] for real-time classification tasks operating under extreme latency constraints. In front-end processing systems, decisions must often be taken within microsecond or submicrosecond timescales, while memory, logic resources, and power consumption are tightly constrained. In this regime, standard deep neural networks—despite strong classification performance—frequently struggle to satisfy strict timing and resource requirements.

Originally developed in the context of quantum many-body physics, TNs offer a structured framework for representing high-dimensional functions through factorizations into low-rank tensors. Architectures such as matrix product states (MPS) [98,99] [see Fig. 11(a)] and tree tensor networks (TTN) [100] [see Fig. 11(b)] can be adapted as supervised learning models operating directly on input features mapped to a higher-dimensional feature space [101–104]. Their key algorithmic novelty lies not merely in parameter reduction, but in the explicit use of network topology and bond dimension as tunable quantities that directly govern model expressivity, computational complexity, and hardware cost. TN applications in particle physics include the study of lattice gauge theories [105] and the tagging of heavy flavor [103] or top quarks [106].

Unlike dense neural networks, inference in TN models corresponds to a sequence of deterministic tensor contractions whose computational cost scales polynomially with bond dimension and linearly with the input dimension, enabling predictable and controllable execution. From a hardware perspective, this structured computation maps naturally onto FPGA and other reconfigurable architectures. The contraction patterns inherent to MPS and TTN models lend themselves to efficient pipelining and parallelization, while fixed-point arithmetic and aggressive quantization can be employed with minimal degradation in performance. As a result, TN-based inference can achieve low latency and reduced memory bandwidth requirements, making such models well suited for deployment in real-time, resource-constrained environments [101,107,108] (see Fig. 12).

*Quantum embeddings.* A central bottleneck in applying quantum computing techniques to particle physics problems is the nontrivial task of mapping high-dimensional classical data from the experiments into the smaller and highly constrained state space of current quantum system. Unlike classical models, quantum hardware natively operates on amplitudes and phases of qubits, making data loading itself a potentially dominant cost and source of information loss. As a result, many existing approaches rely on hybrid quantum-classical pipelines, where classical neural networks are first used to preprocess, compress [109,110], or learn task-relevant representations of the data [111], which are then encoded into quantum states via angle, amplitude, or basis encodings. While this strategy makes near-term implementations feasible, it also shifts much of the representational burden back to classical models, making it challenging to identify where genuine quantum advantage can emerge. Recently, an embedding scheme that uses the raw experimental data (e.g., without prior compression) has been proposed that establishes a correspondence between a reconstructed particle and a single qubit [112]. Designing quantum embeddings that preserve structure, correlations, and symmetries of classical data—while remaining efficient, expressive, and compatible with noisy intermediate-scale quantum devices—remains an open and active area of research and a key determinant of the practical impact of quantum-based data processing approaches.

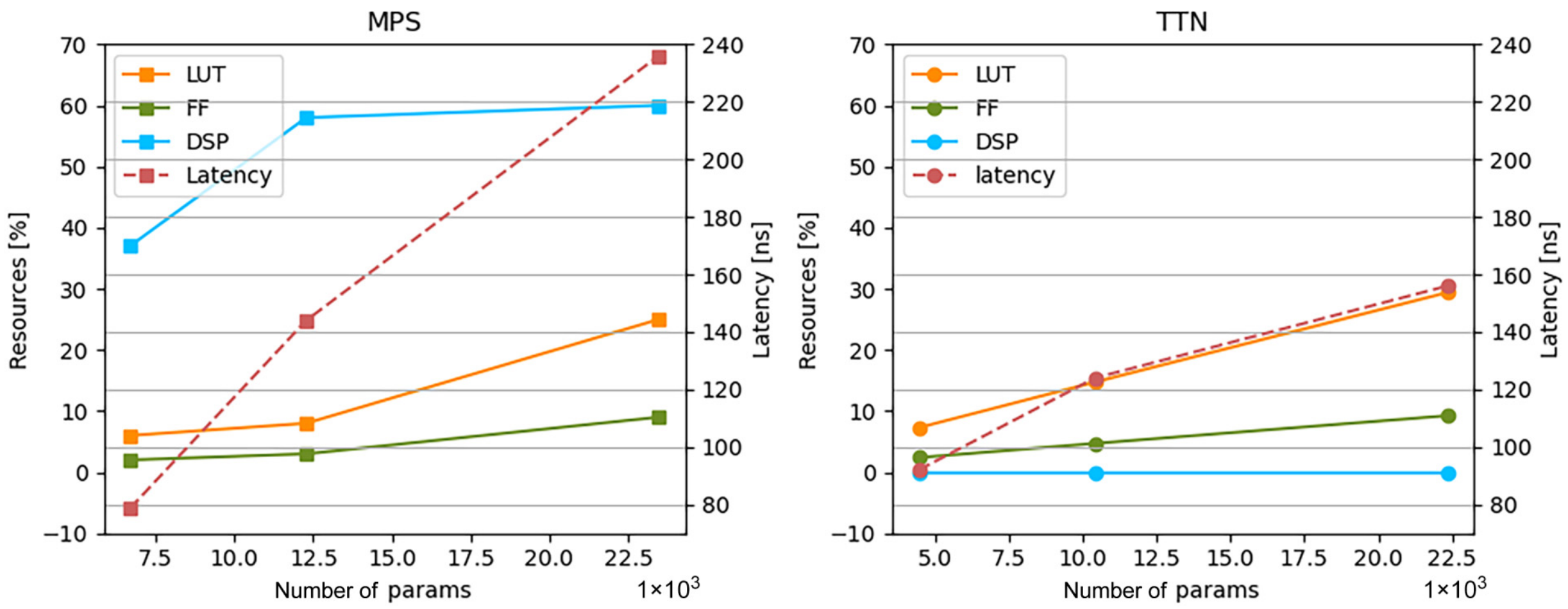

FIG. 12. FPGA resource utilization and inference latency for quantized MPS (left) and TTN (right) models as a function of the number of parameters. Shown are the percentages of LUTs, FFs, and DSPs used on an XCVU13P FPGA, along with the corresponding inference latency (right, $y$ axis). While MPS implementations exhibit high DSP utilization and rapidly increasing latency with model size, TTN implementations avoid DSP usage entirely and show more favorable scaling in both resource consumption and latency. Models were trained using the dataset from Ref. [184] and synthesized as reported in Ref. [101], following HLS and VHDL firmware, respectively, for MPS and TTN.

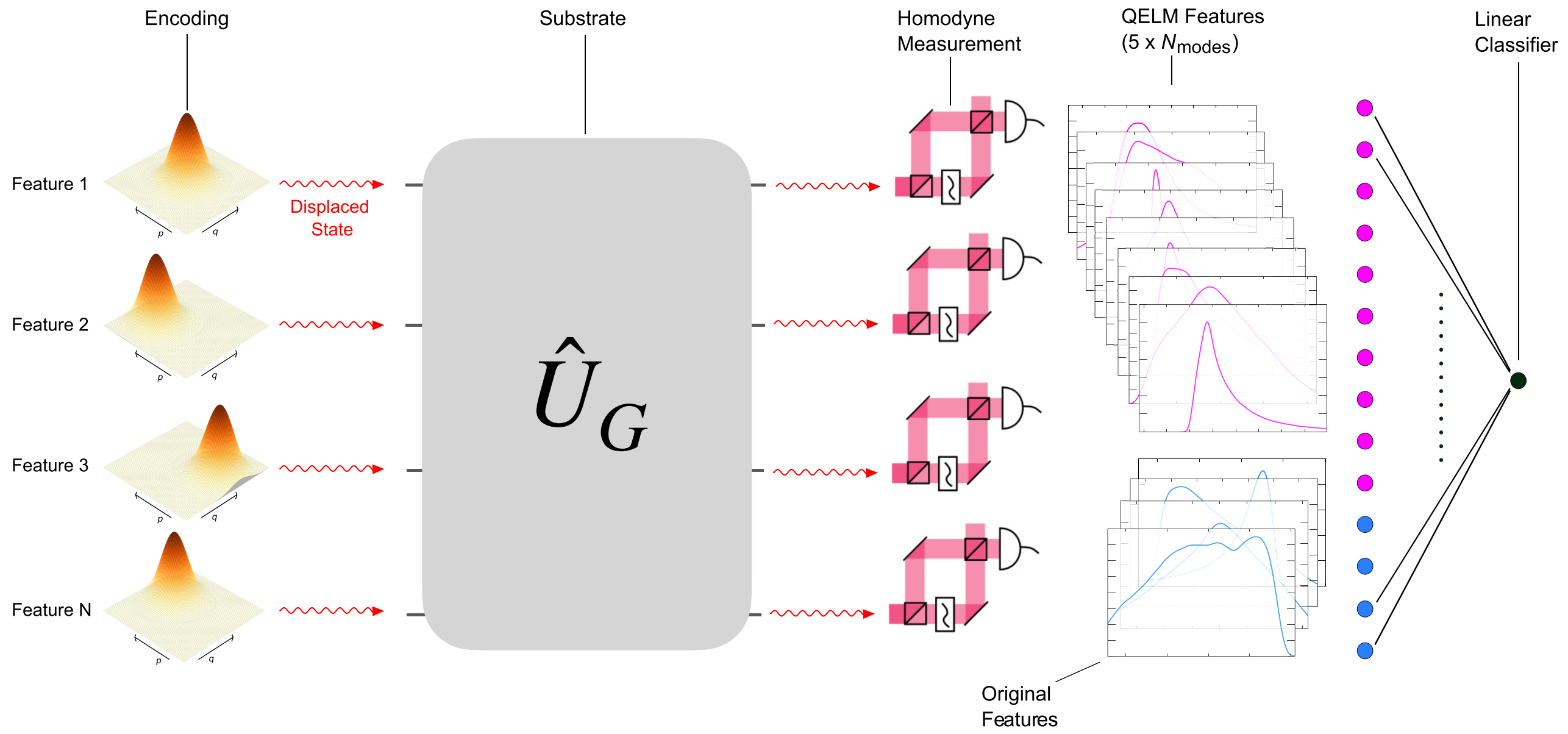

FIG. 13. Visualization of the QELM. The features of the data get encoded onto optical modes via phase-space manipulations. A fixed quantum circuit $\hat{U}_G$ followed by optical measurements creates a random, high-dimensional feature map, which is used, together with the original data, in a linear classifier.

*Quantum extreme learning machines*. Recent work has demonstrated that continuous-variable (CV) photonic quantum extreme learning machines (QELMs) provide a compelling hardware-native approach to low-latency machine learning for collider experiments [113]. In this architecture, illustrated schematically in Fig. 13, classical event features are encoded into the quantum state of multiple optical modes and processed by a fixed photonic circuit. The circuit acts as a random feature generator, transforming the input into a higher-dimensional representation that can enhance class separability. Information is then extracted through optical measurements and passed to a classical linear classifier. Importantly, only the final linear readout is trained, allowing the model to be retrained through a single deterministic linear solve (e.g., ridge or logistic regression) rather than iterative gradient-based optimization.

*Quantum machine learning*. Quantum machine learning (QML) models are typically trained by optimizing the parameters of a VQC using hybrid quantum-classical optimization loops [114], closely mirroring backpropagation in classical deep learning. In this paradigm, a parametrized quantum circuit prepares a quantum state, measurements define a loss function, and a classical optimizer updates the circuit parameters to minimize this loss [115]. Gradients can be computed either via automatic differentiation (autograd) frameworks, which treat the quantum circuit as a differentiable computational graph and propagate derivatives through expectation values, or via explicitly quantum methods such as the parameter-shift rule [116,117], which evaluates gradients by executing the circuit at shifted parameter values on quantum hardware. These gradients are then used in classical optimization schemes (e.g., gradient descent, Adam), or, in more formal settings, in quantum gradient descent, where gradient information is estimated directly from quantum measurements. While this training loop enables principled optimization of quantum models, it also introduces challenges absent in classical backpropagation, including shot noise from finite measurements, barren plateaus in the loss landscape [118], and the high cost of repeated circuit evaluations, making efficient gradient estimation and optimizer design central research questions in QML.

### *2. Distributed quantum systems*

The recent vision and push toward a quantum internet has transitioned from a theoretical ambition to a concrete engineering reality through the advanced quantum networks for scientific discovery (AQNET) project. Led by Fermilab in collaboration with Argonne National Laboratory,

Caltech, Northwestern University, and University of Illinois at Urbana-Champaign (UIUC), AQNET represents the "second generation" of quantum networking. While its predecessor, IEQNET, successfully demonstrated that quantum information could coexist with classical data on existing fiber-optic cables, AQNET is exploring a more complex mission: achieving metropolitan-scale scalability while integrating distributed quantum computing and multinode sensing arrays. For HEP, this project is not merely about communication; it is about building the quantum fabric that will allow geographically separated quantum resources—whether they be processors or sensors—to function as a single, coherent instrument.

A primary technical bottleneck in quantum networking is the high rate of signal loss in fiber optics, which causes entangled photon pairs to "decohere" or disappear over long distances. In late 2025, AQNET researchers achieved a milestone by successfully integrating a high-rate entanglement source, with an order of magnitude improvement in information density. By using highly dense encoding mechanism enabled by state-of-the-art superconducting nanowire single-photon detectors (SNSPDs), the team demonstrated a dramatic increase in the rate of entanglement distribution. This breakthrough is critical toward the goal of building quantum repeaters, which act as "amplifiers" for quantum signals. Without this technology, quantum networks would be limited to a few dozen kilometers; with it, the vision of a transcontinental quantum internet becomes feasible.

Equally vital is the project's success in picosecond-level synchronization. To perform an "entanglement swap"—the process that allows two nodes to connect through an intermediate station—clocks at both ends must be synchronized with extreme precision. AQNET has achieved a world-record synchronization of under 3 ps across 50 km of fiber. This level of timing accuracy ensures that photons arriving from different sources can be measured in the exact same temporal window, a requirement for any scalable quantum architecture.

Beyond computing, AQNET is revolutionizing the field of distributed quantum sensing. By entangling sensors at distant locations, the network can function as a single, ultra-high-resolution detector. This has immediate implications for fundamental physics, namely:

(1) *Dark matter detection*: Entangled sensors can distinguish between local noise and the global "wind" of dark matter particles passing through the Earth. AQNET is creating the necessary conditions to allow distributed axion searches such as the proposed broadband reflector experiment for axion detection (BREAD) experiment.

(2) *Gravitational wave detection*: Emerging theories suggest that detecting photons (e.g., with SNSPDs) at synchronized geographically distant locations can detect gravitational waves in frequencies currently out of reach for laser interferometer gravitational-wave observatory (LIGO).

### D. Novel paradigms

This section explores emerging computational paradigms that promise order-of-magnitude improvements in energy efficiency, latency, and integration density by fundamentally rethinking how computation, memory, and communication interact. These approaches span the full technology spectrum from reconfigurable logic embedded directly within custom silicon (eFPGAs), to optical interconnects for energy-efficient high-bandwidth data transmission (photonics), to computation performed in the analog domain exploiting physics rather than digital logic (analog compute), to heterogeneous integration of specialized chiplets optimized in different process nodes [System in Package (SiP)], and democratized access through open-source hardware design toolchains eliminating commercial licensing barriers.

Each paradigm addresses specific bottlenecks in the HEP computing pipeline: eFPGAs enable algorithm evolution after detector deployment while maintaining ASIC-level power efficiency; analog compute performs inference and signal processing with drastically reduced energy consumption; chiplet-based integration combines optimal technologies for sensing, processing, and communication within compact packages suitable for detector constraints; and open-source compilation infrastructure democratizes custom accelerator development for the broader HEP community without proprietary tool dependencies. Collectively, these technologies represent the pathway toward sustainable computing for future HEP experiments where detector instrumentation must process, filter, and compress petabytes of data per second at the sensor interface under extreme constraints of power budget (milliwatts per channel), material budget (radiation lengths), radiation tolerance (megarads), and latency (nanoseconds).

The subsections that follow detail specific implementations, recent prototypes demonstrated for HEP applications, and critical research directions required to mature these technologies for large-scale detector deployment in the 2030s and beyond.

#### *1. Embedded FPGAs*

While existing workhorse technologies provide many key functionalities for the design specifications of next-generation front-end electronics, in some ways they fall short. Traditional ASICs satisfy many of the demands, but are fixed in functionality once fabricated. Commercial FPGAs provide reconfigurability but generally exhibit higher power consumption, lower radiation tolerance, and limited suitability for cryogenic operation.

eFPGAs integrate reconfigurable logic directly within ASICs, enabling a promising combination of flexibility and efficiency. By supporting FPGA-like programmability inside a custom chip, eFPGAs allow in-field updates to algorithms, firmware patches, and evolving ML-based data processing without redesigning the underlying ASIC. This architecture enables real-time feature extraction, intelligent triggering, and data reduction at the sensor interface. A conceptual illustration of an eFPGA integrated inside a larger ASIC is shown in Fig. 14. Furthermore, open-source eFPGA design frameworks such as FABulous [119] and OpenFPGA [120] have matured due to the expiration of key FPGA patents and the growth of community-driven toolchains. These platforms support customizable and royalty-free logic fabrics, making it easier for research institutions to prototype reconfigurable front-end systems tailored for specific detector environments.

*eFPGAs for intelligent front-end electronics in high-energy physics.* SLAC National Accelerator Laboratory has recently demonstrated two eFPGA prototypes fabricated in 130 and 28 nm CMOS ASICs [71]. These devices validate the feasibility of incorporating small eFPGA fabrics with approximately 500 LUT equivalent logic elements within HEP ASICs. One initial physics application is a smart pixel proof of concept in which a BDT classifier was deployed on an eFPGA to identify high transverse momentum tracks directly from pixel level sensor data. The demonstration achieved full agreement with the quantized software model while emphasizing the need for increased logic density and improved power optimization.

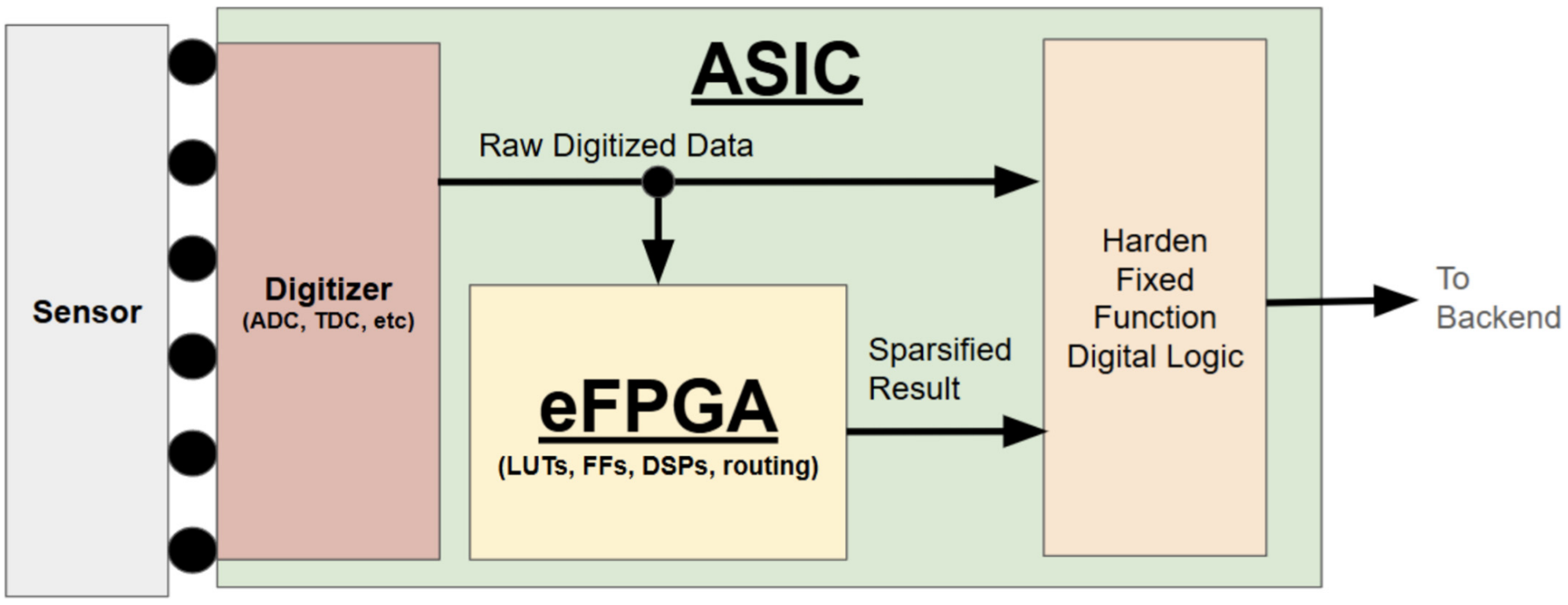

FIG. 14. Conceptual illustration of an embedded FPGA integrated within an ASIC.

Looking forward, eFPGAs create opportunities for intelligent detectors, front-end coprocessors, and mixed-signal architectures that combine digital logic with emerging analog compute techniques. Their reconfigurability supports evolving AI/ML workflows, adaptive calibration, and real-time AD. Ongoing efforts aim to advance open-source design tools, evaluate radiation and cryogenic performance, and scale architectures for next-generation, low-latency, and upgradeable front-end systems.

*Reconfigurable pulse-shape discrimination algorithm implementations using eFPGAs.* Emerging neutron imaging applications, such as single-volume scatter cameras (SVSCs) [121], require high-channel count readout ASICs with low-power dissipation to enable mobile application. Neutrons can be imaged using organic scintillators that provide a pulse-shape discrimination (PSD) capability. With PSD, the scintillator's pulse response for gamma rays and neutrons is slightly different, and this difference can be used to discriminate between gamma rays and neutrons. This is critical in mobile applications where the gamma rate may be 2 orders of magnitude larger than the neutron rate. The separation of the calculated PSD for the two scintillators shown in Fig. 15 allows classification of incident radiation as gamma or neutron.

Traditional PSD signal processing techniques, based on high-speed digitization and digital postprocessing of the data, are highly effective, but they dissipate too much power to be practical in a mobile application [122].

ML algorithms such as BDTs and NNs hold great promise in reducing the energy needed to perform PSD with practical scintillators and radiation environments. Typically, neutron imaging modalities require different triggering, filtering, and PSD settings. Emerging applications such as SVSCs can significantly benefit from ASIC technology. However, providing sufficient tunability to meet the needs of various scintillators and imaging systems is impractical. eFPGA technology allows a readout ASIC to be reconfigured after fabrication, allowing a single ASIC to support various imaging systems. This reconfigurability is critical when applying ML to PSD applications. This is because the input features and NN or BDT parameters can vary greatly across scintillators and energies. While weights in an ASIC-based NN can be adjusted via training after fabrication, the hyperparameters (e.g., number of hidden layers, fixed-point accuracy, optimization algorithms, etc.) are often fixed. In an eFPGA, however, these hyperparameters can be readily adjusted via reconfiguring the eFPGA core.

Applying eFPGA technology to neutron imaging systems allows a single readout ASIC to be more flexible and applicable to a wider range of imaging modalities, energies, and scintillator types [72]. Recently, developments in open-source eFPGA development frameworks, such as FABulous [119], have made the inclusion of eFPGAs on radiation detection ASICs developed by small teams in academic or the National Lab complex feasible for the first time. A prototype eFPGA fabricated in a low-cost CMOS process and capable of implementing a reduced-complexity ML model for PSD is shown in Fig. 16. eFPGA technology holds significant potential to simplify readout electronics for neutron imaging and other radiation detection modalities by allowing application-specific tailoring and ability to implement reduced-complexity ML models on chip, reducing the data rate and lowering power.

*Implementation vision.* For the most area- and power-efficient solutions, ML models can be mapped to RTL code and implemented using the usual custom digital ASIC design flow. Our goal is to use eFPGAs in such ASICs, rather than a custom digital implementation for the ML models in order to maintain configurability of the algorithm, architecture details, and weights/biases.

When it comes to eFPGA targets for ML model deployment, resource constraints inevitably require sparse model deployments. One straightforward application that is being initially considered for ML algorithms on an eFPGA is neutron/gamma classification for neutron imaging systems. Ideally, an event can be identified as a single or multiple scatter event, pileup, accidental coincidence, or dark count/noise and be able to classify a neutron from gamma for single or multiple scatter event sequences. This is mostly done in "offline" processing of recorded events in many current de-tector systems but eFPGAs can potentially be used for this application instead. ML algorithms can be trained to use features of the input pulses to perform particle identification between neutron, gamma, or pileup/other events and deploy the model on an eFPGA fabric. The output classification of each inference per channel/pixel per event can determine whether that channel's hit is considered valid and contributes to a global event reconstruction (i.e., will the info be sent off chip), but the output can also serve as an input feature to a more complicated secondary stage ML model in a master (commercial) FPGA. This stage can use multiple channel info (across chips if needed) to, in theory, make a determination of constituents in a pileup event (nn, gg, ng, gn) and also reconstruct the time of flight in identified multiple scatter events from using position (and particle classification) information of all channels hit within a delta-T window (i.e., tens of nanoseconds). For low-power and real-time applications, this kind of event filtering will be immensely useful compared to "offline" processing, which consumes considerably more power due to having to record and save all waveform events above some global triggering requirement and then performing event filtering on a PC/laptop. In addition, active real-time vetoing can potentially reduce dead time of the system (especially in high-rate gamma background applications). This allows for more efficient triggering on neutron events to provide source reconstruction to a useful position resolution. More complicated models with multiple stages can be the target for a combined eFPGA + commercial FPGA deployed system architecture.

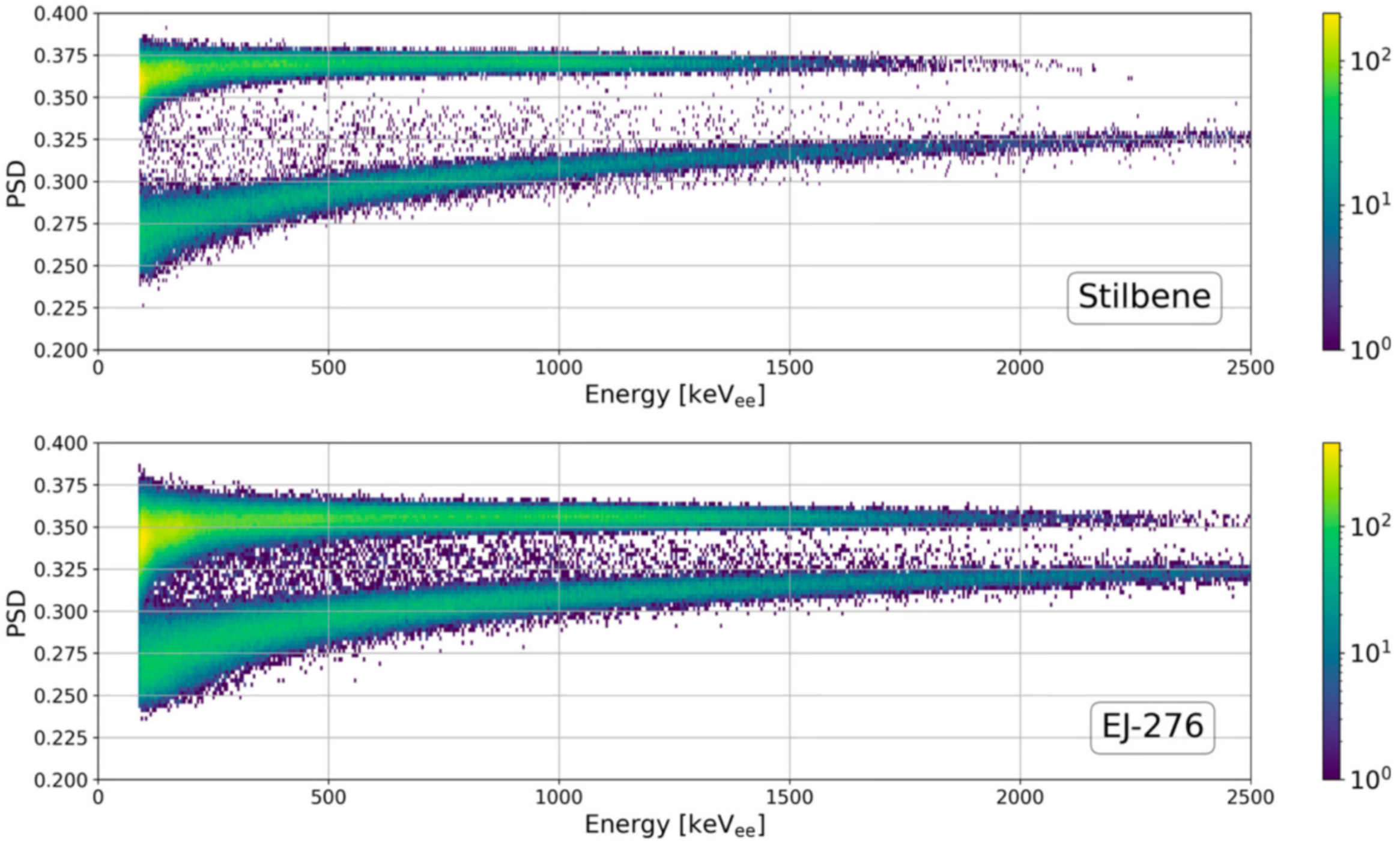

FIG. 15. PSD performance for two organic scintillators. In each trace, the PSD measurement shows two concentrations: the top for gammas and the bottom for neutrons.

The key limitation for an eFPGA ML model deployment is the available input features; these are much harder to change, so a lot of effort will have to go into studying and selecting an optimal set to use. Depending on the level of configurability built into other parts of the chip design (i.e., integration windows), inputs can be adjustable to a degree.

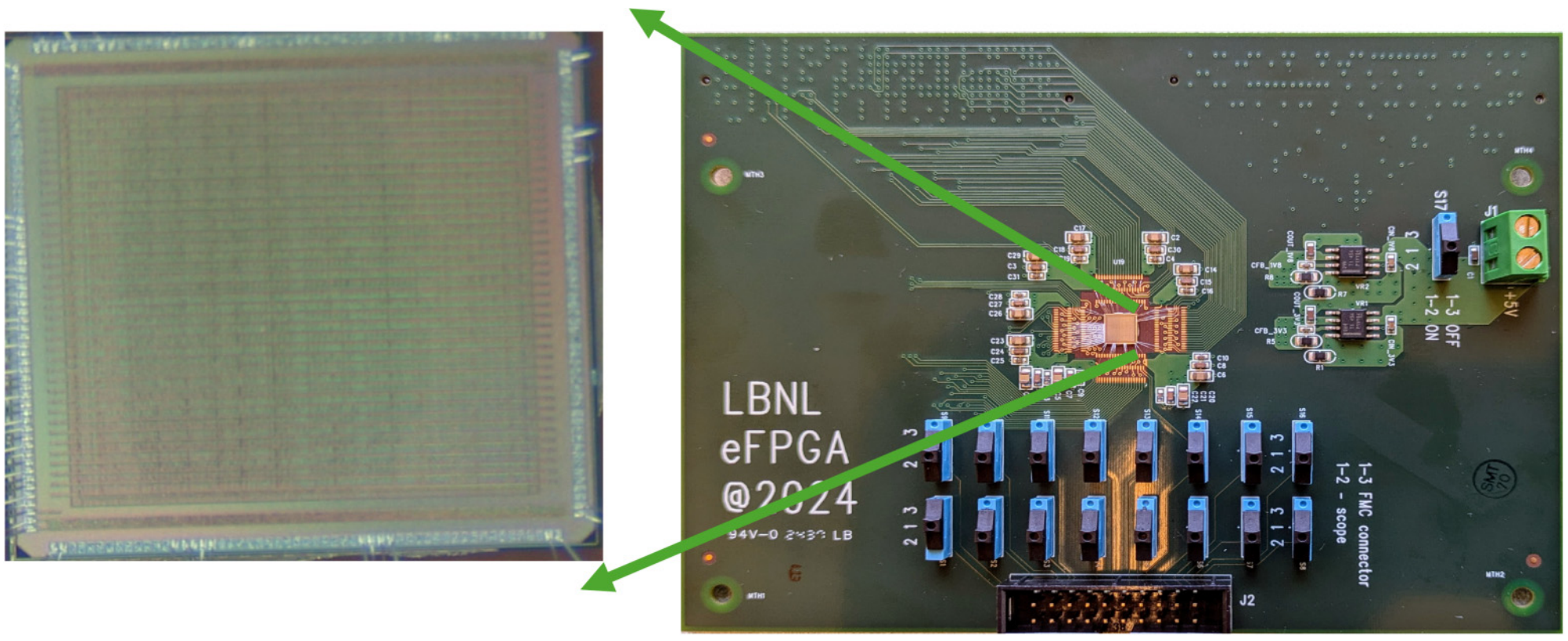

FIG. 16. Prototype eFPGA developed at LBNL. The right shows the eFPGA attached to a test board and the enlarged image on the left shows the eFPGA die.

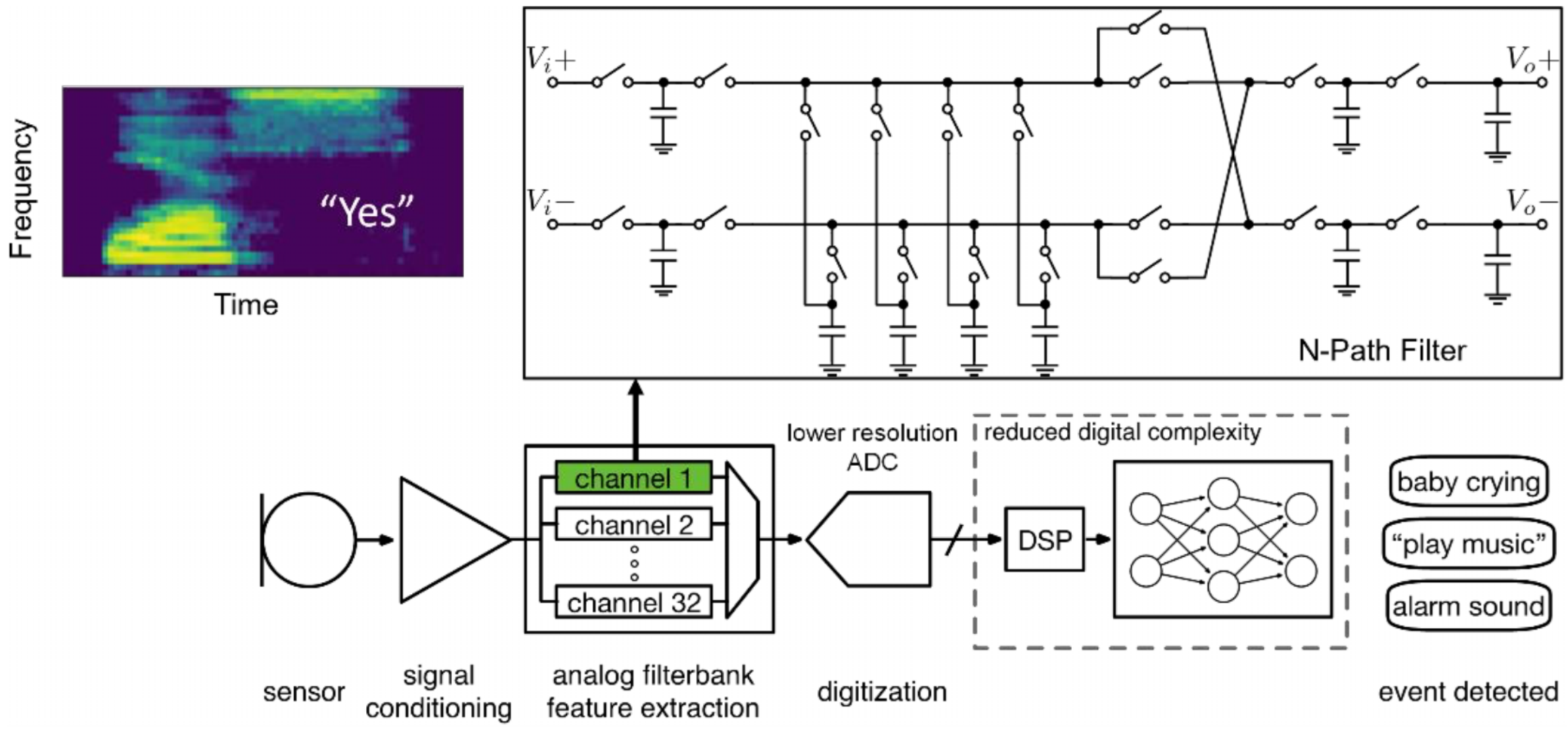

FIG. 17. Analog preprocessing using *N*-path filters in a sound classification system [123].

### 2. Analog compute

*DeCoDe at PNNL.* While current HEP applications employ primarily digital processing, emerging analog computing paradigms warrant consideration for future detector systems. Recent workshops sponsored by DOE-SC, Advanced Scientific Computing Research Program Office on analog computing has identified key challenges for integration into scientific applications:

(1) *Hybrid partitioning*: System-level codesign must quantify ADC and digital-to-analog converter conversion overhead when partitioning algorithms between analog and digital domains. Interface costs (area, power, and latency) may dominate gains from analog acceleration for fine-grained partitioning.

(2) *Noise characterization*: Analog accelerators require noise models accounting for HEP detector conditions—radiation-induced parameter drift, temperature variations, supply voltage fluctuations. Nonstationary noise models reflecting time-varying detector conditions differ from static laboratory characterization.

(3) *Parameter variability*: Device mismatch and process variation impact analog circuit precision. Strategies include device averaging (using larger arrays to statistically reduce variation), algorithmic redundancy (error-correcting codes, ensemble methods), and run-time calibration (periodic recalibration against digital reference implementations).

(4) *Radiation effects*: TID and SEE cause threshold voltage shifts, leakage current increases, and transient upsets in analog circuits. Codesign tools must incorporate radiation-aware device models for detector proximity applications.

(5) *Scalability*: Analog circuit performance degrades with scaling due to wire parasitics, increased crosstalk, and reduced device matching. System architectures must account for these physical limitations when projecting performance to future technology nodes.

PNNL's DeCoDe project support for heterogeneous codesign provides infrastructure for exploring these analog-digital integration challenges as the technology matures for HEP applications.

*Mixed-signal interfaces and compute fabrics for tinyML systems. Analog/mixed-signal preprocessing.* As AI/ML algorithms are marching closer to physical sensors, the design of analog and digital subsystems is becoming more intertwined. Signal acquisition and processing pipelines in integrated circuits have been heavily influenced by the needs of communication systems, which use "human-designed" signals (e.g., QAM-16) to convey a relatively large amount of information per processed sample. On the contrary, in AI inference for sensors, the amount of information conveyed per sample is very small. For instance, Nyquist-sampled speech signals produce a data stream of hundreds of kbit/sec, while only yielding 39 bits/sec of relevant information. This mismatch between the physical bandwidth and information rate motivates many opportunities for analog preprocessing and data reduction prior to sampling and digitization [123]. A specific example catered toward an ML sound classification system is shown in Fig. 17 [124]. Here, the input signal is preprocessed by an *N*-path filter bank before low-rate digitization with reduced bit precision. The filter bank is composed of only passive components and consumes merely 800 nW. The circuit's unwanted harmonic mixing products are absorbed by the machine learning model during a one-time training (independent of manufacturing environmental variations, due to the stability of switched capacitor circuits). Another example is the object detection imager described in Ref. [125]. It computes logarithmic gradients in the analog domain to enable data reduction and compression of the succeeding convolutional neural network (CNN) [126].

*Efficient hardware for neural network inference.* Once a sensor signal is properly conditioned and preprocessed, ML inference for sensors typically relies on deep neural networks like CNNs, or (more recently) transformerlike models. The main challenge here is to minimize data movement across the system's memory hierarchy [127]. One approach for amortizing memory accesses has been to create compute arrays that can "unroll" the filter kernels to amortize memory access over many multiply-add operations performed in parallel. One way to achieve this is through analog arrays using capacitive or resistive crossbar approaches. While the block specifications of such compute macros are often impressive [128] (measured in tera-operations per second per watt (TOPS/W)), their benefit at the system level tends to be less pronounced due to their limited bitwidths and resulting penalty on the required network size. As a compromise, a fully digital implementation was explored with custom-designed compute cores and efficient, closely interspersed custom memory banks. A snapshot of this prototype is shown in Fig. 2 [129]. Its 8-bit architecture achieves state-of-the-art inference energy across a range of tinyML tasks. Its custom latch arrays achieve a read energy of only 15 fJ/byte. The architecture is engineered such that these efficient local memories are hit most frequently, while access energy to system-level memory is reduced by up to 9.5× over conventional architectures.

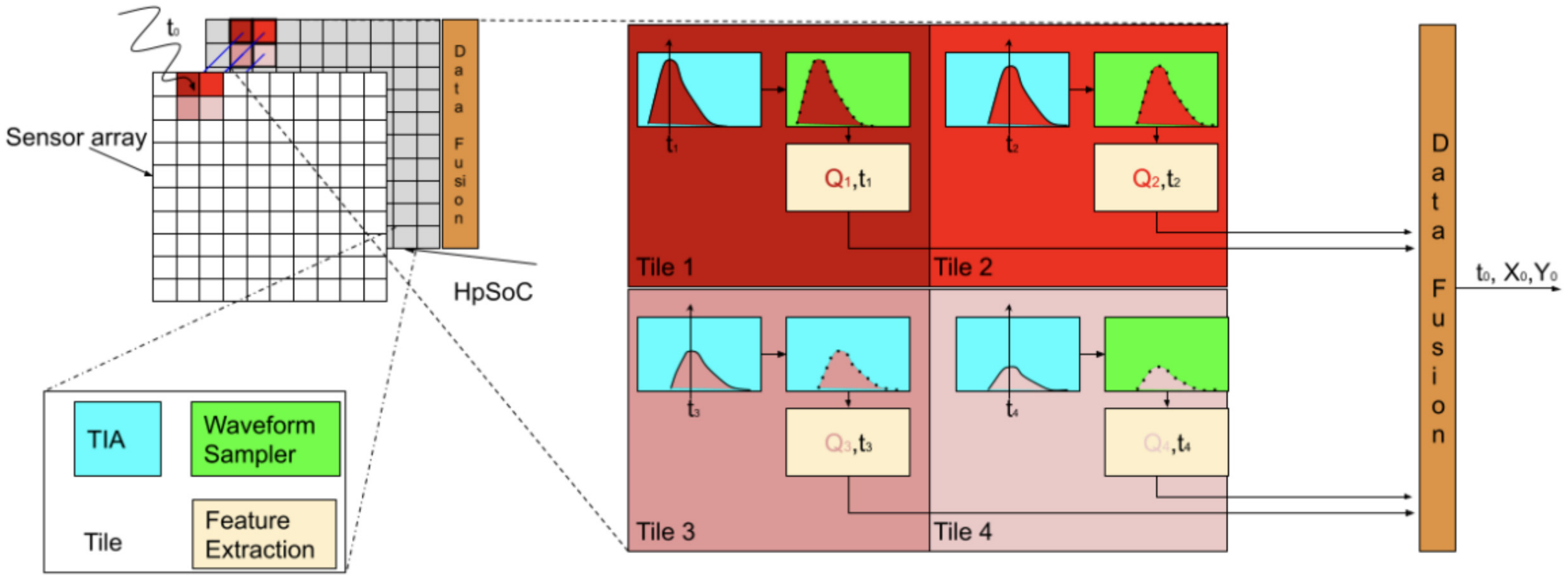

FIG. 18. Architecture of Nalu Scientific HPSoC for ML-based scientific readout applications.

*Chiplet-based heterogeneous integration.* SiP technology enables seamless integration of multiple chiplets—analog accelerators, digital processors, memory, specialized intellectual property (IP) blocks—within a single package, creating modular heterogeneous systems optimized for energy efficiency [130]. Recent advances in open chiplet ecosystems and standards combined with open-source codesign tools make this approach increasingly accessible. The DOE Office of Science Microelectronics Science Research Center project "Democratization of Co-design for Energy-Efficient Heterogeneous Computing" (DeCoDe) integrates MLIR [131] based compiler infrastructure, architectural simulation, and thermal analysis to support codesign of heterogeneous SiPs [130]. For HEP applications, chiplet-based approaches could enable integration of analog sensor interfaces with digital ML accelerators and high-speed serializers, each fabricated in optimal process nodes and technologies, while managing power/thermal constraints through advanced packaging (2.5D interposers, 3D stacking).

*Nalu Scientific HPSoC.* Traditionally, time of arrival gets estimated using TDC (time-to-digital-conversion) on a pulse-shaped signal derived from the sensor. Such a solution is simple and amenable to good integration, but it relies on single point measurement, which makes it prone to many sources of error, such as noise on the threshold voltage and on signal, variable signal amplitude (time walk), jitter on reference clock, and measurement jitter on TDC logic (variable delay in internal logic), all of which degrade the performance. On the other hand, with a full waveform sampling, the entire shape of the signal can be used to average out the effect of most of these noise sources (e.g., DC or slow frequency noise using prepulse baseline estimation, high frequency noise via multisample interpolation or convolution). Even if a full waveform recording is available, the large variability and interdependence of interchannel and intrachannel events can make it hard to predesign an estimator that is adapted to all foreseeable situations (e.g., channel crosstalk, event pileup, channel-dependent parameter variation, etc.). In such cases, an AI/ML approach (inference at the edge) is the most promising and adaptable estimation methodology. However, this requires a joint design of the readout mechanism, and possibly the incorporation in the ROIC of the mechanism to predistill relevant information to perform a first-layer data reduction while at the same time maintaining flexibility to handle diverse signal combinations.

As an example of a possible readout solution that could empower ML-based experiments, Fig. 18 shows the architecture of the High timing Precision System on Chip (HPSoC) series of ROICs designed by Nalu Scientific. The architecture incorporates a dense sensor array readout with modular, independent ("tiles") per pixel, on-chip signal amplification (TIA + gain), continuous waveform sampling and triggered digitization (with internal buffering of past triggers to avoid dead time), on-the-fly timing and amplitude (charge) extraction. The output of any of these components could be the target for the application of a more or less sophisticated exploration of ML extraction, operating directly on the raw samples, the individual extracted Q and Ts, or the output of the fusion logic. The implementation of the latter as an inference engine is a good candidate for first experimental evaluation of the methodology and is being considered as a good demonstrator in a follow-up study.

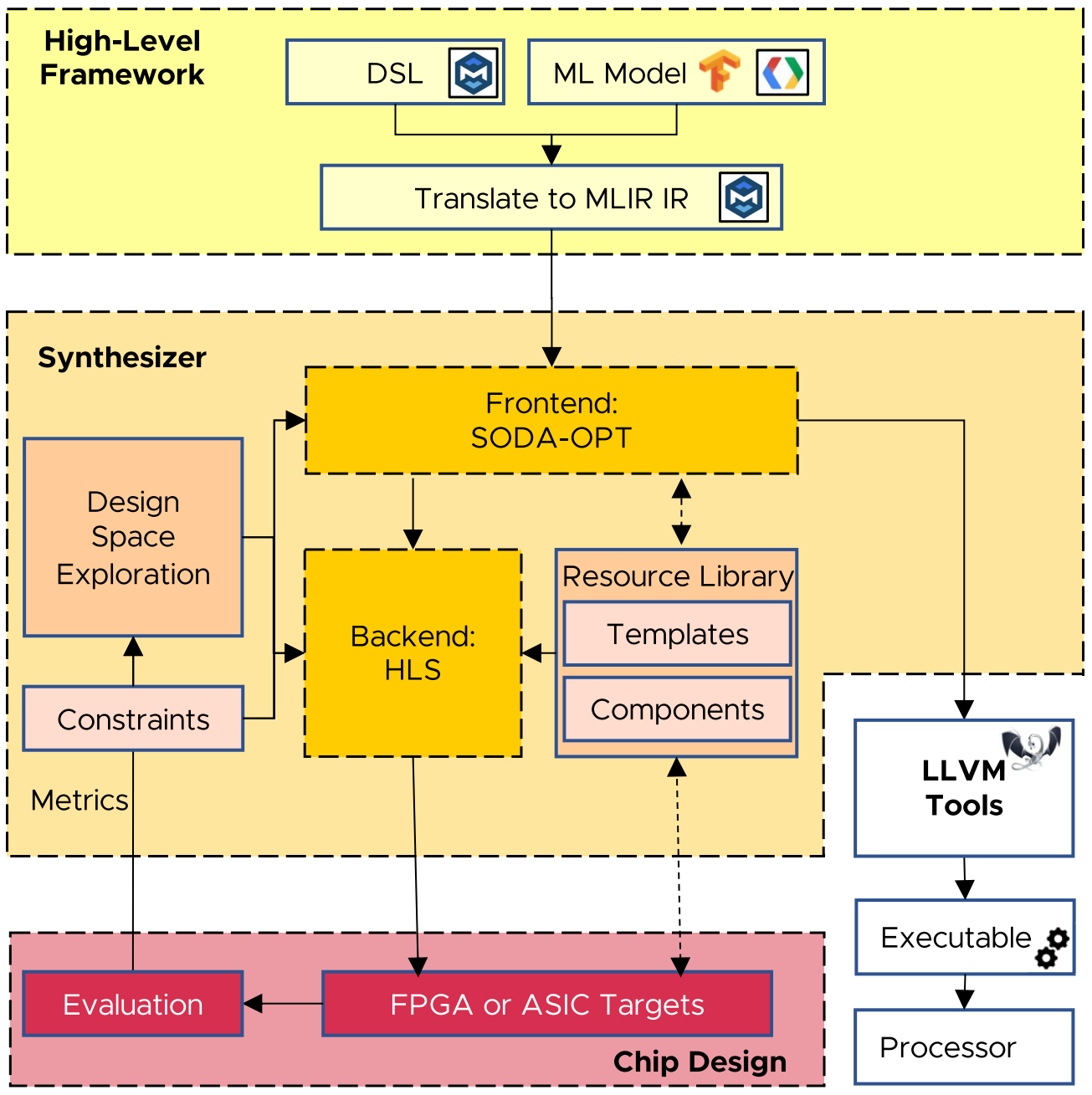

FIG. 19. SODA open-source toolchain architecture enabling Python-to-GDSII compilation without commercial EDA tools. The flow integrates Python ML frameworks (TensorFlow, PyTorch) through MLIR-based intermediate representation, SODA-OPT hardware-software partitioning, PandA-Bambu high-level synthesis, and OpenROAD physical design. The complete open-source flow supports multiple PDKs and enables custom ASIC development accessible to academic institutions without commercial licensing barriers.

### 3. *Open-source tools for hardware-software codesign*

The HEP community has benefited from open-source tools that democratize hardware acceleration (see, e.g., `hls4ml` and others in Sec. II B). Complementary efforts in computing research provide capabilities for automated hardware generation from arbitrary Python algorithms, domain-specific compilation, and system-level codesign.

*MLIR-based compilation infrastructure.* PNNL's open-source compilation framework leverages MLIR to provide unified infrastructure spanning high-level algorithm descriptions through heterogeneous hardware generation. The COMpiler for Extreme Targets (COMET) compiler [132,133] targets domain-specific languages for tensor algebra, supporting computational chemistry (NWChem) and graph analytics with optimizations for CPUs, GPUs, and dataflow accelerators. The SODA Synthesizer [32] interfaces with Python frameworks (TensorFlow, PyTorch), performs hardware-software partitioning and hardware-oriented optimizations through its SODA-OPT front end [134], and then generates custom accelerators via PandA-Bambu open-source HLS [135]. For end-to-end flows, SODA connects to OpenROAD [136] for physical design, enabling complete Python → MLIR → Bambu → OpenROAD → GDSII compilation without commercial tools. Recent extensions support custom instruction sets for pattern-specific accelerators and integration with OpenCGRA for CGRA mapping [137].

Figure 19 illustrates the complete open-source toolchain architecture. SODA provides general-purpose compiler infrastructure for arbitrary Python algorithms—including ML frameworks (TensorFlow, PyTorch), scientific computing kernels, data preprocessing, control logic, and complete algorithm pipelines. SODA supports flexible optimization targets (low latency, high throughput, or balanced trade-offs) and multiple back ends: FPGAs and ASICs with access to commercial process development kits (PDKs) or multiple open PDKs (FreePDK 45 nm, ASAP 7 nm, SkyWater 130 nm, GlobalFoundries 180 nm). This enables end-to-end detector readout pipeline synthesis from high-level Python specifications. Reported results include 364 giga floating-point operations per second per watt (GFLOPS/W) energy efficiency on CNN operators and dramatically reduced turnaround time for hardware prototyping compared to manual HDL development [32].

These MLIR-based tools provide (1) broad scope from ML models to arbitrary scientific algorithms including data preprocessing, feature extraction, control logic, and data formatting, (2) unified compiler infrastructure enabling cross-optimization between different hardware targets (CPU, GPU, FPGA, CGRA, ASIC), (3) direct ASIC synthesis paths with flexible optimization targets (latency, throughput, energy efficiency), and (4) integration with architectural simulation for design space exploration. The modular MLIR architecture allows domain experts to extend the framework with custom dialects for specialized operations, enabling HEP-specific extensions for detector algorithms, trigger logic, or data compression without rebuilding entire toolchains.

*Spatial computing and reconfigurable architectures.* OpenCGRA [137] provides the first open-source unified framework for CGRAs, offering multilevel modeling (functional level, cycle level, RTL) using PyMTL with low-level virtual machine based compiler integration. For HEP applications with structured computation patterns—convolutions in calorimeter processing, nearest-neighbor searches in tracking, streaming data reduction—CGRAs offer efficiency between FPGAs' fine-grained flexibility and ASICs' fixed functionality. The dynamic rebalancing of pipelined streaming extension [138] adds partial dynamic reconfiguration, demonstrating 1.46× throughput improvements for workloads with data-dependent execution times—directly applicable to variable HEP event characteristics, pileup conditions, and adaptive triggering scenarios requiring run-time algorithm adjustment.

*Hybrid analog-digital considerations*. The DeCoDe framework extends MLIR-based compilation to support computing paradigms including analog accelerators. While current HEP applications remain primarily digital, the codesign methodology for integrating analog components (handling precision constraints, interface costs, noise-tolerant algorithm formulations) may inform future mixed-signal detector systems as analog computing matures.

*Key message.* Open-source compilation frameworks provide the international HEP community with accessible infrastructure for general-purpose hardware development from Python algorithms: (1) broad algorithmic scope from ML to complete detector pipelines, (2) flexible optimization targets (latency and/or throughput), (3) unified MLIR-based optimization across multiple targets (CPU, GPU, FPGA, CGRA, ASIC), (4) fully open synthesis flows including ASIC paths, and (5) extensible architectures supporting custom domain-specific extensions. These capabilities democratize custom accelerator development for academic collaborations worldwide.

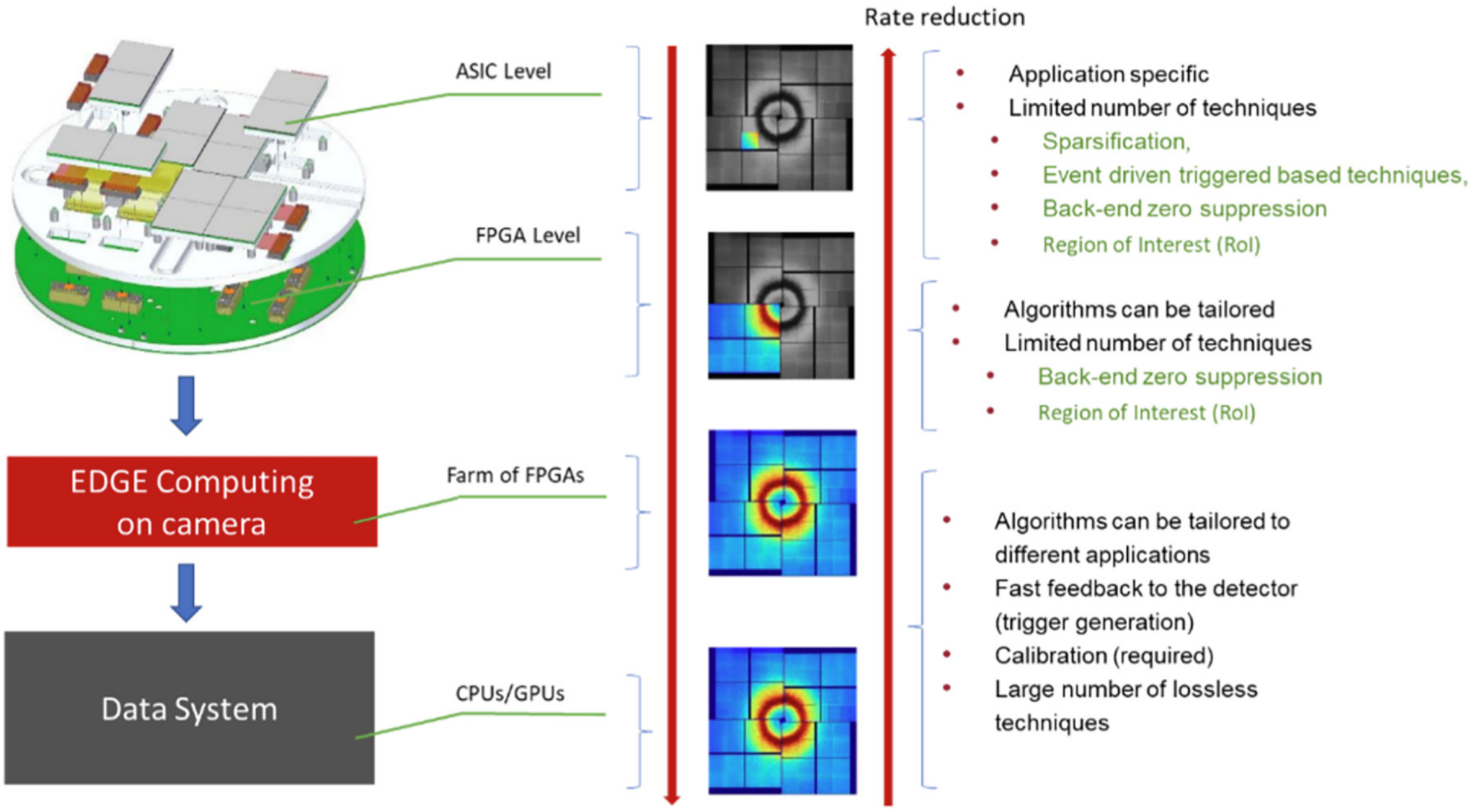

FIG. 20. Distributed data processing from the detector to the back end in the case of the LCLS-2 facility at SLAC.

#### *4. Heterogeneous pipelines*

Heterogeneous data pipelines are processing frameworks that interconnect a wide range of computing technologies, such as FPGAs, embedded processors, GPUs, IPUs, Groq devices, and custom ASICs, to efficiently manage and analyze the enormous dataflows generated by modern scientific instruments. A core principle in designing these systems is the mapping of technologies onto front-end and back-end processing categories. Front-end technologies, often deployed directly on or near the sensors (e.g., FPGAs, ASICs, eFPGAs, and SoCs), handle immediate high-speed data capture, localized processing, and initial feature extraction. Back-end technologies (GPUs, CPUs, etc.) manage more complex analysis that is not time or resource limited. Defining these roles with an eye toward optimization for a particular instrument's conditions and environment ensures each device is used where its strengths best match the physics-driven specifications. Furthermore, the introduction of agentic tools for a faster and more efficient traversal of complex optimization space represents a future of generic end-to-end DAQ design that can be easily optimized for different scientific and practical criteria (e.g., physics measurements, cost, etc.) Figure 20 shows an example mapping of hardware to data task for LCLS.

Optimizing these heterogeneous pipelines involves careful design not just of processing stages, but also of the high-performance transmission and linking mechanisms that relay data from the detectors through the pipeline. Efficient, low-latency communication, utilizing high-speed links and advanced networking protocols, is vital for ensuring that key scientific features extracted by front-end devices arrive at the back end without bottlenecks, maintaining data integrity and enabling real-time insights.

A practical example of heterogeneous optimization in practice is the SLAC streaming artificial intelligence (S3AI) test bed (Fig. 21). It serves as a comprehensive integration platform that connects sensors, heterogeneous accelerators, midscale HPC resources, S3DF systems, and IRI orchestration frameworks. The test bed enables realistic assessment of streaming pipelines, cross-hardware interactions, security considerations, and algorithm deployment under live data conditions. Ample opportunity exists for industrial benchmarking and collaboration by introducing modern commercial computing hardware in a plug-and-play manner. S3AI thus functions as a valuable proving ground for validating the full edge-to-exascale workflow envisioned in ML-HEQUPP.

### E. Analysis facilities

Analysis facilities (AFs) serve as critical infrastructure connecting the development of machine learning models with their deployment on front-end hardware systems. The vision of AI-native high-energy physics—embedding AI end-to-end across the entire experimental lifecycle, from design and optimization through simulation, data processing, and discovery—requires a computational infrastructure that bridges the gap between model development and hardware deployment. AFs provide the computational resources, data access, and collaborative environments necessary to develop and validate the ML algorithms that ultimately run on detector front ends. For example, the SmartPixels project (Sec. III A 1 a) was fully supported by the Fermilab and Purdue AFs, and the analog compute work (Sec. II D 2 b) by the BNL and UChicago AFs—both leveraging community resources available to the respective teams.

The HSF Analysis Facilities White Paper [139] documents the R&D status and future directions for these facilities, noting that in the HL-LHC era (beginning 2029), analysts may need to process up to an order of magnitude more data than currently. The white paper emphasizes requirements for fast

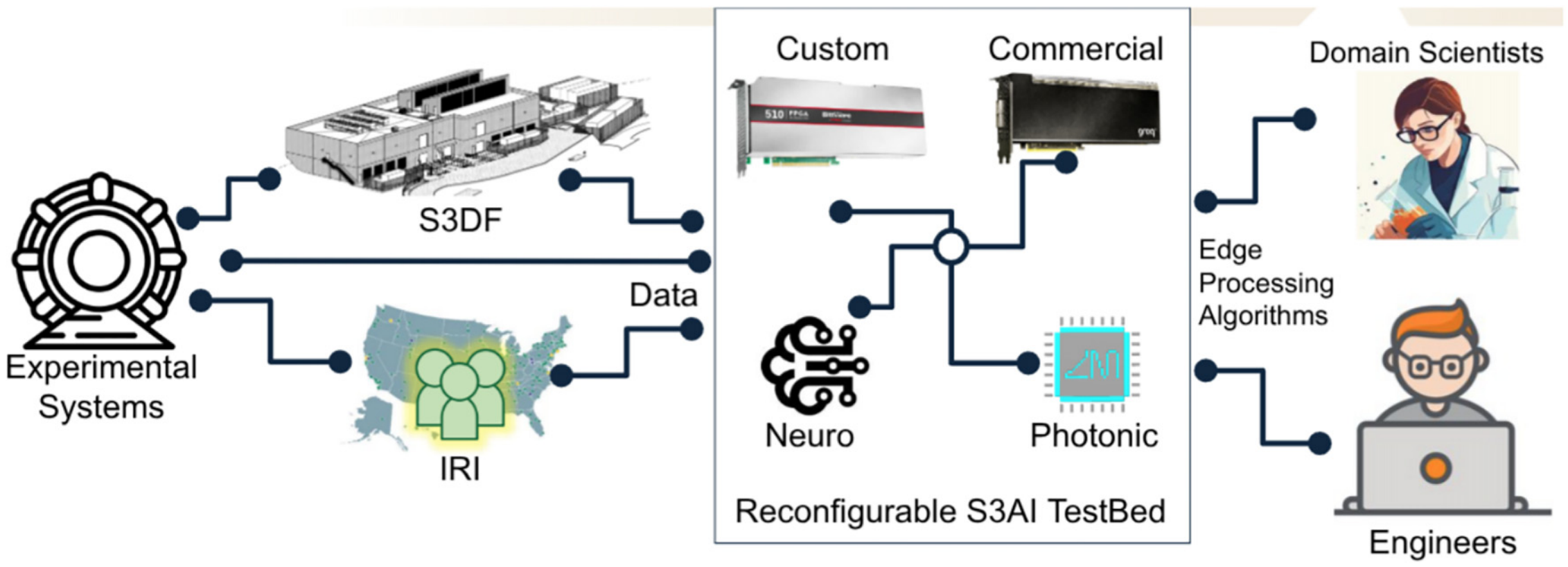

FIG. 21. S3AI enabling development of heterogeneous streaming workflows.

"time to insight" through interactive access, the ability to scale out to distributed resources, and integration with the broader Worldwide LHC computing grid (WLCG) infrastructure. In the context of ML-HEQUPP, AFs play a dual role: They enable the computationally intensive training of ML models destined for resource-constrained front-end hardware, and they provide platforms for validating that deployed models perform correctly at scale.

*Training infrastructure for edge deployment.* Machine learning models destined for front-end deployment face unique constraints: submicrosecond latency requirements, limited memory, fixed-point arithmetic, and radiation tolerance—that differ fundamentally from models running in offline analysis. However, the development cycle for these models necessarily begins in traditional computing environments before translation to hardware implementations. This is particularly true as the field moves toward physics foundation models that can be fine-tuned for specific front-end tasks, and as architectures such as learnable logic networks (Sec. II D) require computationally expensive training despite their efficient inference.

Analysis facilities provide several essential capabilities for this development cycle:

(1) *GPU access for model training*: Training neural networks for front-end applications, including those requiring quantization-aware training for FPGA deployment, demands significant GPU resources. The AF white paper notes that AFs "may not have sufficient GPUs" for training large machine learning models, highlighting the need for AFs to either provide adequate GPU resources or integrate with external GPU-rich resources, such as HPC centers. For applications highlighted in this white paper, this includes training foundation models that can be fine-tuned for specific detector configurations, as well as the computationally intensive training required for multiplication-free architectures.

(2) *Interactive development environments*: The R&D phase of front-end ML development requires rapid iteration—testing different architectures, quantization schemes, and pruning strategies. AFs supporting Jupyter-based interfaces with scale-out capabilities (such as Dask integration) enable physicists to interactively explore the trade-off space between model accuracy, latency, and resource utilization.

(3) *Access to representative data*: Training front-end ML models requires access to realistic detector data or high-fidelity simulations. AFs integrated with experiment data management systems and caching infrastructure ensure developers can access the large datasets needed for training without managing complex data logistics. As trigger-less architectures emerge, the volume of data available for training front-end models will increase substantially.

*Model optimization workflows.* The path from a trained floating-point model to a deployable fixed-point implementation involves multiple optimization stages that benefit from AF infrastructure:

(1) *Quantization studies*: Systematic exploration of bit-width configurations (e.g., using QKeras) to find optimal precision-performance trade-offs.

(2) *Pruning and sparsification*: Identifying and removing redundant network connections while maintaining physics performance.

(3) *Architecture search*: Automated exploration of network topologies subject to hardware constraints (e.g., using SNAC-Pack [140–142]).

(4) *Hardware-aware training*: Iterative refinement incorporating feedback from synthesis tools like `hls4ml` (e.g., using surrogate models for fast resource estimation [143]).

These workflows are computationally demanding and benefit from the batch-mode capabilities that AFs provide for transitioning from interactive development to large-scale parameter sweeps.

*Analysis facility landscape.* Several prototype and production AFs currently support HEP analysis workflows with varying degrees of ML infrastructure. The U.S. ATLAS analysis facilities provide a concrete example of the capabilities that exist today and illustrate both what is working and what gaps remain for edge hardware development.

*Example*: U.S. ATLAS analysis facilities.

The U.S. ATLAS collaboration operates analysis facilities at BNL, SLAC, and UChicago that are available to collaboration members. These facilities provide GPU resources, ML software environments, data caching, and batch computing capabilities representative of what the broader HEP community requires for front-end ML development.

(1) *GPU resources*: Modern AFs increasingly provide access to high-end GPUs essential for ML training. For example, the UChicago facility includes nodes with NVIDIA A100 (40 GB), V100 (16 GB), and consumer-grade GPUs (RTX 2080

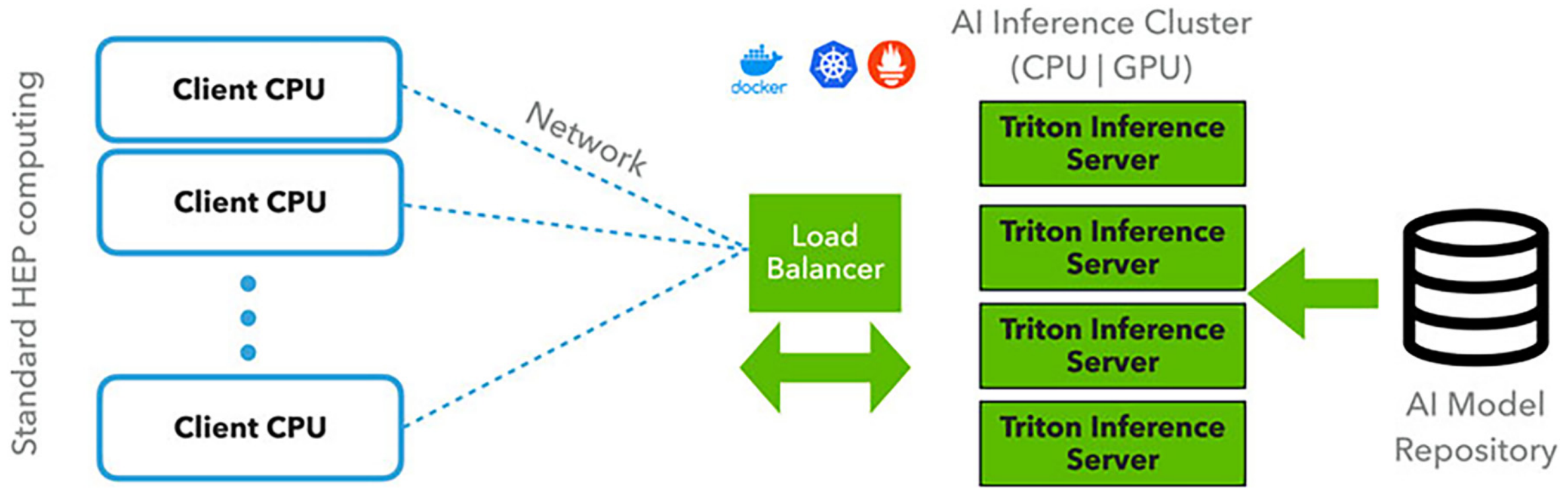

FIG. 22. SuperSONIC: Shared server infrastructure for HEP and MMA experiments.

Ti, GTX 1080 Ti), as well as recent additions of H200 nodes (141GB HBM3 memory). This range of hardware supports everything from interactive prototyping to large-scale training runs. JupyterLab interfaces allow users to request GPU resources interactively, with MIG (multiinstance GPU) support enabling efficient sharing of high-end accelerators.

(2) *Batch computing*: AFs integrate with batch systems (HTCondor, SLURM) that allow scaling from interactive development to large parameter sweeps. This is essential for hardware development workflows that require systematic exploration of quantization configurations, architecture search, and hyperparameter optimization.

(3) *Containerized environments*: Standardized ML containers (available via CernVM File System (CVMFS) and container registries) provide reproducible environments with common ML frameworks (TensorFlow, PyTorch, scikit-learn) and HEP-specific tools. This standardization enables workflows to move between facilities and ensures reproducibility.

*Inference-as-a-service.* NVIDIA Triton Inference Server [144] deployments at several facilities enable transparent CPU and GPU access for inference workloads without requiring users to manage compute scheduling directly. This architecture allows model serving at scale with automatic batching and load balancing, supports multiple ML frameworks and model formats, enables validation of trained models against large datasets before hardware deployment, and provides a model for how front-end inference could be validated at scale. For edge hardware development, inference-as-a-service (IaaS) provides a test bed for validating model performance characteristics before committing to hardware synthesis.

A related community effort within fast machine learning (see Sec. IV) is focused on making it easier to deploy IaaS capabilities across both HPC centers and tier-3/analysis facilities for HEP and multimessenger astronomy applications. In particular, SuperSONIC (services for optimized network inference on coprocessors) [145] (Fig. 22) provides a server infrastructure designed for deployment at Kubernetes clusters equipped with GPUs. Experiments using SuperSONIC include the ATLAS and CMS experiments at the LHC and IceCube at the South Pole.

*Data access infrastructure.* AFs integrate with distributed data management through caching proxies (XCache/XRootD) that enable transparent access to remote data while caching locally for faster repeated access—essential when iterating on ML training with the same datasets. Global namespace support (via Rucio and the WLCG file transfer service) allows users to reference data by logical name rather than physical location, simplifying workflows that span multiple facilities and abstracting away the complexity of multisite storage federations.

For ML development workflows specifically, data access patterns differ substantially from traditional HEP analysis. Training runs repeatedly read the same files in randomized order across epochs. This places different demands on caching infrastructure than the single-pass reads characteristic of event processing. Emerging approaches such as columnar data formats (ROOT RNTuple, Parquet) and analysis-level derived datasets (e.g., ATLAS DAOD_PHYSLITE) reduce the data volumes that must traverse the network, but also require caches and prefetch strategies tuned to column-selective access rather than sequential file reads. AFs that colocate storage with GPU resources can further reduce data-movement bottlenecks, ensuring that accelerators are not starved of input during training.

Looking forward, as triggerless and triggerless-adjacent detector architectures substantially increase the raw data volumes available for training front-end ML models, the data management layer will need to scale accordingly. Intelligent tiering—maintaining hot copies of actively used training datasets on fast non-volatile memory express (NVMe), dynamic random access memory (DRAM)-backed storage while archiving older datasets to tape—will become a standard AF capability rather than an exceptional one.

*FPGA development tooling integration.* A critical capability for edge hardware development is the integration of FPGA synthesis and development tools within AF infrastructure. Some facilities have begun deploying Xilinx/AMD development tools [Vitis, Xilinx Runtime (XRT)] alongside standard analysis software, enabling researchers to perform the complete model-to-firmware workflow within a single environment.

Such integration requires substantial storage resources—FPGA development suites can require 150+ GB of disk space per installation, plus additional space for synthesis outputs. The computational requirements for synthesis can also be significant. However, this investment enables the following:

(1) *Synthesis-in-the-loop development*: Researchers can iterate between model training and hardware synthesis without transferring files between systems.

(2) *Codesign workflows*: Tools like `hls4ml` [28,34,146] and open-source frameworks can be integrated with vendor synthesis tools, enabling automated translation from trained models to deployable firmware.

(3) *Validation against hardware constraints*: Synthesis feedback (resource utilization, timing closure, and latency estimates) can inform model architecture decisions early in the development cycle.

The next frontier is integrating actual FPGA hardware into AF clusters, enabling researchers to validate firmware on physical devices as part of their development workflow. Such integration also introduces licensing and compliance considerations: Xilinx/AMD toolchains (e.g., Vivado/Vitis) often require node-locked or floating network licenses, and some advanced synthesis/implementation features may be separately licensed. AF deployments therefore need license-server infrastructure and access controls (and, in some cases, coordination with institutional procurement/export-control policies) to ensure tools are available to users while remaining compliant.

*Bridging analysis facilities and front-end deployment.* A critical gap exists between models developed in AF environments and their deployment on front-end hardware. The hardware community requires infrastructure that spans this gap, enabling the kind of differentiable, end-to-end optimization that is increasingly recognized as essential for maximizing physics performance. The typical workflow involves the following:

(1) *Interactive development*: Physicists use JupyterLab environments at AFs to prototype ML models using standard frameworks, with access to GPUs for rapid iteration.

(2) *Scale-out training*: Once a promising architecture is identified, training scales to batch resources for hyperparameter optimization and full-dataset training. Distributed computing frameworks (Dask, Ray) enable seamless scaling from interactive notebooks to cluster-scale computation.

(3) *Quantization and optimization*: Models are converted to fixed-point representations using tools like QKeras, with systematic studies of bit-width configurations to find optimal precision-performance trade-offs. For architectures like learnable logic networks, this phase involves the computationally expensive differentiable relaxation training.

(4) *Hardware translation*: Optimized models are translated to hardware implementations via `hls4ml`, FINN, or similar tools, generating firmware for FPGAs or specifications for ASICs. With FPGA development tools integrated into AFs, this step can occur within the same environment as training.

(5) *Validation at scale*: AFs provide the compute resources to validate that hardware implementations produce results consistent with the original floating-point models across representative datasets.

*Infrastructure requirements for codesign.* Enabling true codesign between ML model development and hardware implementation requires infrastructure that goes beyond traditional analysis facilities:

(1) *Integrated synthesis environments*: FPGA development tools must be accessible within AF environments. This requires substantial storage allocation (150+ GB for toolchains) and may require specific operating system environments. License server access must be managed across distributed users.

(2) *Hardware-in-the-loop development*: The natural extension of software-based synthesis is access to physical FPGA hardware for validation. AFs could integrate FPGA accelerator cards that researchers can target for testing, enabling validation of timing, resource utilization, and actual inference performance before deployment to detector systems.

(3) *Heterogeneous computing support*: Future detector systems will employ heterogeneous architectures combining different processor types. AFs should support similar heterogeneity—x86 and ARM CPUs, various GPU architectures, FPGAs—enabling researchers to develop and test on representative hardware.

(4) *Automated workflows*: As the field moves toward agentic systems that can orchestrate complex workflows, the infrastructure must support automated pipelines that traverse the full development cycle. This includes integration with version control, experiment tracking, and continuous integration systems that can validate deployability at each stage.

*Distributed computing for ML workflows.* Distributed computing frameworks like Dask provide a bridge between interactive development and large-scale computation. Users can develop code interactively in Jupyter notebooks, scale computations to hundreds of workers via integration with batch systems (SLURM, HTCondor), and monitor execution through dashboards. This capability is particularly relevant for front-end ML development, where systematic parameter sweeps (quantization studies, architecture search, hyperparameter tuning) require significant computational resources. The same infrastructure supports the computationally expensive training required for architectures like learnable logic networks, where training costs are high but inference is exceptionally efficient.

*Validation and monitoring.* AFs also play a crucial role in validating deployed front-end ML systems:

(1) *Scale testing*: Processing large data samples through both AF-based software implementations and hardware emulations to verify consistency.

(2) *Performance monitoring*: Comparing online (front-end) and offline (AF-based) inference results to detect degradation.

(3) *Retraining workflows*: When detector conditions change, AFs provide the infrastructure to rapidly retrain and validate updated models.

*Recommendations for optimal analysis facilities.* To maximize the effectiveness of AFs for front-end ML development, stakeholders need to consider the following:

(1) *GPU and accelerator resource planning*: Ensure sufficient allocation of modern GPUs (A100, H100/H200 class) or establish pathways to HPC resources for training workloads that exceed AF capacity. Plan for integration of FPGA accelerator cards for hardware-in-the-loop development.

(2) *FPGA tool integration*: Expand the deployment of FPGA development environments across AFs. This requires dedicated storage allocation (150+ GB per toolchain version), license management infrastructure, and potentially containerized environments to manage OS dependencies.

(3) *Inference-as-a-service*: Deploy and standardize inference server infrastructure (e.g., Triton [144]) across facilities

to enable scalable model validation and provide a test bed for front-end inference patterns.

(4) *Heterogeneous computing test beds*: Develop AF resources that mirror the heterogeneous architectures being deployed in detector systems. This includes not only GPUs but also FPGAs, and potentially ARM-based systems for testing the portability of ML inference code.

(5) *Workflow automation*: Ensure AF infrastructure supports automated pipelines that span training, synthesis, and validation. This includes robust APIs, containerized execution environments, and integration with experiment tracking tools.

(6) *Cross-experiment collaboration*: Leverage multiexperiment facilities to share front-end ML techniques and tools across collaborations. A common infrastructure for model development, synthesis, and validation benefits the entire community.

(7) *Benchmarking and validation infrastructure*: Establish standardized benchmarks that include front-end ML components, measuring both physics performance and hardware deployment metrics (latency, resource utilization, power consumption).

(8) *Training and documentation*: Develop tutorials specifically targeting front-end ML development workflows. Existing HEP training materials provide starting points, but specific guidance on the training-to-synthesis workflow is needed.

*Future directions: quantum computing integration.* Looking further ahead, analysis facilities may need to provide access to quantum computing resources as these technologies mature for HEP applications. The types of quantum technology most relevant for particle physics and the ML-HEQUPP scope are described in Sec. II C. DOE's QuantISED program and initiatives like the HEP Quantum Pattern Recognition (HEP.QPR) project at Berkeley Laboratory are already exploring quantum algorithms for pattern recognition in tracking applications. While current quantum hardware remains limited in scale, the infrastructure developed for heterogeneous classical computing—including job scheduling across diverse accelerators, containerized environments, and inference-as-a-service architectures—provides a foundation for eventual quantum-classical hybrid workflows.

Several national laboratories and cloud providers now offer access to quantum processors, and analysis facilities could serve as gateways to these resources for HEP researchers. Key requirements include the following: integration with existing batch and workflow systems; standardized interfaces for submitting quantum circuits; and hybrid execution environments that combine classical pre/postprocessing with quantum computation. The codesign principles emphasized throughout this document (e.g., tight integration of algorithms with hardware constraints) will be equally critical for realizing practical quantum advantage in HEP applications.

## III. PHYSICS APPLICATIONS

The past decade has seen a rapid expansion in the capabilities of AI/ML, resulting in a broad array of potential applications in high-energy physics. While the impact of AI/ML methods has been felt across the field, the constraints imposed by front ends and data acquisition systems limit the possibilities for model deployment in these critical areas. These constraints vary across experiments, but encompass restrictions imposed by low-latency inference requirements, radiation tolerance needs, and minimal resource availability, among others. Successfully enabling AI/ML solutions in these regimes is challenging, but can greatly enhance the physics performance of experiments. These enhancements are not confined only to planned or current experiments, or only to collider experiments or neutrino experiments, but cover a broad range of experiments and physics goals. In some cases, modern AI/ML offers the possibility for improving the performance of standard tasks such as particle identification, while elsewhere they could potentially enable techniques that have previously been impossible, such as smart data compression. While some challenges for these application are the same as those faced in the broader use of AI/ML in physics, there are also some challenges that are unique. In this section, we detail potential applications that could benefit from advances in HEQUPP ML. While these applications are grouped according to the relevant physics and type of detector technology, we stress that principles such as the need for codesign, effective general tools, and preservation/validation of models and solutions are shared across applications [147]. We believe that solutions for these shared problems represent ideal targets for future work, with specialization for the given application possible.

### A. Collider physics

Collider physics provides a rich test bed for hardware-based machine learning, presenting some of the most unique and stringent constraints of any big data system in terms of latency, bandwidth, radiation tolerance, and power. Applications of the ML-HEQUPP paradigm span current facilities (including both hadron and lepton colliders) and several future collider concepts, highlighting how ML deployed in specialized hardware can enhance real-time event selection and offline reconstruction [148,149]. The discussion is roughly organized by typical detector technologies, namely, vertexing, tracking, calorimetry, triggering, and quantum, to clarify common challenges, opportunities, and technology pathways across detectors and experiments.

#### *1. Vertex detectors*

*High luminosity LHC.* At the HL-LHC, on-device ML for vertex detectors is being extensively studied by the Smartpixels collaboration [150–153] from the standpoint of implementation as a full detector system. The sensor technology being targeted here is the same as for the phase 2 upgrade, but with a smaller sensor pitch of $12.5 \times 50\,\mu\mathrm{m}^2$, using a 28 nm technology node for the ASIC. The vertex detector is the last remaining subsystem of the CMS and ATLAS detectors that is not part of the level-1 trigger systems of those detectors. Using ML in the front end to compress data either to a latent space or useful physical cluster features provides the means to stream pixel data at the HL-LHC collision rate, providing the data necessary for heavy flavor tagging and detailed tracking, and thereby enabling the level-1 triggers of these experiments to significantly improve background rejection for events with displaced vertices across a range of $c\tau$ values relevant to $\tau$-leptons, b-hadrons, a variety of beyond-standard-model long-lived-particle scenarios, and narrow mass gap supersymmetry with soft-particle based signatures.

The Smartpixels team, within CMS, is developing a program to study the feasibility of implementation and deployment of such a detector within the next ten years, aiming for roughly half-way through HL-LHC. (Note that this is not yet ratified by the CMS collaboration and is purely an investigation of the possibility.) Of particular note is that if such an upgrade were to be realized, there would be a corresponding upgrade of the CMS level-1 trigger system as well as the pixel back-end system to handle the increased data rate and telecom technologies required to execute such a system. This notably increases the complexity of delivering such a detector, but the improvements could also be staged for improved feasibility. The group is also developing a detailed physics case of the HL-LHC to support such an aggressive addendum to the CMS phase 2 program that stands to improve the physics legacy of the HL-LHC.

Thinking to further future colliders, this sort of technology could be extended to entire tracking systems. ML-based processing elements that link hits, and then tracks segments, together could be integrated into support structures and detector bulkheads embedding a extremely low-latency track trigger into future tracking detectors. Moreover, a calibration strategy even only reading out the tracks is possible through statistical sampling of regions of interest around found tracks. Managing the material budget and cooling of such a detector will be challenging, but the resulting simultaneous reduction of data rate while preserving a robust calibration strategy yields an instrument capable of scaling through orders of magnitude in luminosity at hadron and muon colliders. Demonstrations in the coming years of such highly integrated technologies has the potential to set a direction in detector design.

*Electron-ion collider.* The Electron-Ion Collider (EIC), the future U.S. Nuclear Physics flagship facility at BNL, will collide high-energy polarized electrons with protons and nuclei at center-of-mass energies of roughly 20–140 GeV, enabling precision studies of quarks and gluons in protons and nuclei in both momentum and coordinate space [154]. The EIC will address fundamental questions in nuclear physics, including the origin of the nucleon mass, the nucleon spin, and the behavior of nuclear matter in regimes of high gluon density.

A key challenge is achieving excellent spatial resolution for charged-hadron tracking over the widest possible pseudorapidity range, including low transverse momentum ($p_T < 1\ \mathrm{GeV}/c$). Reconstructing and tagging c- and b-hadron decays central to the EIC program requires resolving decay vertices displaced by only a few tens of micrometers from the primary interaction point. This demands highly segmented sensors for optimal single-hit resolution and a minimal material budget to limit multiple Coulomb scattering and preserve impact-parameter resolution.

Monolithic active pixel sensors (MAPS) have been identified as the ideal technological choice to equip the future vertexing and tracking experiments at the EIC [155]. State-of-the-art MAPS integrate the sensitive pixel structure and full CMOS readout on the same silicon die, building on commercial technology. Compared to conventional “hybrid” pixel detectors [156–158] widely used in the community, in which a separate sensor and a readout ASIC are integrated via bump bonding, MAPS achieve reduced material budget, elimination of interconnects, reduced assembly complexity, and smaller pixel sizes.

Despite clear advantages, current MAPS face challenges under electron-ion collision conditions and EIC-driven constraints. Because it was designed for the extreme hit densities in ultrarelativistic lead-lead (PbPb) collisions at the LHC, MOnolithic Stitched Active pIXel (MOSAIX) employs eight high-speed links (5.12–10.24 Gb/s) to cope with a massive aggregate output bandwidth per sensor segment of $\mathcal{O}(30)\ \mathrm{Gb/s}$ [155,159]. Since the rate of truly hadronic PbPb collisions at the LHC is relatively low (about 50 kHz), the sensor was designed to operate with a relatively long integration time of 2 μs [160]. Finally, MOSAIX was designed for a small-sized tracking detector (ITS3) composed of only three small layers ($R < 4$ cm) with no forward disks, providing ample space for the large amount of services and cabling needed to power the sensors, and to route the dataflow out of the experimental area. In electron-proton and electron-ion collisions at the EIC, the typical event multiplicity is about 2 orders of magnitude smaller than one hadronic lead-lead collision at the LHC. On the other hand, the typical interaction rate of deep-inelastic scattering events considered in the ePIC design is about 500 kHz, which calls for shorter integration times for the vertexing and tracking detectors. The need for differential measurements as a function of both transverse momentum and pseudorapidity forces the experimental collaborations to build a much larger, nearly hermetic detector with forward disks, and substantially strengthens the constraints on minimizing the material budget introduced by cabling, services, and cooling systems, especially at forward pseudorapidities.

By substantially reducing the dataflow at the source and optimizing the data-aggregation process at the sensor periphery, and potentially at the matrix level, on-chip AI addresses the principal experimental limitations of MAPS in electron-ion collisions, namely, the need to minimize the material budget of tracking detectors, reduce the reliance on services, and shorten the integration time associated with signal collection and processing. Ongoing activities focus on the design, optimization, and physical implementation of an AI-enabled chip capable of performing local on-chip reconstruction and adaptive data reduction and compression. A fully scripted, end-to-end automated flow for neural-network training, quantization, and ASIC synthesis has recently been developed, enabling the first realistic estimates of power, area, and latency. Future work would establish the experimental boundaries for deploying AI devices, benchmark algorithmic performance against hardware constraints, and guide global optimization of future data-compression strategies.

Developing such techniques within the design of a real-life MAPS sensor, such as MOSAIX, has unique experimental value, as it enables a direct comparison between traditional on-chip data compression strategies adopted in the most advanced MAPS sensors and the potential benefits of AI-based approaches. Developing on-chip MAPS solutions for the EIC would build on the nuclear physics community’s MAPS expertise and benefit future MAPS-based experiments such as the FCC.

*FCC-ee.* MAPS are a leading candidate technology for future collider vertex detectors due to their fine granularity, low material budget, and cost effectiveness. At the FCC-ee, a MAPS-based inner vertex detector with $20 \times 20$ μm pixels is foreseen in major detector concepts such as innovative detector for $E + e^-$ accelerator (IDEA) [161] and CLIC-Like detector (CLD). Initial projections predict an occupancy in the inner layers that is at the upper end of estimated readout capabilities, primarily driven by low-$p_T e^+e^-$ pairs from beam-induced background (BIB) processes. This necessitates the exploration of a lightweight trigger system that reduces rates without compromising the stringent physics goals. Alternatively, if BIB can be mitigated during readout, an innermost pixel layer that is closer to the interaction point could be afforded. Given that backgrounds and data-rate requirements partially overlap with EIC requirements, it is highly likely that this work can benefit synergistically from the MOSAIX work toward a similar goal.

*Muon collider.* A muon collider represents a fantastic opportunity to explore the 10 TeV scale in a compact and power efficient machine. However, BIB due to muon decays near the interaction region pose a significant challenge for the experiment [162]. Tungsten nozzles placed in the forward region of the detector convert high-energy muon decay products into a shower of $\mathcal{O}(10^8)$ low-energy neutrons, photons, and electrons that leak into the detector each bunch crossing. BIB tends to produce hits that are out of time with respect to particles produced in collisions and generally do not point back to the interaction region.

Extreme data rates and complex spatiotemporal hit patterns strongly motivate on-chip AI for real-time front-end filtering [163]; without on-detector reduction, conventional trigger/DAQ systems cannot buffer or transmit the data within latency and bandwidth limits. Hit rates in the innermost layers of the vertex detector are expected to reach $\mathcal{O}(10\text{–}100)\,\text{hits/mm}^2$, calling for $25 \times 25\,\mu\text{m}^2$ silicon pixels with 30 ps temporal resolution. With an anticipated event rate of 30 kHz, data rates between 20 and 70 Gb/s per front end are expected.

Possible handles for reducing data volume include hit timing and pixel cluster shape. Selections based on hit timing alone, while possible with offline reconstruction and calibrations, cannot be implemented with sufficient precision online. However, signal particles produce clusters whose size follows the incident angle (e.g., short in the central region and longer forward) whereas most BIB hits come from ~10 MeV electrons curling along the beamline and form elongated clusters largely independent of detector position [164,165]. Several promising efforts to demonstrate data reduction based on pixel cluster shapes are ongoing. Simple selections on pixel cluster size have been demonstrated to reject approximately 20%−30% of pixel clusters from BIB with <5% signal cluster loss, resulting in an overall bandwidth reduction of up to 50%. Preliminary studies using a digital neural network for cluster classification further enhance background rejection with moderate signal efficiency loss. These studies exploit features such as the cluster charge profile or charge image, using lower-level simulation inputs, and are inspired by the *smartpixels* concept [45,151,166]. Sensor-level optimizations are also being explored; for example, increased sensor thickness may improve signal-background separation [152]. Together, these efforts motivate continued development of detailed simulations and hardware-algorithm codesign tools to fully exploit data reduction capabilities at the front end.

### 2. *Tracking detectors*

*Cluster counting in drift chambers.* Drift chambers are a proven detector technology capable of precision tracking over large volumes with minimal material budget. In these detectors, an incident charged particle ionizes atoms in a gaseous volume, and the freed electrons are attracted to a central wire by an electric field where they are measured. Traditional particle identification in a drift chamber relies on a measurement of the total energy deposited per unit length ($dE/dx$), but it has long been known that this method suffers from a long tail in the energy distribution as a result of the secondary electrons freed as the initial primary ionized electrons drift toward the collection wire. A measurement of the number of primary electrons collected per unit length ($dN/dx$) instead of $dE/dx$ can provide a significantly more accurate particle identification.

A crucial component of $dN/dx$, or "cluster counting," would be the ability to resolve individual electron arrival times and amplitudes on each wire with fine detail, which traditionally requires high-bandwidth digitization and large off-detector bandwidth. For high-rate environments such as future Higgs factories or $Z$-pole runs where sampling rates and triggers would generate data streams approaching terabits per second. The development of ML algorithms capable of analyzing the raw waveforms in real time and extracting $dN/dx$ directly on the front end would alleviate major bottlenecks in both data volume and bandwidth. Preliminary studies with simulated data have already shown promising capabilities to both distinguish primary and secondary electrons as well as fit within resource/latency constraints, as shown in Fig. 23 [167]. Future work is expected to explore the potential of more advanced AI/ML techniques and their ability to meet the front-end needs of particle identification in future experiments with drift chambers.

### 3. *Calorimetry*

Most physics analyses use some form of AI/ML to identify physics objects such as jets and electrons and/or for whole event classification. However, such an approach has generally been taken a long time after the detector was designed and constructed. It is therefore relevant to question whether a proposed design of a future calorimeter is optimal for the application of AI/ML techniques and, if such techniques are also to be used onboard the detector, what are the potential benefits of an integrated online/offline AI/ML approach.

The first area to consider is what aspects of calorimeter design would benefit the use of AI/ML for object and/or whole event topological recognition. These aspects include the level of granularity, inclusion of timing and tracking layers, interlayer communication, and traditional ECal/HCal separation or a more integrated design. There are trade-offs in terms of complexity versus efficiency for online or offline ML algorithms, increased information and reduced misassociations (e.g., in PFAs) versus increased power requirements and cooling/material needs, complexity of design versus calibration and systematic errors determination, and size of data structures and propagation through online/offline systems.

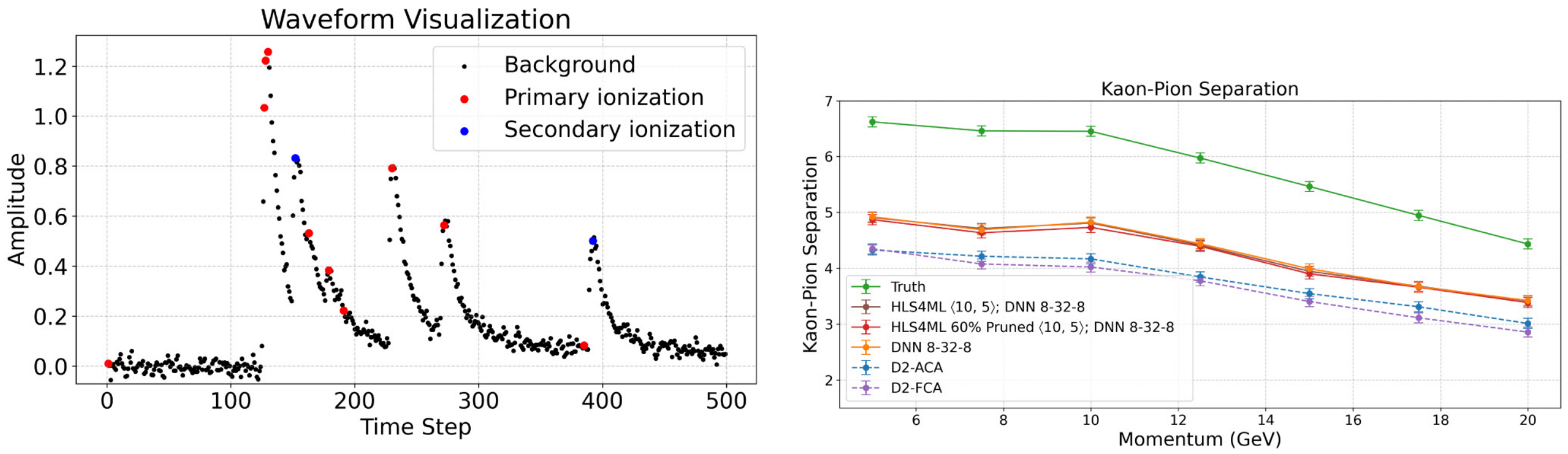

FIG. 23. Example drift chamber simulated waveform used to develop ML methods (left), and corresponding PID performance in terms of pion-kaon separation across particle momenta (right).

An integrated online/offline AI/ML calorimeter workflow must consider both present front-end and DAQ constraints and realistic projections of future latency, bandwidth, and on-detector processing capabilities on the deployment timescale. Relevant topics include the feasibility of low-latency neural particle-flow reconstruction combining tracking and calorimetry, the potential performance and operational advantages of ML/topology-based pattern triggers relative to traditional trigger menus, the value of persisting compact outputs of online ML as seeds or constraints for offline reconstruction, and the role of the DAQ as a feature-extraction and data-organization stage that structures inputs for downstream offline AI/ML algorithms. This paradigm also covers operations such as monitoring, calibration, fault detection, and data-quality assurance, for which a fully ML-based system will require very careful design and verification due to the potentially irrecoverable nature of decisions or actions. This section covers a variety of first efforts toward addressing such challenging questions of an end-to-end AI/ML approach to calorimeter design, implementation and operation.

*EIC forward hadron calorimeter.* The Longitudinally segmented forward Hadron Calorimeter (LFHCal) is designed as a plastic scintillator-steel sandwich calorimeter, which is transversely segmented into $5 \times 5\,\mathrm{cm}^2$ tiles and each tile is coupled to a silicon photomultiplier. The aggregate of the signals from all the photomultipliers are routed out of the detector and digitized by external readout electronics based on the H2GCROC3, a front-end readout ASIC, developed for the CMS HGCAL project.

There are three construction units of the following type. There are 8M modules $10 \times 20 \times 140$ cm$^3$, 4M modules $10 \times 10 \times 140$ cm$^3$. Insert modules which will be installed in the two halves surrounding the beam pipe. The 8M and 4M modules consist of tungsten and 61 layers of steel interleaved with scintillator material with a transverse tower size of $5 \times 5\,\mathrm{cm}^2$. Multiple consecutive tiles are summed to seven longitudinal segments per tower. The insert modules will consist of 10 layers of tungsten and 54 layers of steel interleaved with scintillator. Hexagonal tiles of 8 cm$^2$ are going to be installed, each of which will be readout individually. The high granularity of the detector will aid in distinguishing particle shower maxima close to the beam pipe. The readout of the detector needs to be designed to cope with its high granularity. There are 60 928 readout channels corresponding to the 8M and 4M modules, while there are 23 400 readout channels for the insert modules.

The LFHCal uses the HGCROC for its readout. This ASIC was developed by the OMEGA group for the CMS HGCAL project. There are several advantages associated with this ASIC, such as its high dynamic range and low-power consumption (2W per chip). There are 78 channels per ASIC and the time of arrival is 25 ps. The HGCROC was chosen due to its strong synergy with ALICE FoCal readout R&D. The test boards are designed and produced and bear interfaces compatible with commercial digitizers.

*Dual-readout calorimeters.* Integrating *dual-readout* (DR) techniques [168,169] with *high-granularity* designs and advanced *timing* capabilities represents a promising R&D direction in calorimetry. Simultaneous measurement of Cherenkov and scintillation light allows the electromagnetic shower fraction ($f_{\mathrm{EM}}$) to be determined on an event-by-event basis, mitigating the dominant uncertainty in hadronic energy reconstruction. Enhanced transverse granularity reveals fine shower structure, resolves overlapping composite showers, and improves particle-to-shower associations crucial for particle-flow algorithms.

*Crystal.* In crystal-based dual-readout calorimeters, scintillation light and Cherenkov light are collected by the same photosensor, typically Silicon PhotoMultipliers (SiPM)-based photosensors. The temporal overlap of the two components, combined with the typically much larger scintillation amplitude, makes signal separation a major challenge. The two components have different wavelength spectra, so optical filtering can be used to enhance the relative Cherenkov contribution. However, waveform-level pulse-shape analysis is still required to further decompose the mixed signal. In a typical crystal, Cherenkov light is emitted promptly, whereas scintillation light follows crystal-specific decay profiles, ranging from several hundred nanoseconds in slow crystals such as bismuth germanate (BGO), to sub-10 ns timescales in fast crystals such as lead tungstate (PWO). A representative waveform is shown in Fig. 24 (left) [170] for a bismuth silicon oxide (BSO) crystal, where the example event contains 150 scintillation photons and 50 Cherenkov photons after optical filtering.

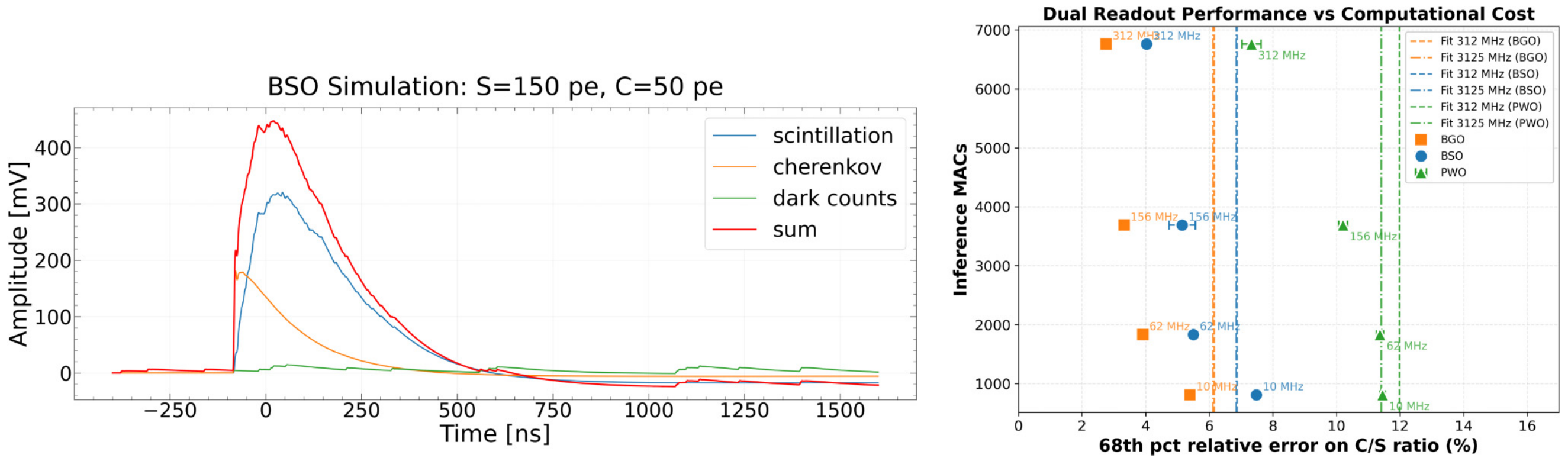

FIG. 24. Example simulated waveform for a BSO crystal dual-readout calorimeter (left). Sixty-eighth-percentile error on the reconstructed Cherenkov-to-scintillation ratio as a function of hardware resource consumption (right). Vertical dashed lines indicate the reference performance from offline template fitting based on a 3 GHz sampling ADC.

The standard template-fitting approach relies on predefined waveform templates for the scintillation and Cherenkov single-photon responses and performs iterative minimization during the fit. This procedure is computationally expensive and impractical for online implementation, thus necessitating full waveform readout and creating significant challenges in power consumption, bandwidth, and data movement. To address this limitation, a machine-learning-based waveform decomposition algorithm [170] has been designed for programmable on-detector hardware, such as embedded FPGAs. The algorithm is optimized to be hardware-friendly, low-latency, and capable of extracting the scintillation and Cherenkov photon yields directly from digitized waveforms, enabling data reduction at the source. In addition, the particle arrival time is extracted as well, providing complementary timing information for online event reconstruction and trigger applications.

As demonstrated in Fig. 24 (right) [170], the proposed ML method achieves better performance than standard template fitting while satisfying real-time latency and resource constraints. Notably, with only a 10 MHz sampling ADC and fewer than 40k LUTs, the ML algorithm achieves 68th-percentile relative errors of 5.84%, 8.14%, and 11.47% on the Cherenkov-to-scintillation ratio for BGO, BSO, and PWO crystals, respectively. The latencies are all below 25 ns. This performance is comparable to that of offline template fitting based on a 3 GHz sampling ADC, demonstrating the strong potential of deploying ML algorithms on programmable on-detector hardware for real-time dual-readout calorimeters.

*Fiber.* In fiber hadronic DR calorimeters, because relativistic particles travel at $v \approx c$ while optical photons propagate at $v \approx c/n$ (where $n$ is the fiber refractive index), precise timing information can be used to reconstruct the longitudinal shower profile. This timing frontier also enables out-of-time pileup rejection, particle identification, attenuation corrections, and full reconstruction of the shower position, energy, and time. Modern AI/ML frameworks such as CNNs and GNNs can readily exploit these highly correlated features to maximize reconstruction precision. Figure 25 demonstrates this potential by showing simulated energy resolution improvements as a function of timing resolution [171].

Achieving high-precision timing in high-granularity calorimeters requires advanced waveform-processing techniques. Preliminary simulation studies have demonstrated that spectral deconvolution using an long short-term memory (LSTM)-based algorithm can effectively extract photon arrival times from raw waveforms, enabling reconstruction of the shower longitudinal development (i.e., energy as a function of longitudinal depth), as illustrated in Fig. 26. A critical next step is to investigate the pruning and simplification of these algorithms for real-time execution. These adaptations must operate within the strict power and bandwidth constraints of front-end readout systems. In doing so, it is essential to ensure that reductions in algorithmic complexity do not compromise the performance of downstream particle and energy reconstruction or high-level physics observables, such as jet energy scale and mass resolution.

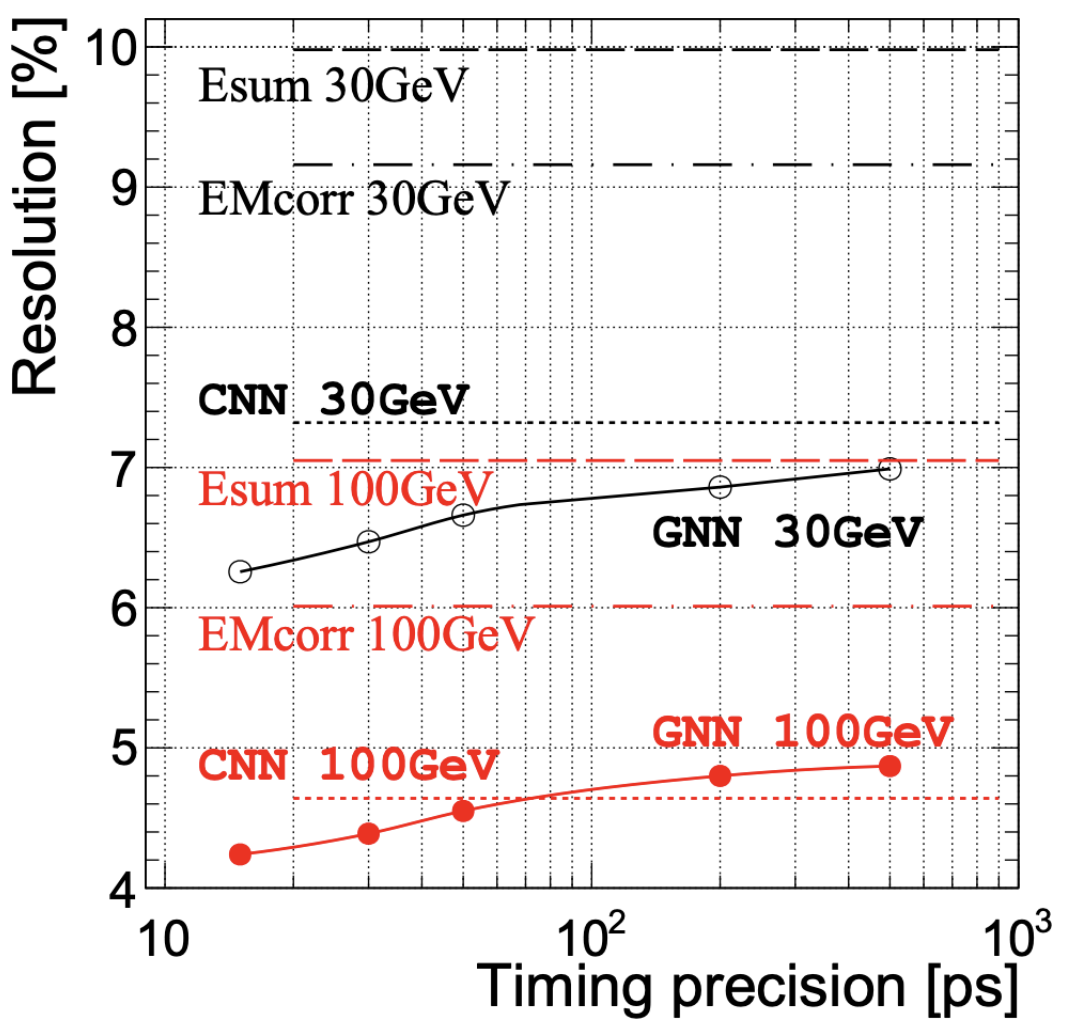

FIG. 25. Energy resolution for pion showers at 30 GeV (black) and 100 GeV (red), comparing a classical hadron calorimeter using energy summation (“Esum”), dual-readout correction (“EMCorr”), and machine-learning-based reconstruction algorithms (“CNN” and “GNN”). The figure is reproduced from Ref. [171].

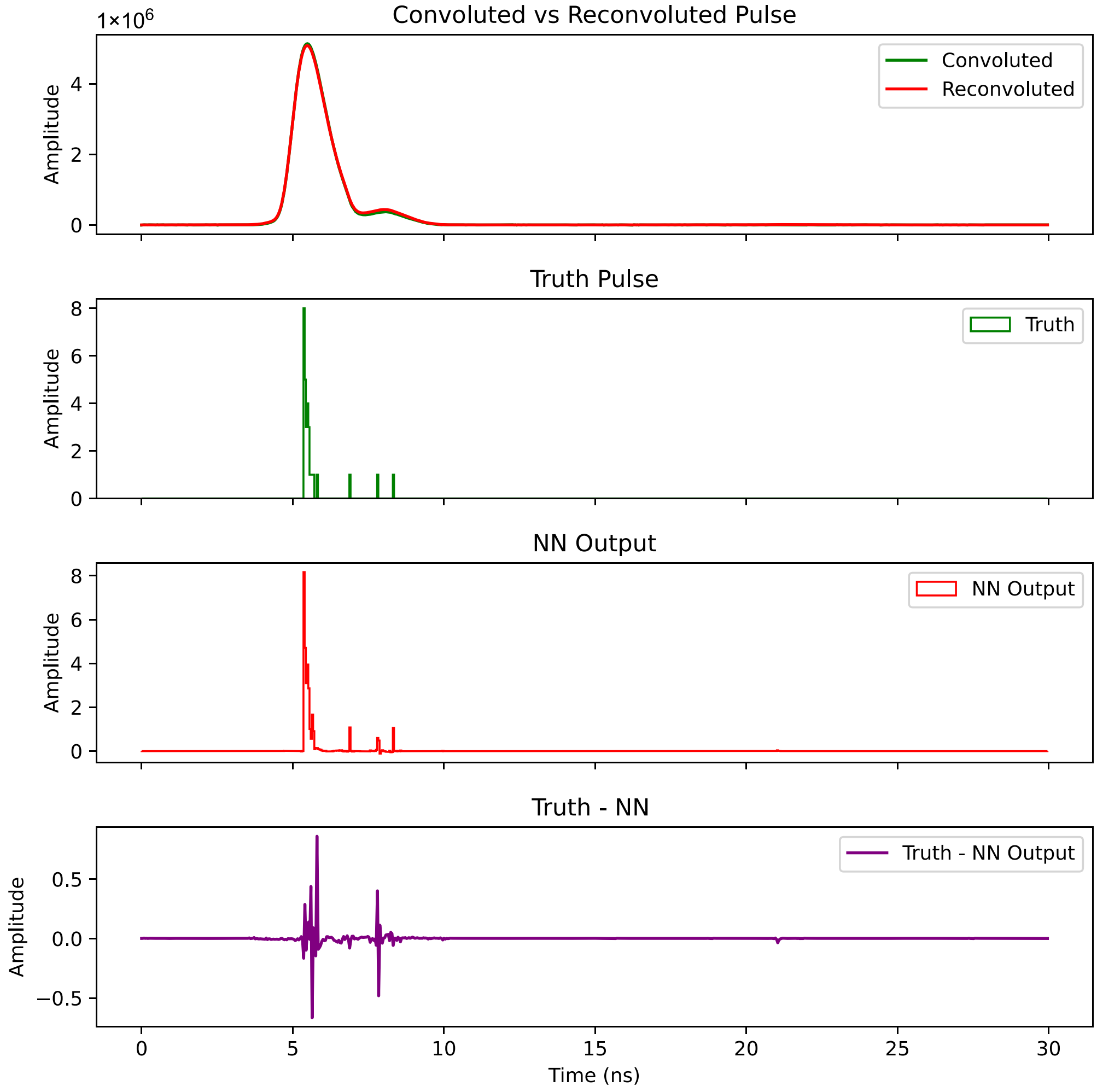

FIG. 26. LSTM-based deconvolution of calorimeter pulse shapes. The reconstructed photon time profile (third panel) is compared with the ground-truth photon time distribution (second panel), demonstrating good agreement (fourth panel).

*Pixelated calorimetry.* Monolithic active pixel sensors used in future sampling electromagnetic calorimeters at collider experiments can provide excellent detection and measurement of gamma- and electron-initiated showers, as well as the energy deposited by hadrons passing through and eventually completing their energy deposition in the hadron calorimeter [172]. This will enable excellent input to the particle flow analysis built from the use of both tracking and calorimetry. There are possible improvements to performance and data collection from application of machine learning techniques.

The hits in the pixels of the electromagnetic calorimeter will be sparse, as illustrated in Fig. 27. This shows two simulations of 10 GeV gamma showers with 25 um × 100 μm pixel sensors sampled every 0.7 radiation lengths. Layer 10 hits show a typical transverse distribution after about 7 radiation lengths, and layer 18 after about 12 radiation lengths. The vacancies are obvious, motivating readout optimized for sparse data. With an example design (silicon detector (SiD) for the linear collider [173]) the calorimeter contains about 5 × 108 pixels, while typically less than 106 will be active during any beam crossing event. For future electron-positron colliders, the event rate is generally moderate enough (with the possible exception of the teraZ factory) that on-detector sparsification and organization can simplify and optimize the operation. The individual layer process is generally straightforward, although multiple threshold applications might be of benefit and need machine learning approaches. Matching of layers would require more sophisticated analysis where machine learning might have an additional impact, with the obvious place for such processing in data collection/compression nodes that collect from the full calorimeter. While the pixels are generally isolated, as the figure shows, cluster formation based on machine learning approaches during readout could benefit on-line analysis. It should be noted that electron-positron linear colliders are expected to be trigger-less.

One critical aspect that needs understanding is any significant increased power required by sensors with readout involving machine learning, and designs for handling any such thermal load without harming performance.

*LHCb ECAL Upgrade II.* The LHCb Upgrade II [174], foreseen to operate after the Long Shutdown 4 period of the LHC, will produce data rates of about 200 Tb/s, necessitating major upgrades of the detector including the front-end electronics. One example where improvements of the performance are needed is the electromagnetic calorimeter, the so-called PicoCal. The readout is based on photomultiplier tubes (PMTs), which produce a voltage signal proportional to the number of scintillation photons that are produced in the active detector material. To cope with FE-back-end data rates, each channel can transmit only two ∼10-bit values per bunch crossing to describe the pulse shape, motivating highly efficient on-detector compression to retain this information. Preserving pulse-shape information improves pileup mitigation and energy resolution.

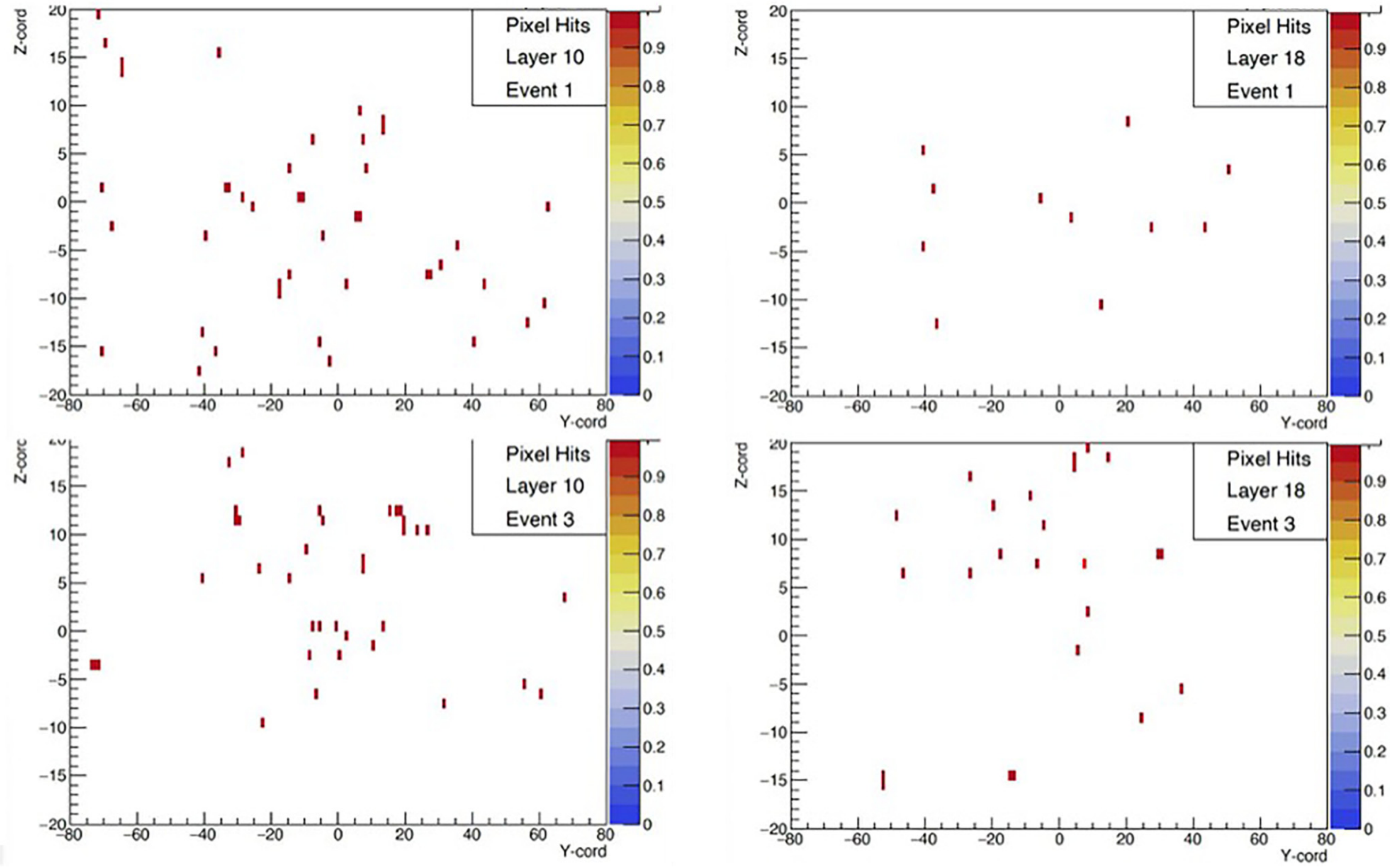

FIG. 27. Transverse distribution of hits for 25 um ($y$) × 100 μm ($z$) pixels around the axis of 10 GeV gamma showers for two events in the SiD ECal. Plots on left show distribution in layer 10 and plots on right in layer 18 (first layer is layer 0). The plots show the hits for 16 $mm^2$ around the core of the shower.

As the on-detector electronics are subject to the high-radiation environment of the LHC, PolarFire FPGAs are used to implement the compression algorithm. The PolarFire FPGAs use a flash-based architecture that inherently makes them less susceptible to single-event upsets. In order to allow for a broader adoption of machine learning on those FPGAs by the community, the `hls4ml` [28,175] framework is extended to also include support for the PolarFire FPGAs.

The compression algorithm is based on an autoencoder architecture, using a latent space comprising two 10-bit floating-point numbers where four bits encode the integer base. This representation is chosen using a quantization-aware training procedure [35,176,177] employing a search over latent-space configurations with different bit representations. Comparing the mean squared error (MSE) as illustrated in Fig. 28 shows that this choice offers the best performance compatible with the constraints imposed by the data rates. The training is performed using a dedicated Monte Carlo simulation of the LHCb PicoCal prototype [174], producing realistic signal responses under the expected conditions. In addition to greatly reducing the data rate, the compression algorithm also improves the timing resolution. For this purpose, a constant fraction discrimination (CFD) algorithm [178] was used to determine the time stamp of a simulated pulse and compare it with the true arrival time. As shown in Fig. 29, a twofold improvement in time-stamp resolution is achievable for compressed pulses, which are smoothed and thus background suppressed. Future work may replace the MSE with physics-driven loss terms (e.g., reconstructed-particle energy resolution), optimizing how information is encoded in the latent space for downstream reconstruction.

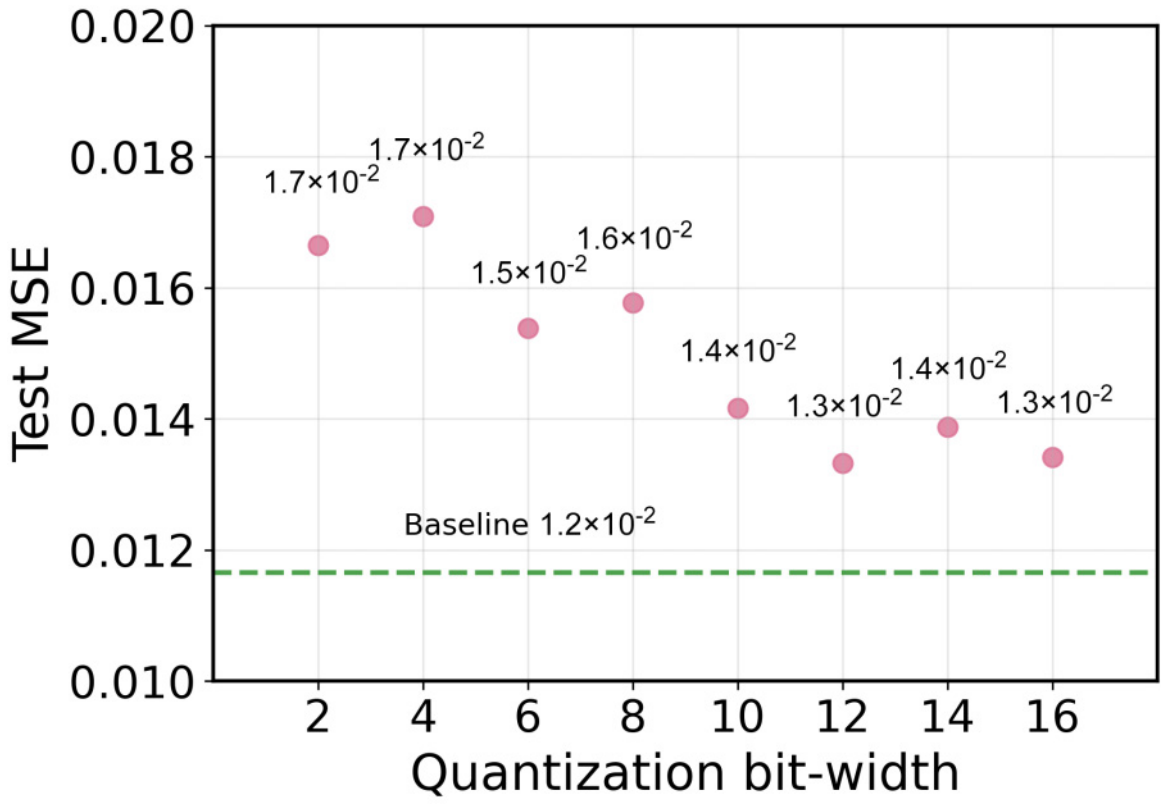

FIG. 28. Impact of the quantization on the autoencoders reconstruction performance measured using the MSE. The baseline corresponds to the unquantized training performance.

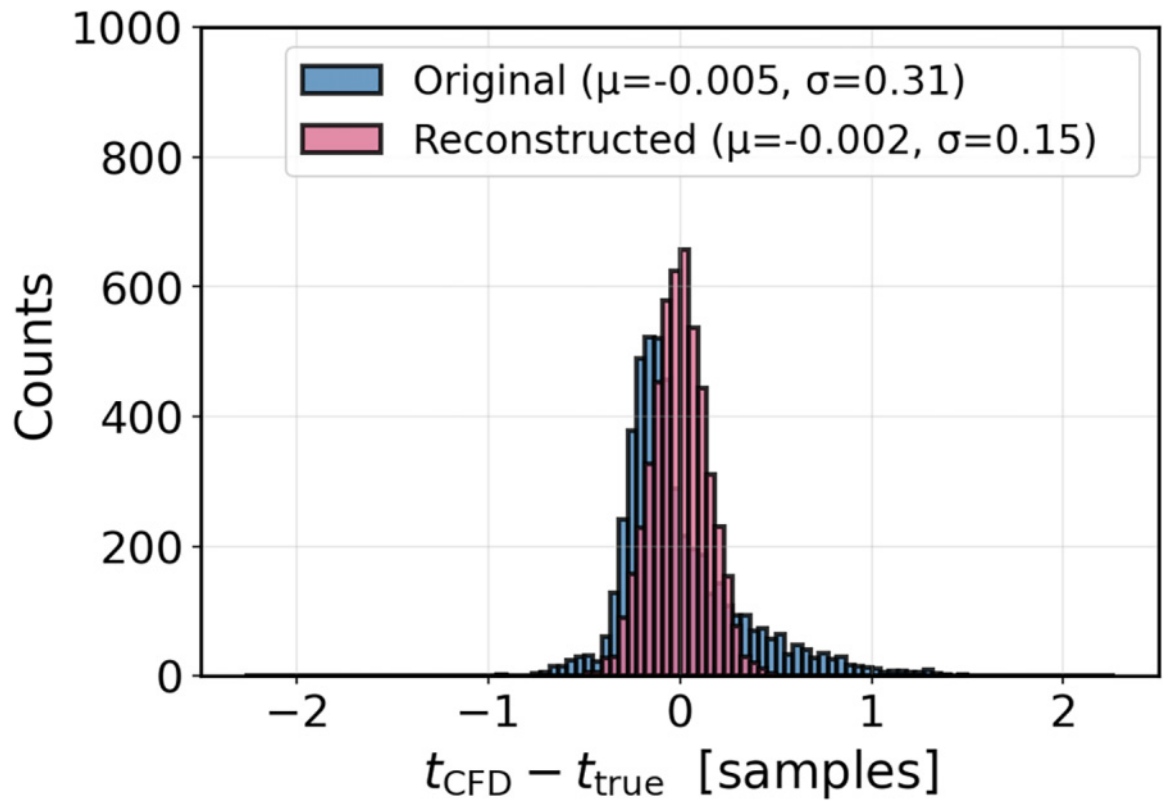

FIG. 29. Residual distribution between the CFD reconstructed timestamps of the original and compressed pulses. The time axis is given in the discrete sampling units of the ASIC discretizing the signal from the PMTs.

With the newly introduced support of PolarFire FPGAs in `hls4ml`, which will be publicly available from version `v1.3.0` onward, the model is synthesized, indicating minimal utilization of hardware resources. The algorithm achieves a maximum frequency of 234 MHz, well above the target of 160 MHz. Using an inference latency of 25 ns corresponding to four clock cycles at 160 MHz and an initiation interval of 4, the implementation only utilizes 3.1% of the lookup tables and 0.3% of the dedicated math blocks. This is well within the bunch crossing rate of 40 MHz at the LHC.

### *4. Triggers*

As a critical low-latency filtering stage necessary for modern high data rate colliders, the trigger is a natural entry point for ML-HEQUPP work into collider experiments. As a result, early conceptualizations of ML-based triggers date back to Tevatron experiments such as CDF [6] and D0 [5]. Current ML-based triggers span standard algorithms such as MET regression [179] as well as specialized algorithms that, for example, target long-lived particle decays in muon detectors [180]. Several specific ML-based trigger efforts focusing on AD are covered below.

The advances covered in this paper with FPGA/ASIC implementations, aggressive quantization and compression, and model-hardware codesign, make it increasingly plausible that ML-based components will become a default choice for many (if not all) triggers in future collider runs. Where ML can provide superior background rejection or improved resolution at fixed rate, an ML-based trigger can offer meaningful gains in physics acceptance within the same latency and power profile. That said, widespread replacement of all trigger algorithms is neither automatic nor guaranteed: Deployment will be driven by demonstrable performance gains, robustness to changing detector and pileup conditions, and the practical constraints of firmware resources, maintainability, and long-term operational stability. In light of these competing effects for different standard trigger paths, in some scenarios a simpler cut-based method will remain optimal.

*AXOL1TL: Anomaly extraction online L1 trigger lightweight.* The Anomaly Extraction Online L1 Trigger Lightweight (AXOL1TL) is a model-agnostic AD trigger at CMS that identifies rare, non-standard model events by analyzing high-level objects at the Global Trigger. Traditionally, triggers rely on fixed object definitions and thresholds, which can overlook complex, unforeseen signatures. AXOL1TL addresses this by processing the three-vector information ($p_T$, $\eta$, $\phi$) of the ten highest-energy jets, four muons, four electrons/photons, and missing transverse energy. By flagging unusual events directly from data, AXOL1TL complements conventional approaches and increases sensitivity to unknown physics.

AXOL1TL employs a variational autoencoder (VAE) architecture trained unsupervised on zero bias collision data to learn the underlying probability density of typical background events. The algorithm was first deployed in 2024 and has recorded significant datasets, including 54.7 $\mathrm{fb}^{-1}$ for its core seeds. In an early version of the algorithm, in order to meet a strict latency requirement of 50 ns (2 FPGA clock cycles), only the encoder portion is deployed during real-time inference. At present, the AXOL1TL algorithm comprises an eight-layer neural embedding and VAE architecture that includes encoder-decoder and the full mean-squared error reconstructed. Analysis of AXOL1TL recorded data shows that while there is some overlap with conventional L1 triggers, AXOL1TL also demonstrates sensitivity to unique events that would otherwise be missed. Performance evaluations using simulated signals reveal that AXOL1TL is particularly sensitive to events containing well-clustered, SM-like decay products. First physics analyses based on the 2024 recorded data are currently in development.

*CICADA: Calorimeter image convolutional AD algorithm.* The Calorimeter Image Convolutional AD Algorithm (CICADA) is a signal-agnostic trigger for CMS designed to identify anomalies using low-level, unclustered calorimeter information. Unlike object-based triggers, CICADA processes raw energy deposits from the Calorimeter Trigger Layer-1, providing sensitivity to unusual energy topologies that might not be reconstructed as standard jets or leptons. The algorithm is implemented as a convolutional autoencoder that captures spatial correlations across the detector's $\eta - \phi$ cylinderical surface. The combined AXOL1TL and CICADA architectures in the CMS experiment trigger system can be seen in Fig. 30.

To ensure efficiency within the L1 hardware, CICADA uses a "teacher-student" training approach where a complex teacher model's knowledge is distilled into a quantized student model optimized for FPGAs. The student model achieves an inference latency of 81.25 ns (13 clock cycles). Deployed in 2024, CICADA has successfully recorded data and operates with a pure rate budget of approximately 150 Hz for its medium seed. Initial analysis of recorded data indicates that CICADA triggers on unique events not captured by standard L1 algorithms. While AXOL1TL focuses on clustered objects, CICADA excels at identifying high-multiplicity, low-energy signatures, such as soft unclustered energy patterns. Rates for both AXOL1TL and CICADA

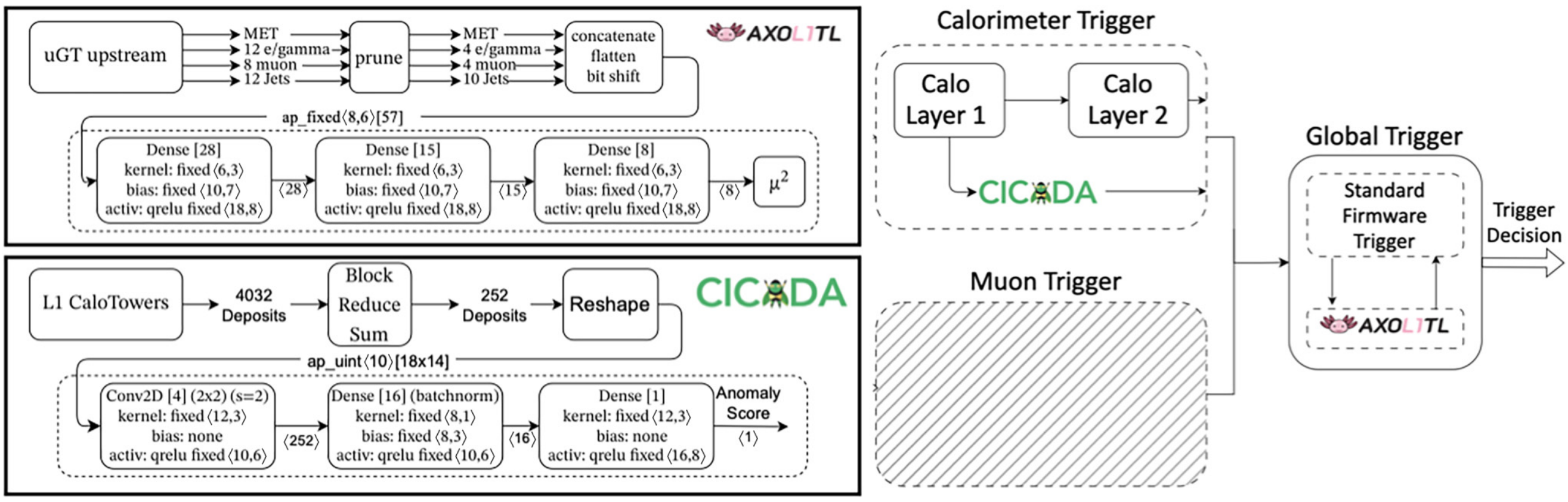

FIG. 30. The left shows the architecture and quantization schemes of the deployed AXOL1TL (top) and CICADA (bottom) models. The right shows an overview of the CMS L1 trigger system and where AXOL1TL and CICADA sit in the data processing chain.

over the course of a 2024 CMS data-taking run can be seen in Fig. 31.

*GELATO: A generic event-level anomalous trigger option for ATLAS in LHC run 3.* The Generic Event-Level Anomalous Trigger Option (GELATO) effort is a model-agnostic AD trigger for ATLAS that identifies rare events without assuming specific signal models. By flagging unusual events directly from data, such triggers complement conventional approaches and increase sensitivity to unknown physics. GELATO came on the heels of AXOL1TL, having been deployed for 2026 data taking and is the first such trigger for ATLAS.

GELATO is a two-level AD algorithm and consists of GELATO L1 and GELATO HLT (Fig. 32). Both are autoencoder-based architectures that train over four vectors of leading particles in the event, with the distinction that the L1 model employs a VAE with a generative adversarial network to refine the latent space during training without adding resources in the constrained scenario of inference at trigger level. GELATO L1 runs on 40 MHz input data and targets an output unique rate of 500 Hz, while GELATO HLT runs only over GELATO L1 events and targets 20 Hz output rate. As shown GELATO L1 and HLT both adopt an autoencoder architecture. In GELATO, each MSE is calculated by ignoring input zero values, which is referred to as "masked MSE."

The GELATO L1 and HLT models are trained on the 2024 enhanced bias (EB) dataset [181], which provides an unbiased event sample with weighted low-rate tails, and closely matches the detector conditions expected for 2025. The performance of the GELATO L1 and HLT pipeline is estimated using an EB test set and simulated samples of a variety of signatures such as dark photon models, heavy neutral leptons, and BSM Higgs. GELATO achieves signal efficiency gains of 5%–70% at L1 and 3%–30% at HLT, demonstrating sensitivity across a broad range of signatures.

The implementation of GELATO into the ATLAS trigger system is conducted using various tools. The GELATO L1 algorithm is converted to HLS using `hls4ml` [27] and synthesized with AMD Vitis. The GELATO HLT model is converted to ONNX format for implementation into Athena, the ATLAS software framework used for collision-event processing, including the HLT system. The GELATO trigger pipeline is constructed by combining GELATO L1 and HLT algorithms with dedicated object preselections and seeding algorithms, which have been tuned to meet the required latency of O(100 ms).

As of August 2025, gradual commissioning steps are in progress for safe deployment of GELATO. As a first step during early LHC ramp-up period in April to early May of 2025, the GELATO L1 rate at the ATLAS Central Trigger system was monitored for stability checks as shown in Fig. 33; however, it was not actually enabled for trigger decisions. It has been confirmed that the total trigger rate of GELATO L1 is as expected and is also consistent with the LHC luminosity and rates of other primary triggers. Following the rate stability checks, in the first runs with stable beams, GELATO L1 was enabled with a prescaled rate of about 1 Hz. In 2026, it is planned to enable the whole GELATO pipeline with full capacity at target unique trigger rates of 500 and 20 Hz for L1 and HLT, respectively. Studies are in progress to deepen the understanding of the triggered events and ultimately use them to search for BSM physics.

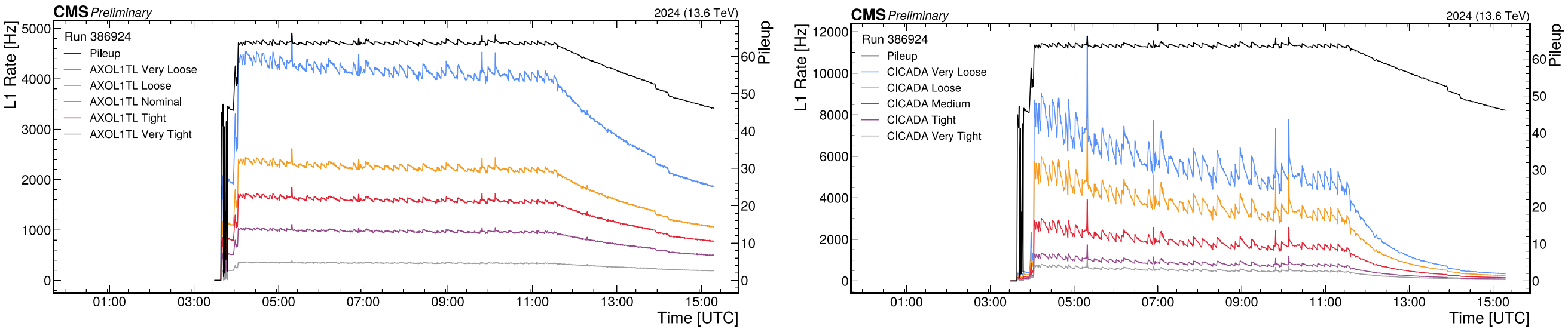

FIG. 31. Trigger rate of various AXOL1TL (left) and CICADA seeds during late 2024.

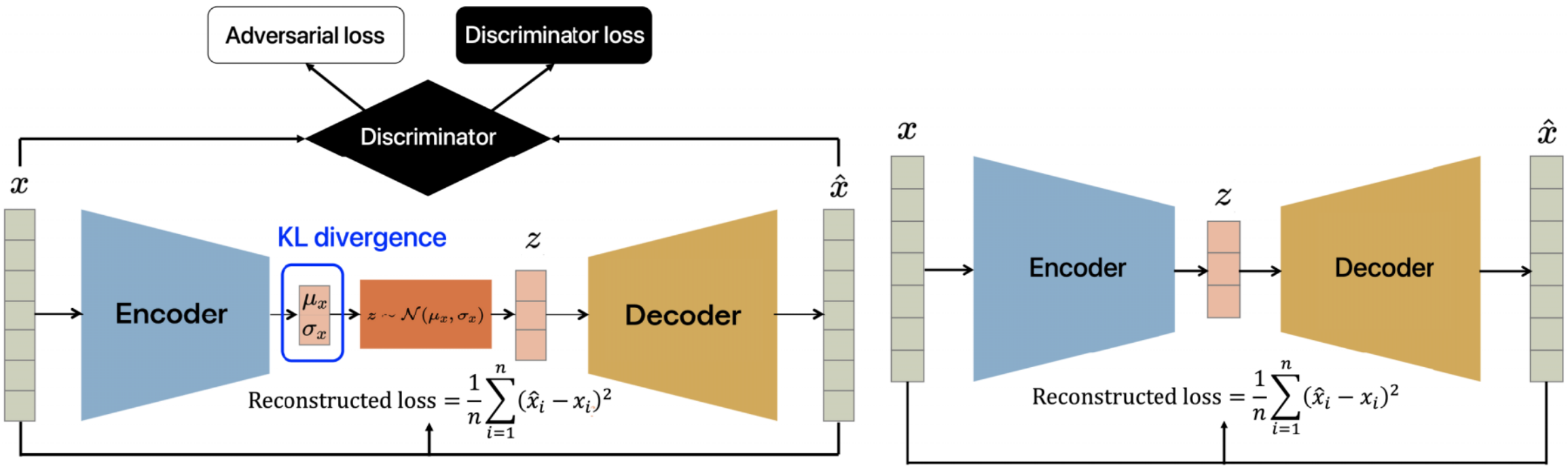

(a) GELATO L1 (b) GELATO HLT

FIG. 32. Model architectures for GELATO L1 and HLT.

*NomAD: Nanosecond anomaly detector at ATLAS in LHC run 3.* Nanosecond Anomaly Detector at ATLAS in LHC Run 3 (NomAD) is an anomaly detector that distills a fully trained VAE using decision trees for implementation on FPGA. The distillation is accomplished by regressing the latent space parameters with the software package and framework FWX-MACHINA, discussed earlier in this document. The training and distilling process is shown in Fig. 34. In the three-step process, the leftmost panel shows the neural network training of the VAE as the "mind." The middle panel shows the regressed decision tree as the "body." The rightmost panel shows the deployed VHDL model on the FPGA.

Implementation is accomplished using muons as input objects, targeting physics scenarios involving multimuon final states. The correlation plot between the VAE anomaly score and the regressed BDT is shown in the left plot of Fig. 35. Increased acceptance of, for example, final states from $B_s$ decays, are shown on the right plot of the same figure. Commissioning of this trigger is ongoing as of this document.

While this particular model targeted multimuon final states for deployment, the approach is general and can be applied to a wider scope of final states and physics scenarios.

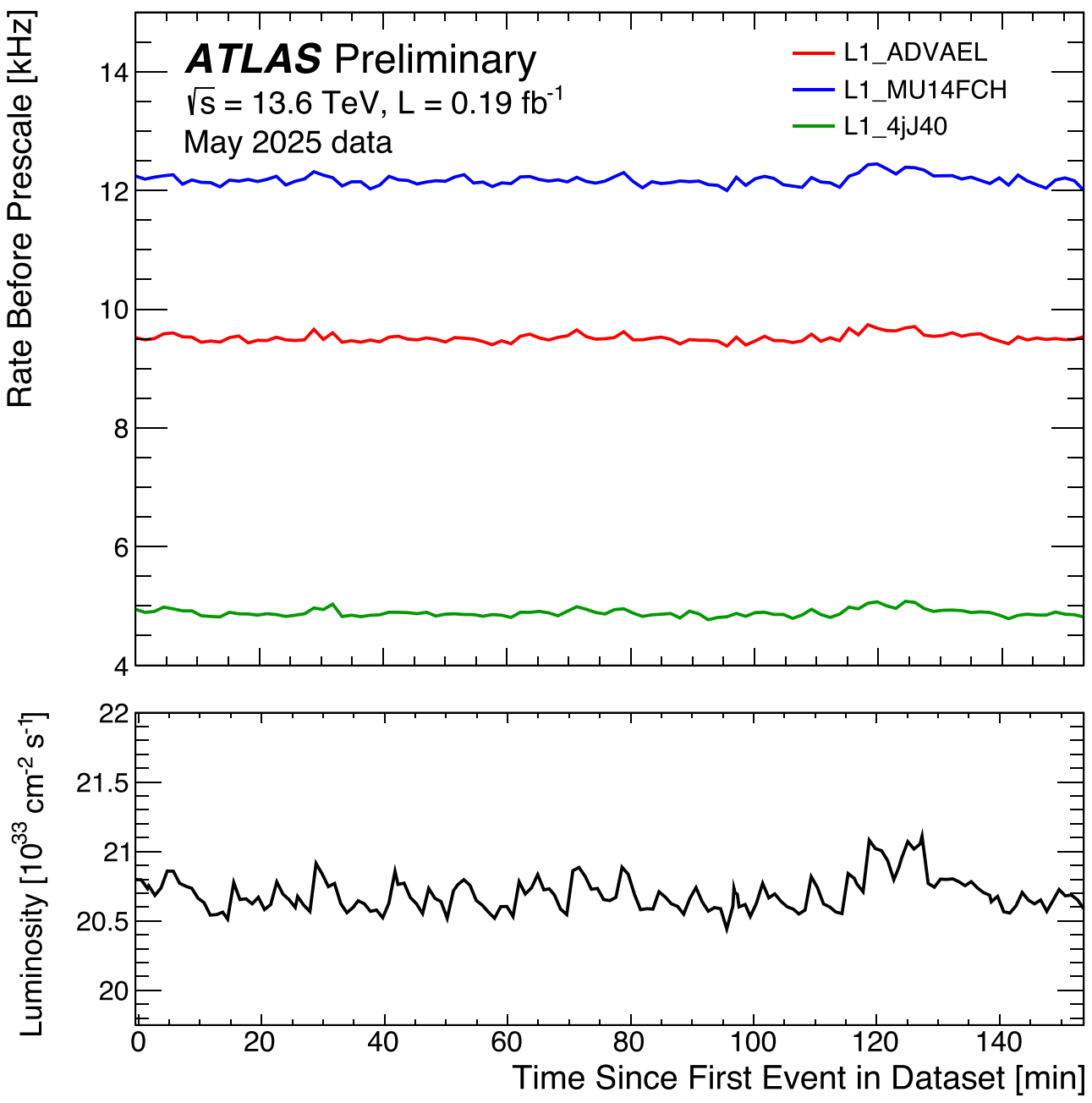

FIG. 33. The trigger rate of the GELATO L1 monitored at the ATLAS Central Trigger in May 2025. The GELATO L1 rate is shown in red, along with the single muon trigger rate in blue and four-jet trigger in green, shown as reference. The GELATO L1 rate is consistent with the reference trigger rates and the LHC instantaneous luminosity shown in the bottom panel.

### 5. *SuperKEKB/Belle II*

The SuperKEKB/Belle II experiment aims to perform precision measurements of CP violation in the *B*-meson sector and to search for physics beyond the standard model using unprecedented dataset of 50 $ab^{-1}$. Achieving the design luminosity requires the nanobeam scheme, but current accelerator operations encounter sudden beam-loss (SBL) events that can damage detector components or accelerator hardware. Fast beam-abort capability is therefore essential. In this context, AI/ML-based AD and real-time diagnostics are becoming increasingly important elements of the machine protection strategy.

(1) *BOR (bunch oscillation recorder)*: The BOR system uses beam position monitors (BPMs) with embedded [e.g., Radio Frequency System-on-Chip (RFSoC)] readout to measure bunch-by-bunch horizontal and vertical beam motion at nanosecond scales, enabling diagnostics of injection noise, beam-beam effects, and precursors to SBL events. Integrating AI/ML for real-time pattern recognition could detect subtle instabilities earlier than loss-based methods, enabling preemptive beam aborts to protect hardware and improve machine stability.

(2) *XRM (x-ray beam size monitor)*: The XRM system measures bunch-by-bunch vertical beam profiles using synchrotron x-rays, coded-aperture imaging, and fast silicon sensors, enabling rapid beam-size diagnostics and predictive monitoring. Hardware-aware ML inference on RFSoC-class embedded platforms could detect early SBL signatures and enable low-latency abort or mitigation decisions.

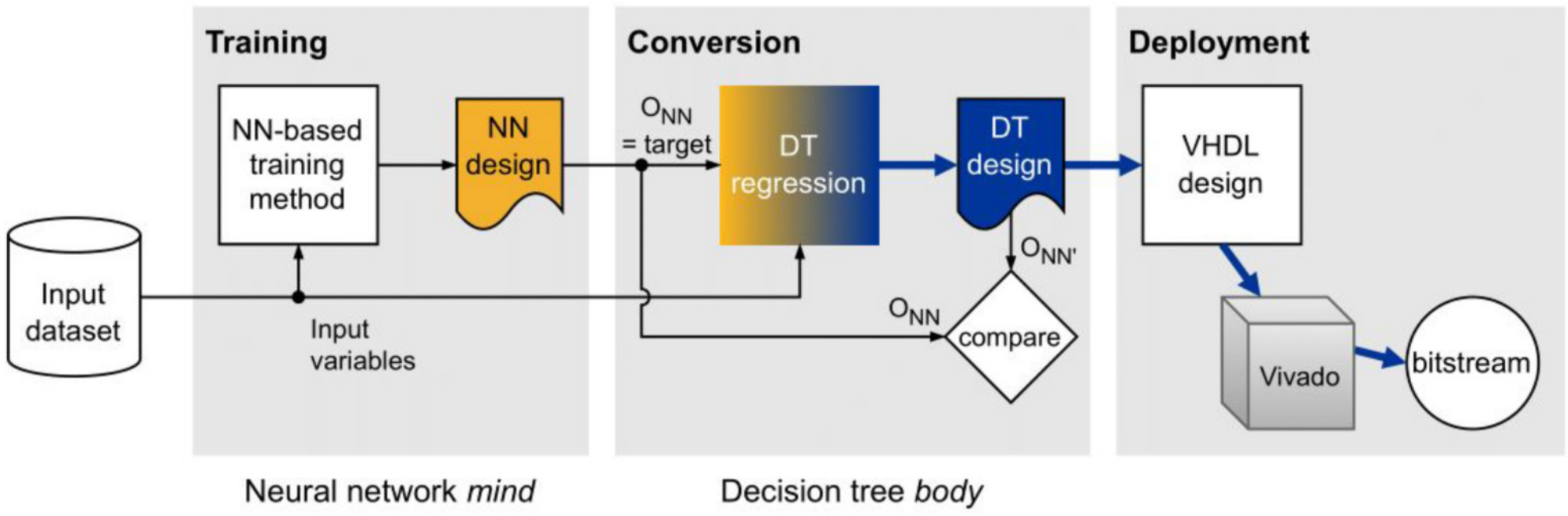

FIG. 34. Distilling process of the variational autoencoder for NomAD at ATLAS.

Belle II also offers opportunities to incorporate AI/ML into detector subsystems. The level-1 trigger uses $\mathcal{O}(100)$ FPGAs to process detector data in real time within a $\sim$5 μs latency set by front-end buffering. Devices such as Xilinx Virtex UltraScale FPGAs reconstruct global signatures (e.g., drift-chamber tracking, calorimeter clustering, and event topology), where ML-designed FPGA inference is increasingly important. The FEE receive ADC waveforms and timing information from smaller subsets of detector channels and are typically built around more resource-limited FPGAs (e.g., Xilinx Virtex-5, Spartan-6). Consequently, front-end ML must operate under tighter constraints in latency, logic resources, and power, motivating the development of highly optimized and hardware-aware inference models.

(1) *CDC (central drift chamber)*: The CDC is the primary tracking detector used for both offline reconstruction and the real-time level-1 trigger. An ML-based noise-reduction method was developed using ADC waveform processing. To avoid introducing additional latency or dead time in the trigger path, a channel-by-channel parallel implementation is essential. Given the limited resources of the Virtex-5 FPGA on the current CDC front-end boards, each ML module must be extremely compact, motivating the choice of a BDT over neural networks for higher resource efficiency. The resulting implementation adds only two clock cycles of latency and occupies approximately 5% of the Virtex-5 resources. An upgraded CDC front end, incorporating an ASIC and a Kintex-7 FPGA, is currently under development and is expected to support more sophisticated ML models to further mitigate crosstalk noise and enhance tracking performance.

(2) *TOP (time-of-propagation detector)*: The TOP detector provides charged-hadron identification in the barrel region, particularly for $\pi/K$ separation. AI/ML-based waveform feature extraction is currently being explored on the back-end readout PCs, where GPU-accelerated systems can support low-latency inference. The planned electronics upgrade over the next seven years aims to migrate these feature-extraction algorithms directly onto the front-end FPGAs. This would enable real-time event selection, reduce data volume, and improve the robustness of the identification performance under high-luminosity conditions.

(3) *ECL (electromagnetic calorimeter)*: The ECL is preparing a major upgrade of its front-end electronics, replacing the Spartan-3–based shaper DSP boards with a modern platform equipped with significantly larger FPGAs (e.g., Virtex UltraScale+). This transition opens the opportunity to incorporate AI/ML algorithms directly into the front end. Hardware-aware ML models could enable real-time waveform analysis for improved $e/\pi$ separation, double-pulse and pileup suppression that are increasingly important under higher background conditions.

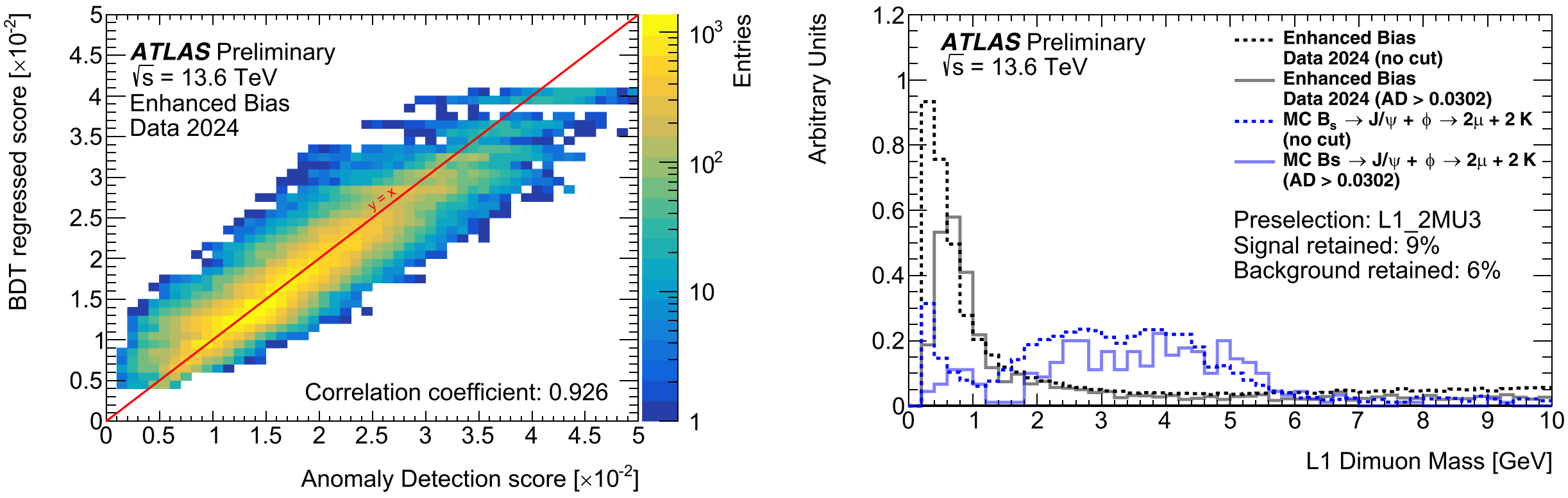

FIG. 35. Results for NomAD at ATLAS. (Left) Correlation of the VAE anomaly score and the regressed BDT score. (Right) Dimuon distributions of data and MC with and without NomAD.

(4) *KLM ($K_L$and muon system)*: The KLM consists of resistive plate chambers (RPCs) in the barrel and scintillator modules in the endcap, providing essential muon, and $K_L^0$ identification. The current level-1 trigger is based largely on hit counting, offering limited rejection of cosmic-ray and beam-induced backgrounds. A neural-network–based trigger that uses muon-track candidates and basic track-parameter information would significantly improve muon identification while reducing fake triggers. In the endcap, ML-based inference of photoelectron counts from the waveforms would improve timing resolution and strengthen muon and $K_L^0$ identification under high-background conditions.

(5) *Dispaced vertex trigger (DVT)*: The search for long-lived particles is a key part of the physics program at Belle II, focusing on neutral particles that decay into charged pairs from vertices displaced from the interaction point. The existing level-1 track trigger is tuned for prompt tracks and would likely reject these displaced-vertex events, motivating a dedicated DVT [182]. Implemented on FPGA boards within Belle II's pipelined, dead-time-free 5 μs level-1 latency, the DVT runs in parallel with current triggers and selects events where two oppositely charged tracks form a common displaced vertex. It performs track finding using multiple Hough transforms over a grid of assumed track origins, and then uses neural networks to identify the correct vertex via shape analysis of Hough clusters. A Hough-map-based preselection reduces the number of origin hypotheses and significantly cuts FPGA resource usage.

#### 6. Quantum technology

*Tensor networks for real-time triggering.* As a practical application of TNs in low-latency, resource-constrained hardware settings, one prominent example is the trigger system of CMS and ATLAS detectors [183]. The first level of the trigger system, the L1T, is an FPGA-based system that requires fixed latency, resource-efficient inference, and strict power constraints. While deep neural networks have demonstrated strong performance in this regime, they often lack intrinsic interpretability and structured control over correlations. These limitations can be overcome using TNs. In addition to offering competitive accuracy (Fig. 36), analyzing the pairwise correlations of the model's inputs via quantum mutual information (QMI) provides insight into the correlations within the model itself. QMI analysis enables reduction in effective long-range correlations across a tensor chain or tree, simply by permuting the input features, thereby lowering the parameter count and arithmetic complexity with minimal performance loss [101].

Further directions include development of TN-specific compression strategies that exploit gauge freedom, coverage of a wide range of trigger-level tasks including AD and particle identification and the establishment of standardized workflows (model ↔ interpretation → quantization → firmware) for integrating TNs into the broader HEP front-end machine learning ecosystem. Quantifiable and broad comparisons to classical ML, taking into account a variety of performance metrics and resource constraint criteria, will further help the particle physics community determine the best use cases for TNs and other quantum-inspired ML in future experimental contexts.

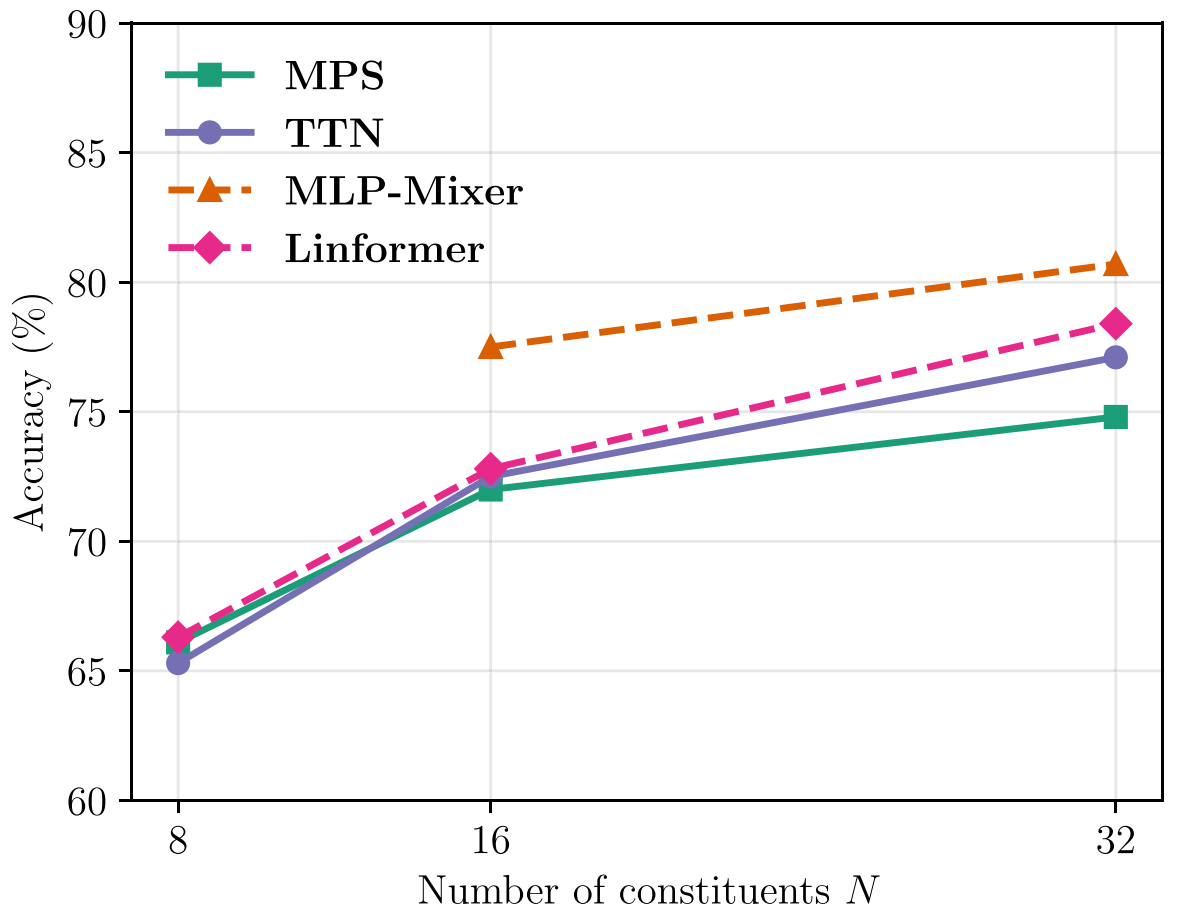

FIG. 36. Accuracy of MPS and TTN models reported in Ref. [101] compared for different number of parameters guided by increased input size and dimension between neighboring tensors. Models are both trained and evaluated using the dataset from Ref. [184] designed for jet tagging (multiclass classification) task. State-of-the-art classical ML models for fast inference on FPGA are shown as dotted lines for comparison [185,186].

*Continuous-variable photonic quantum extreme learning machines for fast collider-data selection.* CV QELMS, as described in Sec. II C 1, are particularly well matched to the constraints of front-end collider data processing and trigger systems. Inference latency is set by the optical path length and detector response and is therefore fixed, deterministic, and on nanosecond scales, independent of event complexity. Furthermore, photonic devices operate at room temperature and do not require cryogenic cooling, while benefiting from long decoherence times. Retraining to accommodate changing detector conditions, calibration drifts, or domain shifts in the data stream can be performed rapidly because no backpropagation, learning-rate tuning, or stochastic optimization is required.

Benchmark studies on representative collider classification tasks, including top tagging (Fig. 37) and Higgs boson identification, show that CV-QELMs consistently outperform parameter-matched classical multilayer perceptrons and match or exceed wider networks at larger training set sizes, despite training far fewer parameters [113]. Selected results are shown in Fig. 2. Notably, the QELM exhibits significantly lower variance across runs, reflecting the stability of fixed photonic random features and analytic readout training. These results indicate that Gaussian photonic substrates can supply compact yet expressive random feature maps without requiring quantum computational advantage, positioning CV-QELMs as a realistic hardware ML primitive for online data selection, and potentially even first-level trigger integration, at future collider experiments.

*Particle tracking and active particle sensing via quantum devices.* Prior work on quantum annealing for track clustering illustrates one of the earliest applications of quantum optimization to a core reconstruction task in collider physics [92], even valid more generally for particle tracking and reconstruction. Therein, the problem of clustering charged particle tracks to identify primary interaction vertices along the beam axis is formulated as a binary optimization problem that is amenable to solution on a quantum annealing device. Using a $D$-wave quantum annealer, track clustering is embedded onto the device's hardware graph, optimize annealing parameters, and benchmark performance against classical simulated and deterministic annealing under equivalent time budgets. Although current hardware limitations constrain the size and connectivity of instances that can be directly mapped, the results demonstrate a measurable performance gain for quantum annealing in modestly sized problems and establish practical strategies for embedding, parameter tuning, and hybrid workflows. By validating this approach on realistic collider topologies, the study provides a foundational proof of concept for quantum-accelerated reconstruction in high-pileup environments expected at the high-luminosity LHC. This approach can be extrapolated to other future collider facilities and their environmental conditions. Furthermore, quantum tomography can provide an efficient method for extracting angular distribution and quantum correlations among the final-state particles from collider data, as demonstrated in Ref. [187].

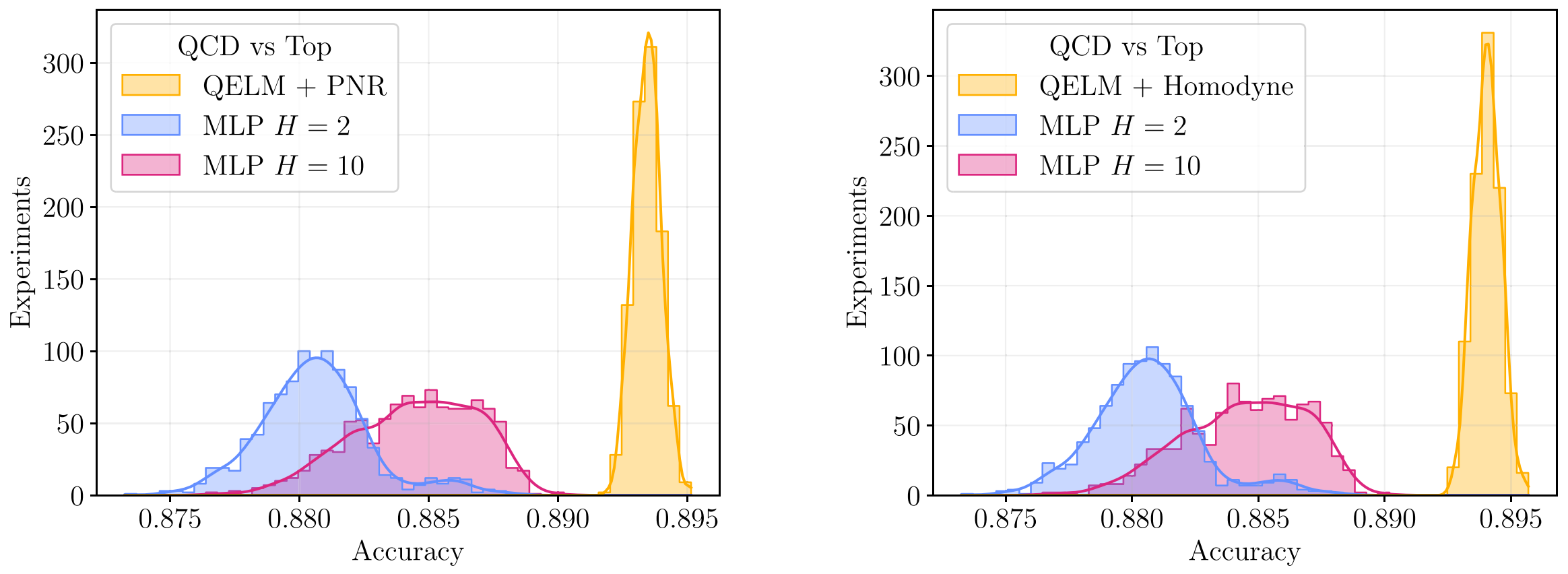

FIG. 37. QCD vs top jet classification with $F = 16$ features and $n_{\text{samples}} = 105$ from the `hls4ml` dataset [184] using logistic regression on the CV-QELM outputs. Photon-number-resolving (PNR) measurements (left) and homodyne measurements (right). Accuracy distributions are compared against MLP baselines with $H = 2$ and $H = 10$. The CV-QELM achieves higher mean accuracy and substantially lower variance than both MLP baselines.

Building on this foundation, two complementary research directions emerge that extend the scope and impact of quantum methods in experimental particle physics. While using different methods, these are synergistic to each other.

First, the vertex clustering paradigm can be extended from one-dimensional (1D) beam-axis clustering to full three-dimensional tracking. The comprehensive 3D reconstruction of charged particle trajectories within complex detector geometries is among the most computationally challenging tasks in high-energy physics. By casting full track finding and fitting as a global optimization over hit associations and track hypotheses, and expressing this as a form suitable for quantum annealers or hybrid quantum-classical solvers, it is possible to exploit global search capabilities to navigate the combinatorial complexity inherent in tracking. Key aspects of this extension include embedding detailed detector geometry and spatial correlations into the optimization representation, integrating physical constraints given by, e.g., magnetic fields, and developing scalable embedding strategies that leverage next-generation annealing hardware.

A second avenue is the exploration of arrays of qubits in gate-based quantum processors as direct sensors for weakly interacting or non-photon-interacting particles, including dark matter candidates. In this conceptual framework, qubits are repurposed from purely computational roles to act as highly sensitive detectors, where rare interactions with passing particles could induce perturbations in qubit states that are amplified and read out with high fidelity. Gate-based architectures, such as superconducting circuits or trapped ions, inherently support high coherence times and precision control, enabling quantum sensing protocols that surpass classical limits. By engineering tunable coupling between qubit arrays and external fields or sensing media, and by leveraging entanglement and error-mitigated readout schemes, such systems could probe minute energy depositions or decoherence signatures indicative of interactions with hidden-sector fields. This direction positions quantum processors not just as accelerators of existing algorithms, but as transformative hardware for particle detection and fundamental physics discovery. Limited access to gate-based systems will allow to establish figure of merits for expected sensitivities in given noisy realm of gate-based systems.

*Quantum-enhanced graph neural networks for particle tracking in high-energy physics.* Charged-particle tracking at the HL-LHC presents unprecedented computational challenges. With proton-proton collisions occurring at bunch-crossing rates up to 40 MHz and pileup approaching $\mathcal{O}(200)$ interactions per crossing [188], tracking detectors generate extremely dense hit patterns that lead to a combinatorial explosion in candidate hit-to-hit connections.

GNNs offer an attractive approach by reformulating tracking as a graph learning problem: Detector hits become nodes, candidate connections become edges, and the reconstruction task reduces to edge classification. While recent GNN-based tracking pipelines demonstrate strong performance, they typically require dense MLPs with wide hidden dimensions and numerous message-passing iterations, resulting in substantial parameter counts and computational overhead [189,190].

QML presents a complementary paradigm through parametrized quantum circuits that can represent nonlinear transformations in high-dimensional Hilbert spaces using relatively few trainable parameters [191]. Motivated by the potential for parameter-efficient expressivity, Ref. [192] explores a hybrid quantum-classical GNN architecture for edge classification in TrackML graphs [193]. In our approach, variational quantum circuits replace some classical linear layers in encoders, networks, and decoder. Quantum modules are implemented using PennyLane [114] integrated into an interaction network [190] baseline via the a charged object reconstruction network (ACORN) framework [194].

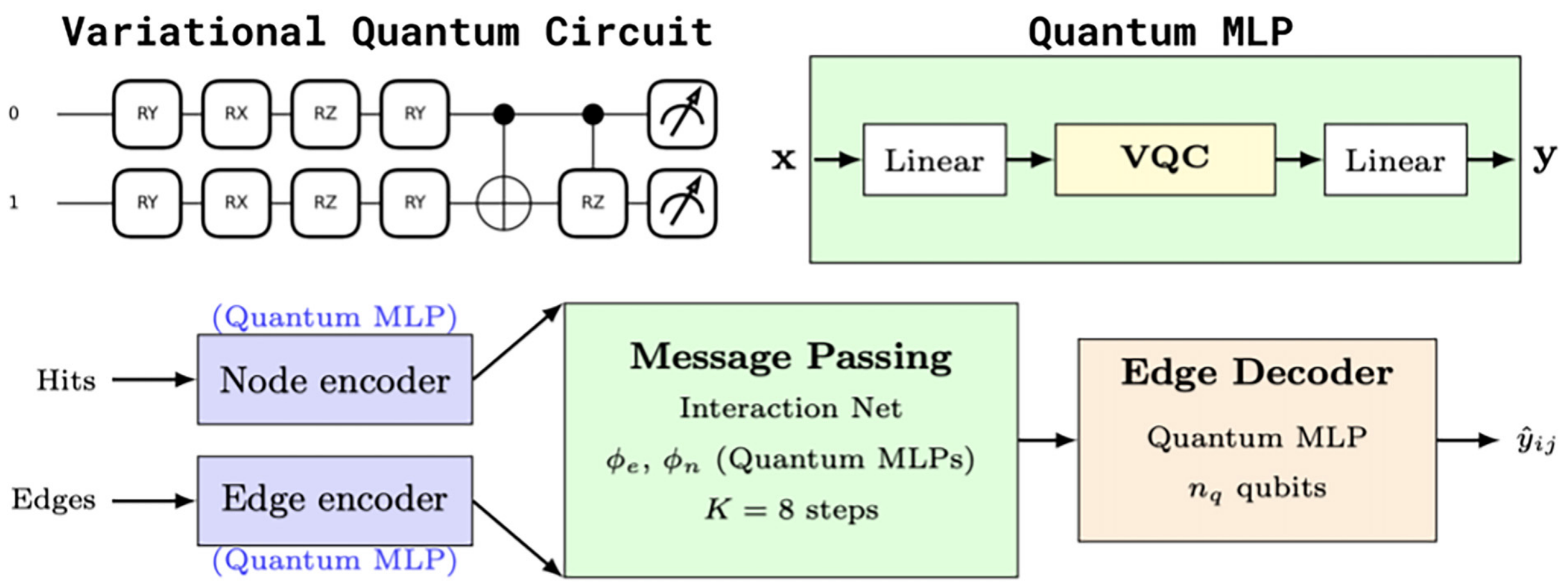

FIG. 38. Hybrid quantum-classical GNN pipeline. All MLPs (node encoder, edge encoder, message passing $\phi_e$ and $\phi_n$, and edge decoder) use quantum circuits in the hybrid model [114,190].

The hybrid architecture shown in Fig. 38 was implemented in which quantum circuits serve as drop-in replacements for classical linear layers within node encoders, edge encoders, message-passing functions, and the final edge classifier. The approach integrates PennyLane-based quantum modules into a classical GNN framework using ACORN and PyTorch Geometric, enabling end-to-end training on TrackML graphs. Quantum circuits with between two and eight qubits are evaluated, using angle encoding, entangling layers, and Pauli-$Z$ measurements to produce classical outputs.

Performance is evaluated using standard TrackML metrics, including receiver operating characteristic (ROC) area-under-curve (AUC), edge-level efficiency, and purity. The results shown in Table I demonstrate that low-qubit quantum models can achieve performance comparable to classical baselines while substantially reducing the total number of trainable parameters. Quantum circuits operate in $2^n$-dimensional Hilbert spaces, which enables compact representations. In our eight-qubit model, the quantum circuit components contribute 320 trainable parameters out of a total of 2553 parameters in the full model. This compares to the classical baseline with 2485 total parameters, all of which are in classical layers. The two-qubit configuration achieves the most dramatic parameter reduction, with only 711 total parameters (80 quantum) versus 2485 in the classical baseline and achieves similar classification accuracy with more than a factor of 3 reduction in parameter count. The larger quantum circuits (four and eight qubits) exceed the classical baseline across multiple metrics. The role of entangling gates in creating feature correlations for edge classification warrants further investigation. While the quantum models use fewer trainable parameters, practical efficiency must also account for the overhead associated with repeated quantum circuit sampling and measurement.

TABLE I. Performance across model configurations [114].

| Model | Hidden dim | Qubits | Total params | Quantum params | AUC (%) | Eff (%) | Purity (%) |
|---|---|---|---|---|---|---|---|
| Classical GNN | 32 | – | 2485 | 0 | 92.1 | 76.92 | 66.09 |
| Quantum GNN | 16 | 2 | 711 | 80 | 91.6 | 77.80 | 63.52 |
| Quantum GNN | 16 | 4 | 1325 | 160 | 93.2 | 79.17 | 66.67 |
| Quantum GNN | 16 | 8 | 2553 | 320 | 95.3 | 84.45 | 69.97 |

This study shows that hybrid quantum-classical GNNs are a promising direction for future investigation in particle tracking, particularly as quantum hardware and hybrid computing infrastructures mature. While the current results rely on classical simulation of quantum circuits and are limited to modest system sizes, the study provides an early demonstration of how QML techniques may contribute to parameter efficiency and expressivity in high-energy physics reconstruction pipelines.

*Superconducting microwire single-particle detectors as quantum trackers.* For decades, silicon-based sensors have been the workhorse of HEP detection. However, looking toward future colliders, traditional silicon is reaching its physical limits regarding radiation hardness, power dissipation, and timing resolution. Recent breakthroughs present the superconducting microwire SMSPD, representing successful migration of superconducting sensor technology from small-scale laboratory quantum optics to large-area, high-energy particle tracking [195,196]. Connecting to the quantum tracking algorithms presented earlier in this section, such hardware capability provides another angle in which particle tracking at colliders can substantially benefit from quantum hardware and the corresponding integration of ML-based operation and readout.

The first step required addressing the primary historical bottleneck of superconducting sensors: the scale-up problem [195]. Although SNSPDs excel at infrared photon detection, their micrometer-scale active areas have limited their use in HEP tracking. This team characterized a $2 \times 2$ mm$^2$ SMSPD array in high-energy test beams at Fermilab (120 GeV protons and 8 GeV electrons/pions), showing that wider "microwires" can still detect MIPs via Cooper-pair breaking. Achieving stable operation and 1.15 ns time resolution over a larger area, the study provides a first proof of concept for scalable superconducting-film tracking architectures for collider experiments.

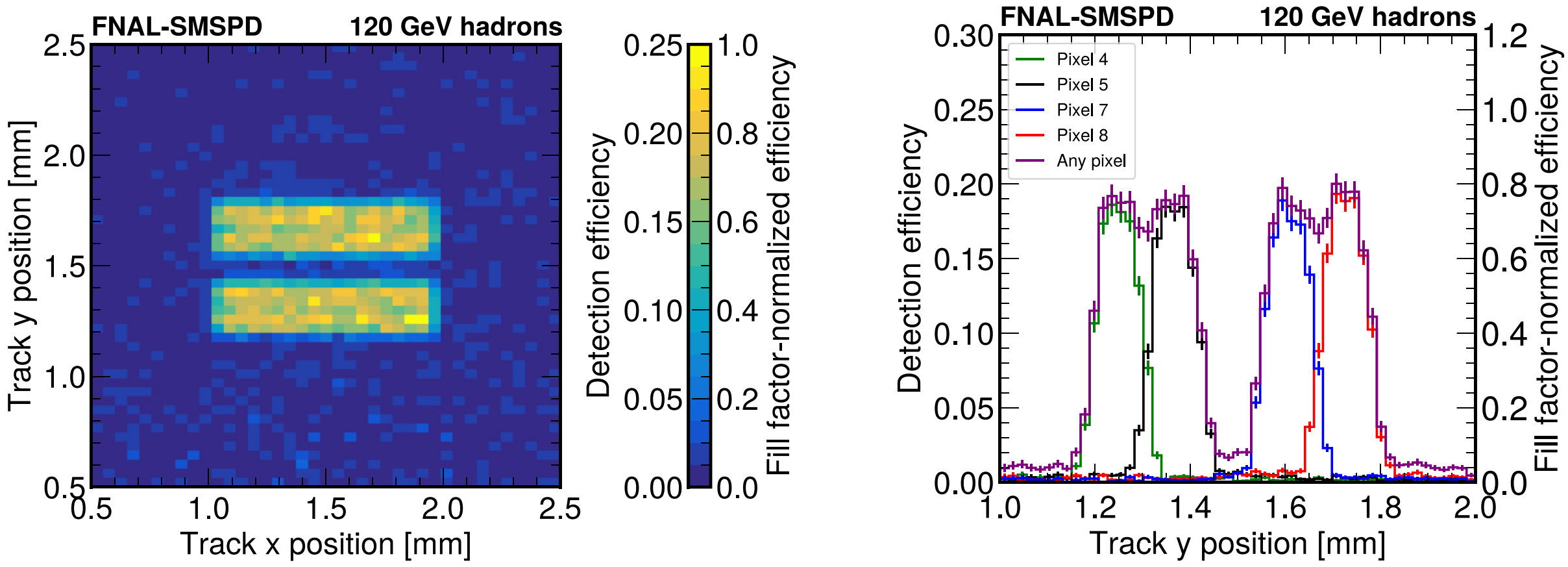

FIG. 39. The (fill factor-normalized) detection efficiency of the four SMSPD pixels is shown. The efficiency as a function of the incident track position in the $x$ and $y$ directions (left) and in the $y$ direction only (right) while integrating along the $x$ axis a 0.7 mm length from the center of each SMSPD pixel. The data in both plots were recorded with the pixels 4, 5, 7, and 8 operating at 8.7, 9.1, 8.8, and 9.1 μA, respectively. The gap observed between pixels 5 and 7 is expected and due to the fact that pixel 6 in between was not read out. Both figures are taken from Ref. [196].

Further work conducted at the CERN super proton synchrotron (SPS) H6 beamline optimized the sensor geometry and utilized a 4.7 nm thick tungsten silicide (WSi) film to achieve unprecedented performance metrics [196]. First, the device achieved a 75% fill-factor-normalized efficiency for muons, marking the first such high-efficiency detection by a microwire array (see Fig. 39). Second, the time resolution was improved by nearly an order of magnitude, reaching approximately 130 ps. In this way, SMSPDs are on a trajectory to provide four-dimensional (4D) tracking capabilities, where spatial hits are coupled with picosecond-level timing to resolve the extreme event pileup expected in future high-luminosity environments.

These results motivate an R&D program in superconducting tracking systems: Beyond performance, SMSPDs offer intrinsic radiation hardness and cryogenic, near-zero quiescent power, reducing cooling infrastructure and tracker material budget, all critical for momentum resolution at the EIC and other precision experiments. They are also especially compelling for a muon collider, where ~130 ps (with a path to sub-10 ps) timing enables tight time gating for BIB suppression. To realize this potential, investment is needed to scale from test-beam prototypes to wafer-scale sensors with integrated cryogenic readout (e.g., cryo-CMOS), leveraging synergies between DOE HEP and the National Quantum Initiative. Recent works move SMSPDs from high-risk R&D to a viable, high-performance option for future collider subsystems, with rapid progress from nanoseconds to picoseconds timing underscoring the urgency of scale-up and systems integration.

### B. Dark matter

As dark matter detectors become more sensitive, they no longer capture only potential dark matter events but also neutrinos from natural sources such as the Sun. This neutrino background, often referred to as the "neutrino fog," is shown in Fig. 40. The figure indicates that imminent experiments should soon attain the sensitivity necessary to detect solar neutrinos [197]. Neutrino signals, which cannot be eliminated or masked, resemble the rare dark matter interactions, presenting an irreducible background that constrains the detection capabilities of dark matter detectors.

This challenge raises the importance of annual modulation searches. Because the Earth orbits the Sun, the relative velocity of Earth through the galactic dark matter halo changes throughout the year, producing a predictable annual modulation in the expected dark matter signal [198]. Solar neutrinos also modulate annually, but, as shown in Fig. 41, their modulation is nearly out of phase with the dark matter signal predicted by the standard halo model [199,200]. This phase difference provides one of the few ways to disentangle dark matter interactions from the otherwise indistinguishable neutrino background. Annual modulation studies are essential for sustaining discovery potential as experiments approach the neutrino fog.

Dark matter experiments are also sensitive to environmental factors. Annual modulation searches, with a one-year period, are fundamentally dependent on the long-term stability of the detectors. If the detector performance fluctuates, the modulation signal may become obscured or misjudged. Additionally, an increase in noise levels establishes a higher threshold for the entire year's data, thereby reducing the modulation search's sensitivity. Real-time data analysis with rapid signal-integrity monitoring can give experiment operators the ability to detect anomalies within minutes rather than months or years later. This awareness enables active, real-time correction of systematic errors and ensures the stability required for annual modulation studies. In contrast, conventional analyses in dark matter experiments often take months

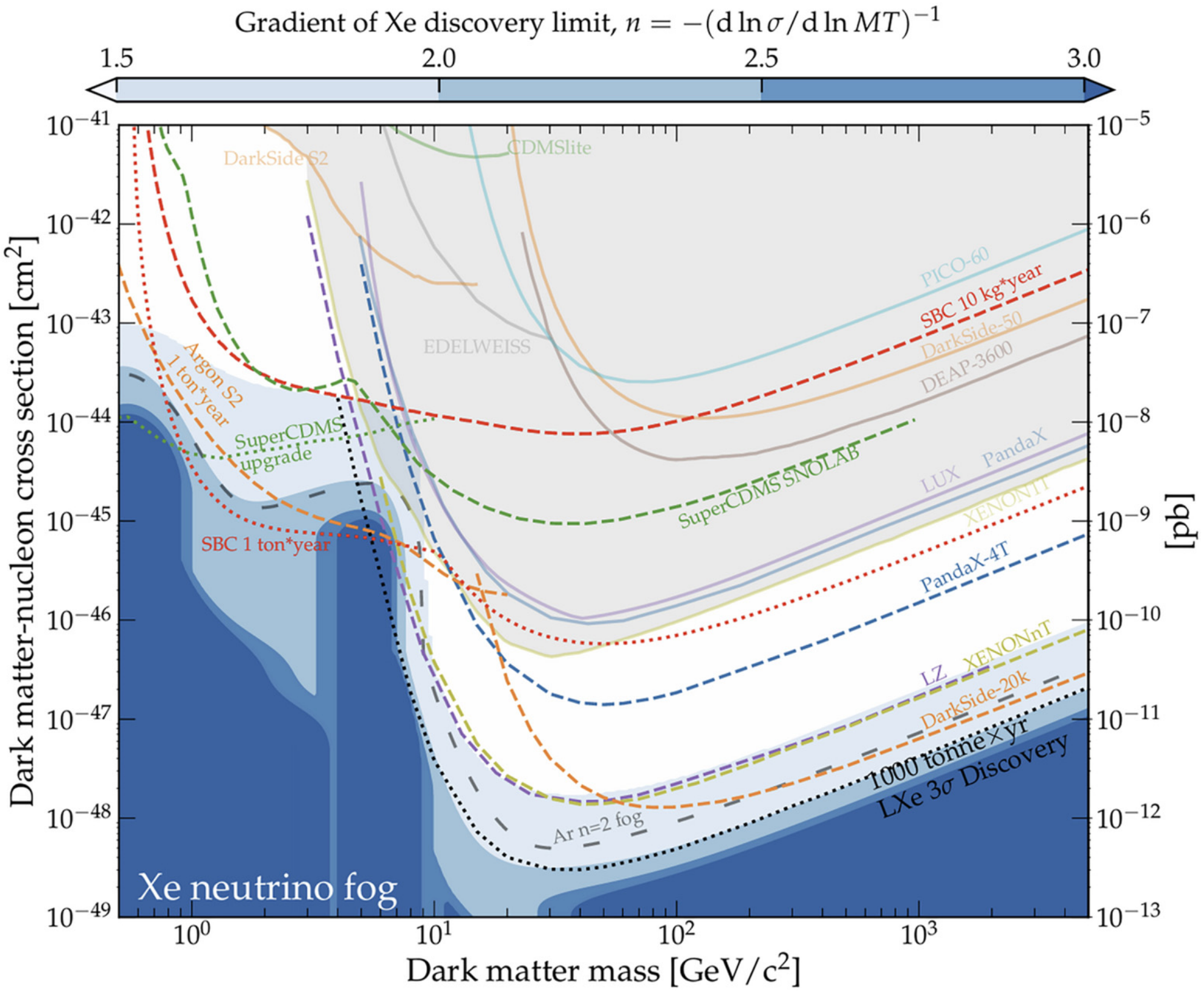

FIG. 40. Experiments are sensitive to solar neutrinos (i.e., the "neutrino fog"), making neutrinos an irreducible background [197]. Maintaining dark matter sensitivity will increasingly rely on directional detectors and annual modulation analyses.

to years to complete, which precludes timely corrective action.

Given the upcoming neutrino background, efforts to address experiment stability are increasingly critical. Lightweight AI models on edge devices have the potential to report analysis-level data-quality indicators (such as noise spectra, pulse-shape trends, and energy spectra) within minutes, operating within the power and space constraints of underground detector environments. Enabling real-time monitoring of dark matter experiments is critical to maintaining the stability experiments need as they approach the neutrino floor and annual modulation experiments become increasingly important.

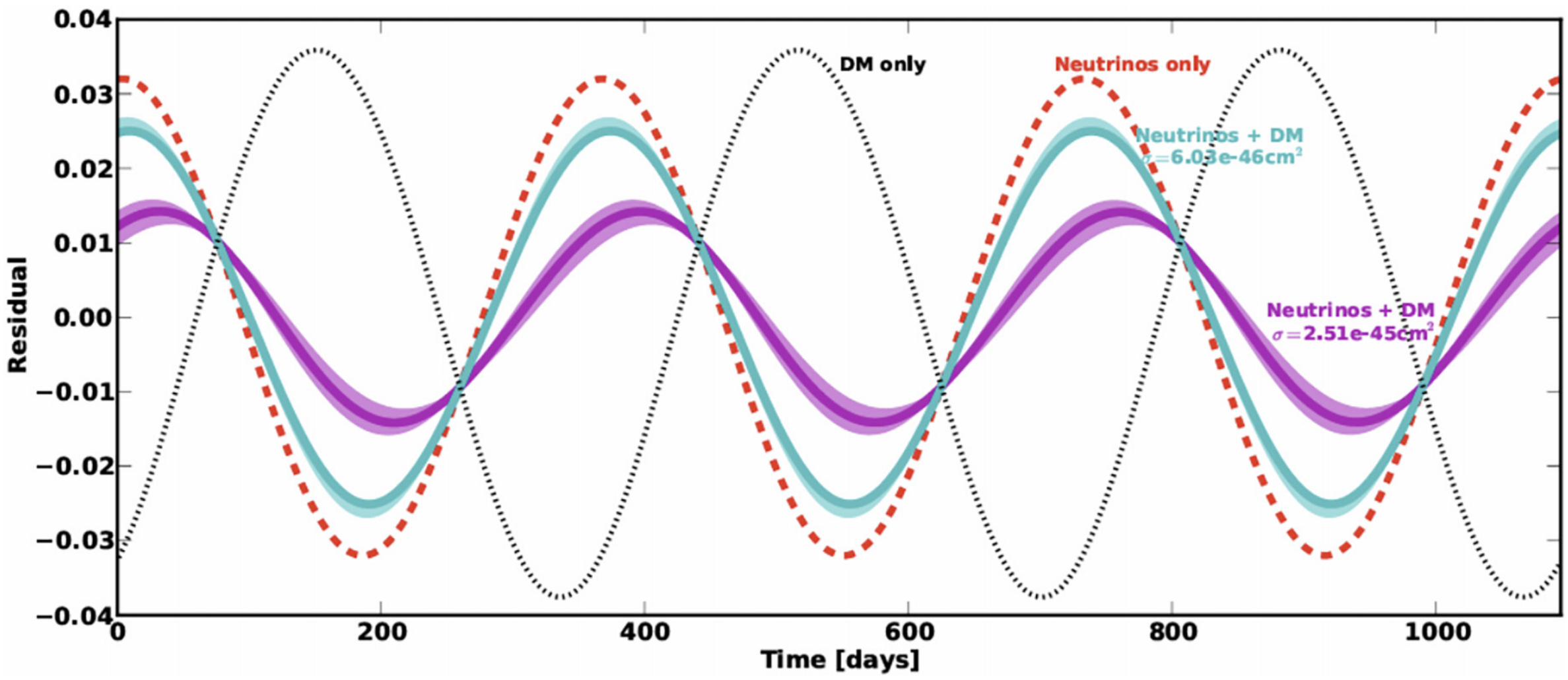

FIG. 41. From Ref. [199]. The annual modulation of solar neutrinos has almost the opposite phase of the annual modulation of dark matter in the standard halo model. However, to be sensitive to even a 3% amplitude will require up to 10 years of stable data from a 50-ton xenon detector [200]—and this assumes a perfectly stable experiment.

For dark matter searches that focus on low-mass candidates, noise reduction is an ever-present concern because the noise floor directly limits the sensitivity to small dark matter signals. Dark matter experiments are already using (non-ML) FPGA processing to lower the signal trigger threshold while minimizing noise triggers [201] and are exploring ML-based approaches on FPGA [202] as part of the hardware triggering system. In this case, FPGAs are used because they can be integrated into the trigger logic and digitization hardware. Variational autoencoders [203,204] are another method that may lower trigger thresholds and may be less susceptible to distorting the signal by deweighting noise frequencies. In both cases, having the ability to update the signal filtering prior to the trigger may offer improved stability. Also of interest are low-power electronics that can act as a full data acquisition system, producing digitized pulses or even final physics estimators. In many dark matter detection experiments, the need to pull signals out to room temperature to read the signals adds significant noise; using low-power AI/ML hardware at the 77 K temperature stage would avoid unnecessary Johnson and pickup noise.

#### 1. Axion experiments

The axion, a compelling dark matter candidate, if discovered, could simultaneously solve the strong CP problem of QCD and account for dark matter. The "axion haloscope" is a detection technique, designed to convert ueV axions into detectable microwave photons in a tunable, high-$Q$, resonant cavity permeated by a strong static magnetic field. These detectors, used by collaborations like the axion dark matter experiment (ADMX), HAYSTAC, CAPP, and ORGAN, are ideal for 0.5–10 GHz photons. However, higher-frequency haloscopelike detectors will face significant challenges. Traditional haloscopes lose sensitivity, due to decreased detector volume and increase in quantum noise at higher frequencies, making axion detection near impossible without noise mitigation and signal boosting techniques. Proposed solutions, such as multicavity and quantum sensing approaches, will increase detector complexity; a complexity only tractable with advanced automated controls. AI/ML techniques for *in situ* radio frequency (RF) response characterization and quantum device control would ease operations in an environment that will change with respect to temperature and magnetic field over time, sometimes voiding standard sensor calibrations. Example potential uses of AI/ML for future axion discovery applications are presented below.

*ML for cold RF characterization.* In anticipation of a future scenario where the periodic ensemble characterization of a scaled-up multiplexed system is eating into axion search data collection, it is critical to seek opportunities to glean more information from fewer measurements. Inspired by the way environmental changes such as magnetic field and temperature change the RF responses of devices, previous work performed a series of measurements designed to tease out the feasibility of using vector network analyzer (VNA) RF scattering parameter ($S$ parameter) responses as dual-purpose measurements. Three experiments were performed investigating the viability of using NNs for mapping $S$ parameters to device environment [205]:

(1) *Learning an RF "fingerprint"*: Two identical circuits with near-identical RF circulators were assembled. A set of switches toggled a VNA to collect $S$ parameters on either device. These data were fed to an NN along with a label of which circuit was being measured for supervised learning.

(2) *Predicting a magnetic field from an RF fingerprint*: A toy model of an axion search, involving a cavity, an antenna, and a circulator, was assembled. A permanent magnet's proximity to the circulator could be adjusted. $S$ parameters through the toy model were captured for different magnet positions, and magnet position was provided to the NN as a label (Fig. 42, left).

(3) *Predicting a temperature from an RF fingerprint*: A circulator was mounted for $S$-parameter measurements on the mixing chamber plate of a dilution refrigerator; temperature information, collected from all fridge stages, acted as labels to the NN (Fig. 42, right).

Each case supported the hypothesis that $S$ parameters hide features of physical phenomena occurring in an RF network that may be recognized by existing NN techniques to aid *in situ* characterization. Figure 42 shows this predictive power. Currently unused wideband measurements may be able to diagnose equipment anomalies and physical conditions.

*Quantum amplifier auto-tuning.* Parametric amplification is a generic phenomenon that occurs when strong and weak waves interact in a nonlinear medium: Energy from a strong "pump" tone is transferred to a weak "signal," resulting in gain. Quantum parametric amplifiers serve as the first stage of amplification in an ultralow-noise receiver, largely determining the noise performance of the system. Unfortunately, optimal performance can only be reached when the operating parameters of the sensitive amplifier are precisely tuned, and automated tuning is currently a time-consuming iterative process. For example, the ADMX receiver relies upon a Josephson parametric amplifier as its primary amplifier, and as a result, the experiment's dead time in data collection is dominated by amplifier rebiasing.

Reinforcement learning (RL) may offer a solution for "auto-tuning" quantum amplifiers. An agent was trained to autonomously tune a kinetic inductance traveling wave parametric amplifier. It controlled the system in real time, by adjusting three parameters: bias current, pump frequency, and pump power. The agent received a reward proportional to achieved gain measured. The system was sometimes intentionally hamstrung and forced to use a fixed value for one of the parameters; preliminary results suggest that the RL is able to swiftly recover. Future reward shaping work will explore SNR, fast tuning, minimal heating, and bias point stability.

#### 2. Quantum sensing radiative decays

Dark matter remains undiscovered despite extensive searches, and the cosmic neutrino background (C$\nu$B)—a cornerstone prediction of Big Bang cosmology—has never been directly detected due to its ultralow energy ($\sim 10^{-6}$ to $10^{-4}$ eV). Meanwhile, quantum sensing has advanced rapidly, with qubits achieving unprecedented sensitivity to weak electromagnetic fields. This convergence motivates a possibility: Elusive particle decays might pro-duce electromagnetic fields strong enough to be detected by quantum devices, opening unexplored regions of parameter space.

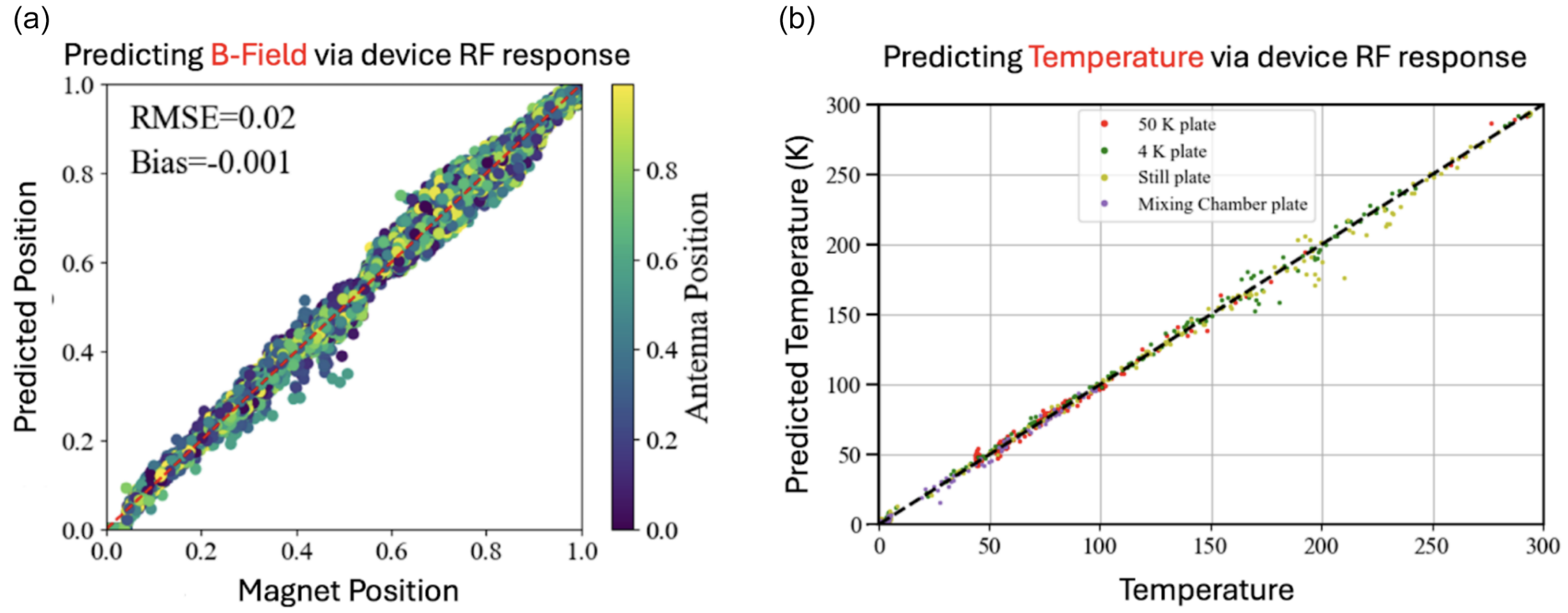

FIG. 42. Two examples demonstrating that an NN can learn an RF fingerprint and then predict device environment, whether it be magnetic field strength (left) or temperature (right).

The extraordinary sensitivity of quantum devices such as superconducting transmon qubits and trapped ions opens up a strategy to detect radiative decays of extremely weakly interacting particles—specifically DM and C$\nu$B [206]. Radiative decays generate extremely faint photons. Although far too weak for traditional detectors, these photons can induce measurable transitions in carefully engineered quantum systems.

A radiative decay $X_2 \to X_1 + \gamma$ from DM or neutrinos produces a coherent photon that induces an effective electric field. This field drives a transition in the quantum device from its ground state to an excited state. Placing the device inside a highly reflective cavity causes the decay photon to become trapped, enhancing the probability of interaction. Two experimental platforms are considered: (1) transmon qubits with tunable gigahertz resonators and well-understood noise properties and (2) trapped ions with long coherence times and complementary frequency ranges.

The study of dark matter radiative decay considers a two-component scenario with a small mass splitting. Using realistic quantum device parameters, the analysis shows that current devices already probe significant regions of dark matter parameter space. Reinterpreting the DarQ experiment [207] sets a bound of $m_2 \geqslant 4 \times 10^{-5}$ eV for fixed $\Delta m^2/m_2$ and constrains the dark matter-photon coupling $\mu$, as shown in Fig. 43. Discovery and exclusion contours indicate sensitivity to mass-splitting regimes for transmon frequencies in the 1–10 GHz range, with trapped ions offering complementary reach in long-coherence regimes [206].

For C$\nu$B, radiative neutrino decays depend on extremely small transition magnetic moments $\mu$. Current devices cannot detect the induced qubit excitations, but even next-generation architectures with dramatically improved coherence and scalability may reach existing reactor bounds on the transition magnetic moment. Nonetheless, this work establishes the first constraint on neutrino magnetic moments using a quantum device and presents a scalable pathway toward future quantum-based C$\nu$B detection [206].

Overall, this work establishes quantum sensing as a powerful emerging tool for exploring physics beyond the standard model. Quantum sensing represents a promising frontier for fundamental physics, with several avenues for dramatic sensitivity improvements, including longer coherence times, higher-frequency qubits in the 10–100 GHz range (albeit with increased decoherence challenges), large entangled qubit arrays, and expanded cavity volumes.

### C. Neutrino physics

#### *1. LArTPCs and DUNE*

Liquid Argon Time Projection Chambers (LArTPCs) combine fine-resolution spatial and calorimetric measurements of ionization deposits across multiple sensing views to produce 3D images of particle interactions. Combined with timing and calorimetric information from photon detectors, they enable low-threshold particle ID and reconstruction in a generation of detectors—like the deep underground neutrino experiment (DUNE)—that are capturing neutrino interactions from a variety of sources across a broad energy range.

The high-resolution and large-dynamic range imaging capability of large LArTPCs comes at the cost of massive data volumes. A single 10-kton DUNE far detector module will generate over 10 Tb/s from its front-end electronics, with individual recorded events exceeding 7 GB. Detector monitoring and data selection systems must handle large data streams to both ensure efficient detector operation and reduce the volume of data to fit within offline data processing and storage resources. Real-time TPC data processing has been demonstrated in both firmware [208] and fast parallelized software [209,210]. Fast parallelized software processing will be used in DUNE to process the stream of TPC data through a hierarchical set of algorithms that identify regions of interest on each TPC channel, cluster them into localized activities, and form final trigger candidates to inform data selection. DUNE's data selection system is designed and expected to efficiently record interactions depositing more than 100 MeV of visible energy, identify single interactions down to 10 MeV, and capture galactic supernova neutrino bursts.

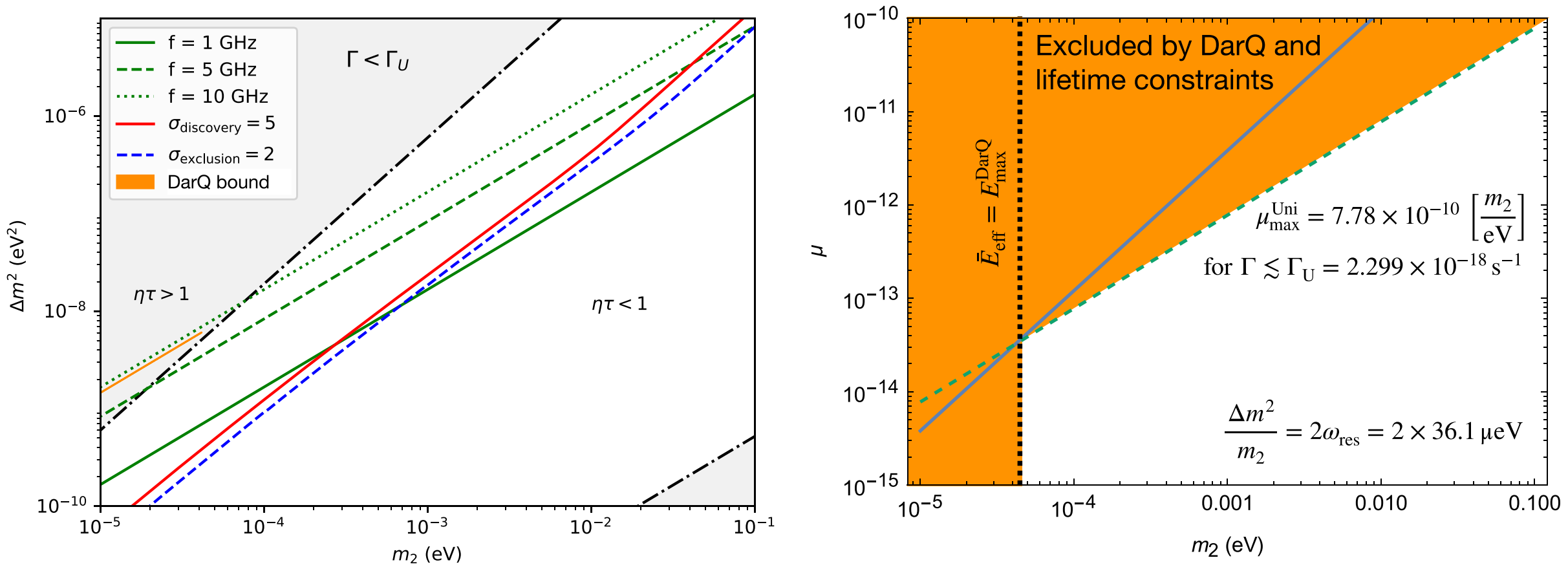

FIG. 43. (Left) $5\sigma$ discovery (red, solid) and $2\sigma$ exclusion (blue, dashed) contours for dark matter radiative decay in the $(m_2, \Delta m^2)$ plane, assuming the maximum allowed coupling, a coherence time of 100 μs for a single qubit, and the experimental parameters described in the text. (Right) Expected bounds on $(m_2, \mu)$ by the DarQ experiment at 95% CL [207], which leads to $m_2 > 4.15 \times 10^{-5}$ eV for $\Delta m^2/m_2 = 2\omega_{res}$ assuming $\Gamma \lesssim \Gamma_U = 2.299 \times 10^{-18}\ s^{-1}$ and coherence time of DarQ, $\mathcal{T} = 1$ μs. Taking the cosmic microwave background (CMB) bound, $\Gamma \lesssim \Gamma_{CMB} = 1.7 \times 10^{-25}\ s^{-1}$, we obtain $m_2 > 3.07 \times 10^{-12}$ eV.

The image-like nature of LArTPC raw data, the success of AI/ML methods in offline reconstruction and analysis, and stringent resource constraints on space, power, and cooling in detector environments like DUNE's motivate exploration of AI/ML closer to the detector, both in heterogeneous computing and "at the edge" in readout electronics. Improvements in AD, data forecasting, and decision agents could significantly improve the performance of LArTPCs with better monitoring and more stable and efficient operation. Data reduction through intelligent compression, filtering, and selection could benefit from AI/ML techniques to enhance the physics reach of current and future LArTPCs. Some specific examples of active research and development are given below.

*Low-latency ML-based supernova burst pointing in DUNE.* The next nearby core collapse of a massive star will yield unprecedented physics and astrophysics information from gravitational wave, photon and neutrino messengers. The intense, few-tens-of-second, few-tens-of-MeV all-flavor core-collapse supernova neutrino burst is expected to precede the electromagnetic observables by some time. The interval could be tens of minutes, to hours, to days [211], depending on the nature of the progenitor star. The prompt neutrino burst means that the neutrino observation can provide an early warning to astronomers [212] of an impending supernova [213].

Directional information from the neutrino burst is clearly of immense value for subsequent early-time observation of the supernova or potential localization of a vanished star. Anisotropic neutrino interactions provide the best prospect for neutrino-based pointing. DUNE's tracking of neutrino-electron elastic scattering (eES) events has been shown to enable several-degree pointing [214], assuming sufficient "channel tagging" (interaction classification) to separate eES from the dominant quasi-isotropic $\nu_e$ charged-current ($\nu_e$CC) events. However, to be useful to the astronomical community, such reconstruction must happen with low latency—ideally within several minutes.

Because the baseline expectation for DUNE is that full reconstruction will require at least several hours in order to transfer data to Fermilab for processing, we are pursuing a faster "*in situ* compute" approach [215,216]. In this approach, the full raw dataset can be reduced (bakground rejection, region-of-interest selection and denoising steps) on buffered data from DUNE DAQ readout units, making use of FPGA or GPU coprocessors, and then sent over the network to a single processor for directional reconstruction of the SN burst data. Recent proof-of-concept studies have demonstrated overall acceleration of reconstruction with machine-learning algorithms [215–217]. Further opportunities for low-latency pointing improvement will address channel tagging (critical for high-quality pointing) as well as potential use of $\nu_e$CC event final-state pattern distributions [214]. Full implementation of low-latency pointing will need to address power and cooling challenges.

*Capturing neutrinos and new physics at ProtoDUNE with an ML online trigger algorithm.* The massive LArTPCs of DUNE will measure the properties of neutrinos with unprecedented precision [218]. Currently, two full-scale prototypes of the DUNE far detectors are at CERN, which are called ProtoDUNE-HD and ProtoDUNE-VD [219]. Recent studies indicate that the ProtoDUNE detectors have the potential to detect neutrinos and BSM particles from one of the targets in CERN's north area that is exposed to the 400 GeV SPS beam [220]. As a proof of concept, initial studies have focused on observing neutrino interactions in ProtoDUNE [221]. A challenge in surface detectors like ProtoDUNE is the selection of interesting signals coming from BSM and neutrino interactions whilst rejecting the overwhelming cosmic-ray background. This has motivated the development of an exclusive trigger algorithm within the context of the DUNE DAQ system.

An online trigger algorithm was developed for ProtoDUNE-HD and ProtoDUNE-VD in order to maximize the potential sample size in a future ProtoDUNE neutrino and BSM program. The ProtoDUNE DAQ runs a hierarchical series of algorithms in real time that determine whether to record an event. At the lowest level, trigger primitives

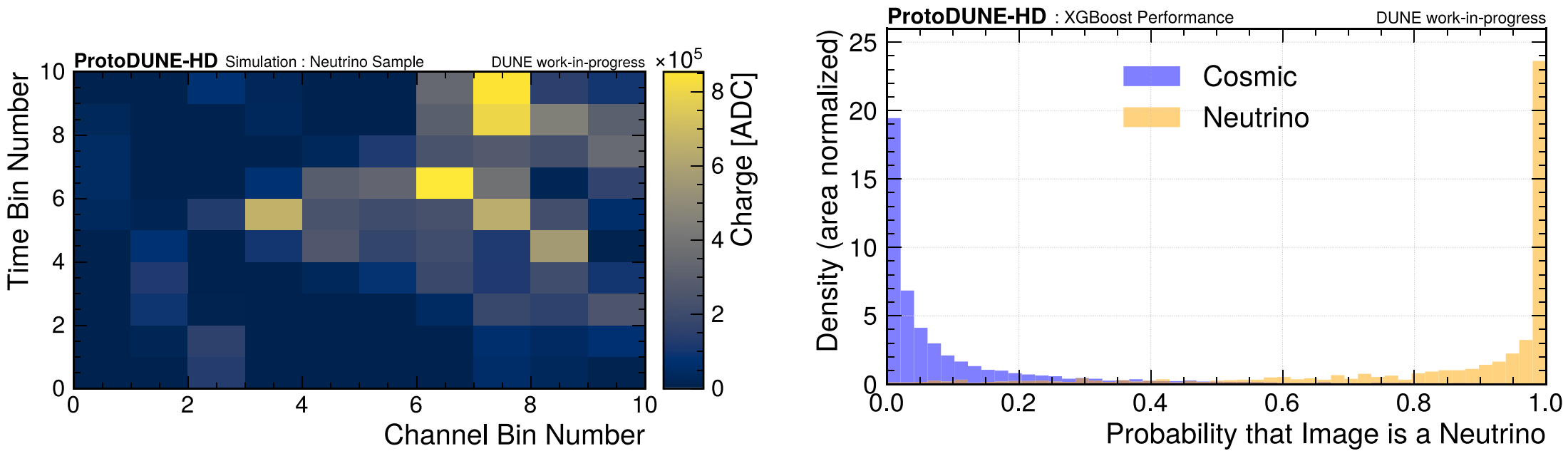

(a) (b)

FIG. 44. Figure 44(a) shows an example of a neutrino image used to train the XGBoost model. The XGBoost model is trained on cosmic and neutrino images that pass a minimum total charge threshold of 200 000 ADC and its performance is shown in Fig. 44(b).

(TPs) are generated, which represent a cluster of charge on an individual detector channel. The TPs from a plane of ProtoDUNE detector channels are then processed by the following algorithm:

(1) Check whether the sum of the charge of the TPs in a 0.3 ms window of time crosses a minimum threshold of 200 000 ADC units.

(2) Generate a coarse pixelated time versus channel number image of the TPs in the 0.3 ms window, where the amplitude of each pixel is the total charge of the TPs in the pixel.

(3) Pass the image to an XGBoost ML model, which calculates the probability that the image is a neutrino. If the probability crosses a threshold, trigger ProtoDUNE.

This trigger algorithm therefore relies on an XGBoost model to classify images as either neutrino or cosmiclike [222]. To train this model, tens of thousands of cosmic and neutrino ProtoDUNE-HD images were simulated using LArSoft [223]. Figure 44(a) shows an example of a simulated neutrino image used for training and testing the model. Each pixel of the image is a feature of the XGBoost model in addition to the total charge across all the pixels. Therefore, for a ten-by-ten pixel image, there are 101 features in the XGBoost model. Figure 44(b) demonstrates that the XGBoost model performs well in distinguishing cosmic and neutrino events. XGBoost was chosen due its fast prediction times on the CPUs available to the ProtoDUNE DAQ. The XGBoost model was trained and then exported as compilable C++ code that was directly built into the DUNE DAQ software.

Figure 45(a) shows the probability of neutrino selection as a function of the trigger rate due to the cosmic-ray background. The trained XGBoost model is compared with applying a threshold on the total charge in a 0.3 ms time window, which is the basis of the best currently available trigger algorithm for a neutrino and BSM search at ProtoDUNE. The XGBoost model outperforms the charge threshold across all potential trigger rates.

The XGBoost-based algorithm was implemented in the DUNE DAQ software, allowing a live test at ProtoDUNE-VD to be carried out. ProtoDUNE-VD successfully took data using this algorithm at a stable trigger rate of approximately 5 Hz. Figure 45(b) shows an example ProtoDUNE-VD event captured during this data run. It is therefore possible to trigger for neutrinos at ProtoDUNE using an XGBoost-based algorithm with efficiencies of approximately 30% or higher. It should be noted that the future DUNE underground detectors will be shielded from cosmic rays, allowing 100% trigger efficiency for high-energy events without the need for ML algorithms. Nonetheless, this can be viewed as a feasibility case study, well motivated for rare event searches at surface detectors such as ProtoDUNE, demonstrating that ML algorithms can be deployed in the online DUNE DAQ while meeting strict latency requirements.

*Real-time AD in DUNE.* DUNE's DAQ and online event selection system is designed to meet the experiment's physics requirements. To further accelerate discoveries through dedicated data streams with improved signal-to-background ratio, we proposed an AD approach in DUNE's AI/ML kickoff workshop in early 2025 [224], which addresses three goals: searching for BSM physics signatures, detecting supernova burst (or presupernova) neutrinos, and monitoring detector malfunctions—all in real time using raw data. To accomplish this, we have been developing a CNN-based autoencoder (AE) model. Autoencoders are well-suited to this task because they learn, in an unsupervised manner, to compress input data into a compact latent representation and then reconstruct it. The model optimizes by minimizing the reconstruction loss—the difference between the original and reconstructed images. Crucially, when the trained model encounters data that differs from what it learned as "normal," the reconstruction degrades, flagging potential anomalies.

We conducted a proof-of-concept study targeting supernova burst neutrino detection in summer 2025 [225]. We trained an AE model exclusively on radiological backgrounds, the dominant background source for supernova burst neutrino searches. We then evaluated the model on three distinct datasets: two supernova burst neutrino interactions (charged current and elastic scattering) on top of radiological backgrounds, and a held-out sample of radiological backgrounds-only not used during training, and obtained promising preliminary results [225]. The mean squared error distributions revealed clear separation between radiological backgrounds (i.e., nominal events) and supernova burst neutrinos (i.e., anomalies), demonstrating the model's ability to distinguish genuine signals from nominal detector activity.

The deployment goal relevant to real-time operation is running the autoencoder directly on the FPGAs in warm interface boards (WIBs) where all raw data from the cold electronics are aggregated before transmission to DUNE's DAQ. The power and latency constraints of the underground environment require model compression through pruning and quantization. We are investigating this using frameworks such as hls4ml [34,175] and cgra4ml [47], with a DAQ-server deployment on trigger primitives as an alternative path. Development of more realistic training datasets and evaluation on additional signal classes are ongoing.

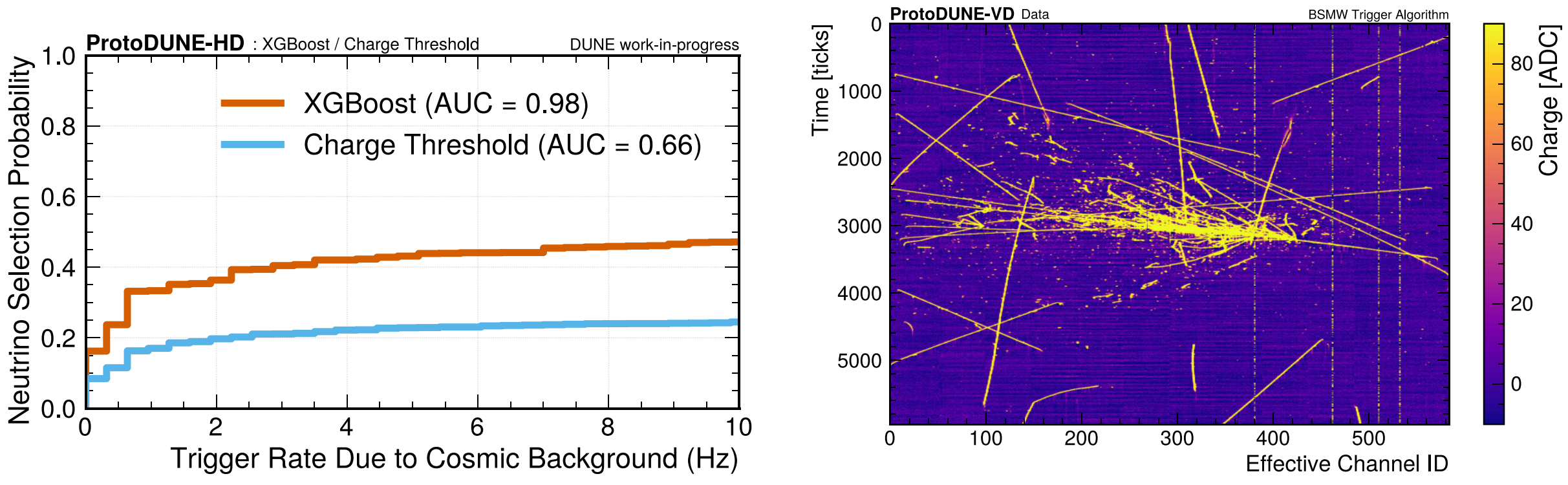

(a) (b)

FIG. 45. Figure 45(a) shows the neutrino selection probability as a function of the trigger rate at ProtoDUNE-HD due to cosmic-ray background. Figure 45(b) shows an event display from ProtoDUNE-VD data taken during a test of the XGBoost-based trigger algorithm.

*Real-time AI for data selection in DUNE.* Real-time and efficient identification of rare and low-energy physics signals among electronics noise and radiological backgrounds, needed to enable an expanded astroparticle physics program in DUNE, faces stringent constraints on latency, power, and computational resources in front-end electronics. ML integrated directly into front-end data paths provides a promising path for increased intelligence and efficiency gains in data selection, compression, and AD, especially when combined with techniques and architectures that allow for reduced resource utilization. Below, we summarize key developments in applying ML to LArTPC front-end electronics, with a focus on recent contributions demonstrating feasibility for FPGA implementations in DUNE.

DNNs have proven effective at classifying raw LArTPC wire-time readouts, greatly reducing data volumes while preserving physics-rich information. As demonstrated in Refs. [216,217,226], CNNs can achieve high accuracy for selecting rare astroparticle and BSM signals in DUNE over a wide energy range, or identifying regions of interest of low-energy signals with increased accuracy. These methods could complement existing trigger algorithms by improving performance in the low-energy regime ($\sim$10 MeV and below), where achieving high efficiency is challenging due to overwhelming radiological background rates, provided overall trigger rates can be maintained at the required level. Furthermore, such 1D or 2D CNNs can be accelerated on resource-constrained hardware such as FPGAs. Refinement using tools such as `hls4ml` [175] enables deployment of quantized networks onto FPGAs under strict resource, latency, and power constraints, in a way that is compatible with the operational environment and performance needs of DUNE's front-end electronics.

Beyond supervised classification, AD can be a game changer for identifying rare or unexpected data patterns. LArTPC waveform denoising and low-energy signal identification using autoencoders have been explored in Refs. [227,228]. A recent study [229] introducing autoencoder-based AD in LArTPC wire-time data, combined with knowledge distillation to allow the model's execution on FPGAs, has also shown that unsupervised identification of outlier behavior such as high-multiplicity interactions in streaming LArTPC data is possible. This offers opportunities to expand front-end ML capabilities not only to adaptive data selection but also to rare-event triggering and potentially to detector monitoring.

In DUNE, ML-augmented front-end systems can interface with modular trigger architectures to make early, physics-informed decisions and reduce data rates by orders of magnitude. Integrating ML into DUNE front-end electronics requires models that are: quantized and compressed for FPGA implementation; optimizable for tens of microseconds to milliseconds latency inference; robust under continuous, zero-dead-time operation; and power efficient for deployment in deep underground environments. Codesign of ML models and hardware architectures, aided by tools like `hls4ml`, is essential.

Machine learning for front-end electronics is emerging as a promising approach for next-generation LArTPC detectors. Demonstrations of FPGA deployment of CNNs and AD networks suggest that real-time ML-enabled data selection can be implemented within representative latency and resource constraints relevant to DUNE's front-end environment. While these studies demonstrate compatibility with FPGA-level resource and timing budgets, full system-level validation, including integration with the broader DUNE data selection architecture, as well as power, cooling, and sustained trigger-rate constraints, remains an important area for continued R&D. Future work should explore alternate network architectures, pursue full-system integration and demonstration in currently operating LArTPC detectors or prototypes such as short baseline near detector (SBND) [230] or ProtoDUNE [231], and develop shared community infrastructure for ML model training and deployment.

*Smart time projection chamber.* The smart TPC project [224] investigates whether simple ML-based pattern recognition deployed at the front-end electronics level can improve triggering and filtering for LArTPC data volume reduction by enhancing signal-to-background ratio before data reaches the DAQ.

DUNE's baseline cold electronics [232] transmit all digitized data to DAQ servers without zero suppression or compression, where software-based algorithms generate trigger primitives and perform data selection. This design is expected to meet DUNE's physics requirements. Smart TPC explores whether performing simple pattern recognition earlier in the readout chain directly on digitized waveforms could reject dominant radiological backgrounds closer to the source, lowering energy threshold, reducing downstream data volume and potentially improving sensitivity to low-energy signals. Optionally, TPC charge and photon detector light coincidences currently performed only in offline software could be moved into the electronics for further background rejection if the physics case warrants it.

We propose two complementary pathways. The near-term approach deploys lightweight classifiers such as boosted decision trees, which have been demonstrated within embedded FPGA resource constraints with sub-100 ns inference [59,175], on existing warm-side FPGAs (WIB and DAPHNE). The longer-term approach explores similar classifiers on custom cold ASICs operating at 87 K inside the cryostat, building on the CMS Smart Pixel group's demonstration of BDT-based inference on 28 nm CMOS front-end ASICs for HL-LHC tracking [71,233]. The FPGA pathway serves as a near-term demonstrator and risk reduction for the ASIC approach.

This project partners with microelectronics engineers and the CMS Smart Pixel group. Partnership with the LArPix [234] and Q-Pix [235] groups would be mutually beneficial to leverage complementary expertise in low-power on-detector inference and self-triggered readout architectures. The primary question smart TPC aims to answer is whether ML-based pattern recognition at the electronics level can meaningfully improve lowering energy threshold and increasing signal-to-background ratio for low-energy physics including solar and supernova neutrinos, beyond what software-based data selection alone achieves.

*LArTPC calibration using pretrigger data.* Many existing LArTPC calibration strategies rely heavily on abundant cosmic-ray muons. For deep-underground detectors such as the DUNE far detector, the cosmic-muon flux is strongly suppressed, significantly reducing the availability of steady, high-statistics calibration samples. While dedicated calibration systems—such as laser and light-emitting diode systems—are designed to mitigate this limitation, they require specialized data-taking runs with data-acquisition configurations that differ from standard physics operations and are mostly having limited spatial coverage except for special devices such as pulse neutron sources [218,236]. As a result, a continuous, spatially distributed probe that samples detector conditions uniformly over time is highly desirable.

Commercial atmospheric argon contains $^{39}$Ar at an activity of order 1 Bq/kg, producing a persistent and uniform $\beta$-decay source throughout the detector volume with a well-defined low-energy spectrum at $\mathcal{O}(100\ \mathrm{keV})$ [218]. This intrinsic radioactivity can therefore serve as a continuous calibration handle for both the TPC readout chain and the photon-detection system. However, relevant data are typically discarded or only minimally exploited due to the prohibitive computing and storage costs associated with processing untriggered, continuous waveforms at the full $^{39}$Ar rate.

A promising path is to move the $^{39}$Ar calibration upstream, computing compact summary features directly from digitized data prior to triggering. This can be enabled by the processing capabilities in front-end electronics such as FPGAs. Machine-learning approaches such as compact convolutional encoders are ideal candidates for these feature extraction tasks with low resource footprint constraints. Deployed as resource-aware inference models, they can provide calibration proxies for the overall charge scale, light yield, their corresponding local nonuniformities, and the gain and timing response of individual electronics channels. Propagating only these lightweight metrics downstream preserves the near-continuous, high-granularity information while avoiding the storage and processing burden of full untriggered waveforms, eventually paving the way toward a time-dependent calibration and reduced detector systematics in physics. In addition, such automated calibration capability could release pressure to experiments for taking data dedicated to calibration purposes and personpower for analyzing the data to monitor the detector status. This helps for experiments to focus more on scientific goals and accelerates the scientific discovery process.

### 2. Neutrino telescopes

Next-generation large-scale neutrino telescopes will operate at data rates and detector scales that increasingly strain traditional trigger and data-handling architectures. These experiments are located in resource-scarce environments, where per-module power budgets are typically at the level of a few watts, limiting the usage of ML techniques. Together, these constraints will force current CPU-based real-time, *in situ* reconstruction and triggering to discard a large fraction of data to satisfy data-transmission limits. Recent advances in edge ML hardware motivate a reexamination of this paradigm [237–239].

*Real-time TPU-based, low-power ML for neutrino telescopes.* In the context of neutrino telescopes, Ref. [240] presents the first demonstration that machine-learning-based neutrino event reconstruction can be performed on Edge devices such as *tensor processing units* (TPUs) [241–243]. These could be installed within each detector module, within a power envelope of only a few watts, while surpassing the accuracy of currently deployed real-time methods. This contribution summarizes the key technical challenges inherent to this problem and the workarounds that make such an approach viable today.

A critical enabler of this effort is the availability of open-source, community-developed software. On the physics side, all detector simulation and data generation were performed using Prometheus [244], an open-source neutrino telescope simulation framework. On the hardware and software side, there also exist communities actively developing software foundations for the hardware architectures [245–247].

Deploying ML models on ultra-low-power edge devices introduces several fundamental challenges for neutrino telescope workflows that differ qualitatively from conventional GPU-based acceleration, mainly:

(1) *Power constraints*. Edge accelerators suitable for in-detector deployment operate at $\mathcal{O}(1-3)$ W, orders of magnitude below GPUs typically used for ML inference.

(2) *Limited numerical precision*. The Google Edge TPU [243] used in Ref. [240], like many other Edge computing platforms across different hardware architecture designs, requires fully integer (`uint8`) inference, eliminating floating-point arithmetic.

(3) *Architectural constraints*. Supported operations are limited: Tensor dimensionality is capped, three-dimensional convolutions are disallowed, and only a restricted set of neural network layers can be executed efficiently. This goes against the traditional four-dimensional convolutions or graph neural networks [248,249] design strategies for neutrino telescopes.

To address these challenges, the reconstruction pipeline needs to be codesigned with the hardware, rather than adapted *post hoc*. For example, in the implementation presented in Ref. [240], the following key features of design were introduced:

(1) *Time-series reformulation*. Instead of conventional convolution-style algorithms, the data can be processed using an early language-model-type approach via RNN. In this encoding, detector photon hits are discretized into a sequence of temporal snapshots. Similar snapshot techniques could be used in other experiments where coarse time graining can be used. The use of RNNs fits the more limited TPU architecture.

(2) *Hybrid CNN-RNN architecture*. Hybrid architectures can be used to encode all available information. For example, TPUs allow for two-dimensional convolutions. Thus, one can perform low-dimensional CNN convolutions that serve as encoders to the RNN backbone. In the application given in Ref. [240], two-dimensional convolutional encoders extract spatial features from each time slice, while a recurrent (LSTM) network integrates temporal information—avoiding unsupported 3D convolutions while retaining sensitivity to event signature. This technique can also be used in fine-grain detectors; for example, a tracker could be split into various planes, where each plane is processed first through a two-dimensional convolution, and then an RNN acts on the whole event.

(3) *Quantization-aware data encoding*. Aside from problem formulation, quantization-aware data preprocessing is also performed. Inputs are constructed to be integer valued by design, enabling full-network uint8 quantization. A staged quantization and retraining procedure mitigates accuracy loss from reduced numerical precision.

Though some of these are specific to neutrino telescopes, many of the insights here can be extended to other neutrino experiments.

Despite these constraints, the fully quantized edge-deployed network introduced in Ref. [240] achieves median angular reconstruction errors of approximately 6° in ice and 7°–8° in water, significantly outperforming currently used real-time CPU-based reconstruction algorithms, which typically achieve $\sim$10° resolution at trigger level, with their performance limited by the power constraint introduced in real-time processing. These first results make TPU-enhanced event reconstruction an interesting avenue of future investigation.

Most importantly, this improvement is achieved at dramatically higher energy computational efficiency, see Fig. 46 (adapted from Fig. 6 of Ref. [240]). For single-event inference—the relevant regime for real-time triggering—edge ML devices outperform GPUs by orders of magnitude in power efficiency, enabling real-time inference at $\mathcal{O}(100)$ Hz/W.

It is worth noting that angular regression is neither an ideal task nor a commonly desired one for quantized inference. Nevertheless, the results strongly motivate future efforts focused on classification, triggering, and intelligent data reduction, where edge ML is expected to perform even more effectively. More broadly, this proof of concept highlights the importance of hardware-algorithm codesign and open simulation tools in enabling the next generation of intelligent, resource-aware particle physics experiments.

*Efficient neural network architectures for real-time neutrino telescope data processing.* Detectors such as IceCube [250], KM3NeT [251], and the forthcoming IceCube-Gen2 [252] instrument cubic-kilometer-scale volumes with thousands of optical modules, generating trigger rates on the order of kilohertz while searching for rare astrophysical signals. Real-time reconstruction is critical for background rejection and for enabling rapid alerts that support multimessenger astronomy campaigns [253]. IceCube operates at the geographic South Pole, where computing resources and data transmission bandwidth are severely limited, making it a natural test bed for developing efficient inference methods.

Beyond the environmental constraints, the data produced by neutrino telescopes exhibit properties shared by many next-generation detectors: extreme sparsity, irregular detector geometries, high dimensionality from nanosecond-resolution timing, and substantial variance across many orders of magnitude in energy. To address these challenges, recent work has explored architectural innovations, including sparse convolutions and learned data compression, enabling high-performance real-time computing suitable for deployment in these resource-constrained environments.

Reference [254] demonstrated sparse submanifold convolutional neural networks (SSCNNs) for event reconstruction in neutrino telescopes. Traditional CNNs prove highly inefficient on sparse data, performing operations across the entire input volume regardless of occupancy. SSCNNs, originally developed for computer vision applications [255], restrict convolution operations to nonzero elements while preserving input sparsity across network depth, maintaining efficiency in deep architectures. This approach has already demonstrated success in liquid argon time projection chamber experiments [256,257].

The SSCNN approach achieves angular reconstruction at rates exceeding 9.9 kHz on an NVIDIA A100 GPU, representing approximately a 16-fold speedup compared to traditional CNNs [258] and substantially exceeding the roughly 3 kHz trigger rates of current neutrino telescopes. This throughput enables real-time processing at the detector site for improved trigger-level background rejection and rapid alert systems.

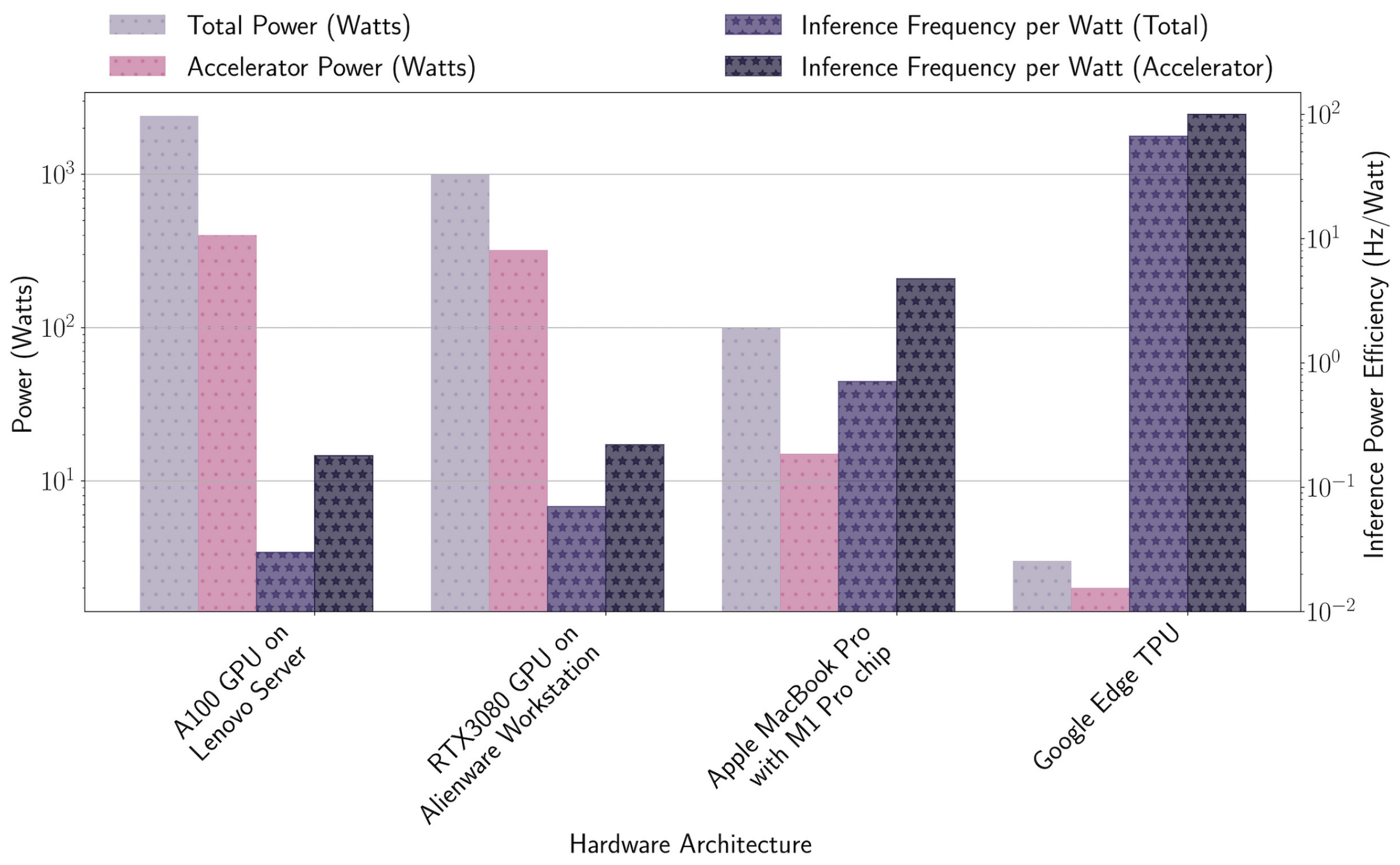

FIG. 46. Inference power efficiency for single-event neutrino reconstruction on different hardware architectures. Edge ML devices achieve the highest inference-per-watt performance, despite operating at drastically lower absolute power.

The architecture naturally accommodates irregular detector geometries by operating on spatial coordinates directly, enabling straightforward adaptation to diverse detector configurations, including water-based telescopes such as KM3NeT and P-ONE [259].

Complementing the sparse convolution approach, Ref. [260] developed `om2vec`, a transformer-based variational autoencoder that learns compact latent representations of photon arrival time distributions (PATDs) recorded by individual optical modules. Raw PATDs span thousands of nanosecond-resolution time bins with highly variable photon counts, requiring careful handling to preserve physics-relevant features such as the "double-bang" signatures characteristic of tau neutrino interactions.

The `om2vec` architecture encodes arbitrary PATDs into fixed-length latent vectors of 32–128 dimensions while retaining sufficient information for accurate reconstruction. Compared to traditional asymmetric Gaussian mixture model fitting approaches [261], `om2vec` achieves order-of-magnitude speedups on CPU and 2-order-of-magnitude speedups on GPU, with substantially lower failure rates. The learned representations enable downstream reconstruction networks to achieve performance comparable to models operating on full timing information while reducing input dimensionality and simplifying model pipelines. This approach integrates naturally with community tools such as GraphNeT [248], facilitating adoption across the neutrino telescope community.

These architectural developments have direct implications for edge computing deployment. The SSCNN's ability to process events at trigger-level rates on GPU hardware, combined with comparable single-core CPU performance to traditional likelihood methods, suggests viable deployment scenarios ranging from GPU-accelerated processing at detector sites to CPU-based reconstruction in bandwidth-limited environments. The `om2vec` representations offer an additional pathway for data reduction by replacing high-dimensional PATDs with compact latent vectors early in the processing chain, potentially relaxing throughput constraints that currently limit timing resolution. The self-supervised nature of the variational autoencoder framework also enables training directly on detector data without labeled simulation, supporting robust performance despite simulation-data discrepancies.

### *3. COHERENT*

The COHERENT collaboration has measured coherent elastic neutrino-nucleus scattering (CEvNS) at the Spallation Neutrino Source [262–264], establishing a powerful neutral-current neutrino probe sensitive to all neutrino flavors. Tonne-scale CEvNS detectors offer a unique opportunity to detect neutrinos from a Galactic core-collapse supernova, with few to $\mathcal{O}(10)$ events per tonne expected for a 10-kpc collapse. However, because COHERENT detectors have little overburden, the expected supernova neutrino signal is overwhelmed by steady-state detector backgrounds by several orders of magnitude. We describe an example ML trigger on front-end electronics to enable supernova neutrino detection for next-generation COHERENT detectors [265].

Core-collapse supernovae release neutrinos with energies up to tens of MeV over a timescale of $\sim$10 s [266]. While next-generation neutrino detectors, such as Jiangmen underground neutrino observatory (JUNO) [267], Hyper-Kamiokande [268], and DUNE [269] are primarily sensitive to $\nu_e$ and $\bar{\nu}_e$ via charged-current interactions, CEvNS provides a flavor-blind measurement of the total neutrino flux. This makes CEvNS detectors uniquely robust against uncertainties from neutrino oscillations in dense supernova environments.

Because the CEvNS cross section scales approximately as $N^2$, compact detectors with heavy nuclei can achieve relatively high interaction rates compared with charged-current interactions. The diverse target materials across the COHERENT array enable cross-validation of signals and constraints on systematic uncertainties, enhancing confidence in a potential supernova neutrino observation. COHERENT CEvNS targets currently deployed or planned for the near future include Ar, Ge, NaI, and cryogenic CsI. In the following, we take a future potential upgrade of CsI as a concrete example; however, the implementation of an ML-based self-trigger can be applied for any COHERENT subsystem.

Cryogenic CsI scintillators exhibit high light yield and low intrinsic radioactivity, enabling sensitivity to few-keV nuclear recoils for CEvNS detection. For beam-triggered mode, the spallation neutron source (SNS) duty factor provides a factor of $10^3$ to $10^4$ background suppression. In self-trigger mode, a cryogenic CsI detector would have background rates of approximately 7 $\mathrm{kg}^{-1}\,\mathrm{s}^{-1}$, dominated by natural radioactivity and cosmic muons [270]. Extrapolated to a tonne-scale detector, this corresponds to raw trigger rates of $\mathcal{O}(10^4)$ Hz, far exceeding the expected supernova signal rate. Backgrounds can be broadly categorized into high-energy events from radioactivity and cosmic muons, as well as low-energy instrumental backgrounds, dominated by scintillation afterglow and PMT dark noise. An effective trigger must suppress both classes while maintaining high efficiency for CEvNS-like events. We consider a multistage trigger architecture suitable for implementation in front-end electronics.

A strong correlation exists between deposited energy and waveform maximum amplitude. Applying an upper cut on the waveform maximum efficiently suppresses high-energy backgrounds, reducing the event rate by 1–2 orders of magnitudes without impacting CEvNS sensitivity. A lower bound on the waveform integral is also applied to reject coincident PMT dark-noise triggers. After these selections, the remaining waveforms are sparse and low amplitude, greatly reducing computational load for subsequent processing.

At low energies, the dominant background arises from CsI scintillation afterglow, which exhibits a distinct temporal structure compared to genuine energy-deposition events. To exploit this difference, we developed a lightweight PointNet-based neural network [271] that operates on sparse waveform representations. Each event is encoded using the time indices and amplitudes of waveform samples exceeding a small threshold, preserving pulse-shape information while minimizing data volume. The network achieves strong separation power, rejecting over 90% of afterglow-induced events at a signal efficiency of approximately 85%. This corresponds to an additional order-of-magnitude reduction in the trigger rate. The trained PointNet model was synthesized into FPGA firmware using the `hls4ml` framework [27,28]. Hardware tests confirm excellent agreement between FPGA inference and software validation, with an inference latency of $\sim$10 μs. This demonstrates that ML-based pulse-shape discrimination can be integrated directly into front-end electronics without compromising real-time operation.

The combination of the waveform-integral cut and the PointNet-based afterglow rejection model provides a highly effective multistage background suppression strategy for the cryogenic CsI detector. Each stage targets distinct sources of background, leveraging both simple pulse-amplitude correlations and advanced ML techniques to achieve efficient signal purification while maintaining high CEvNS detection efficiency. We note that self-trigger methods can be used also for other searches for beam-correlated BSM signals outside of the standard beam-trigger window.

#### *4. KamLAND2-Zen*

KamLAND-Zen currently sets the world-leading limit of neutrinoless double beta decay ($0\nu\beta\beta$) and is the first experiment to probe a portion of the inverted mass ordering parameter space for a subset of nuclear matrix element calculations [272]. With the forthcoming KamLAND2-Zen upgrade, the experiment is positioned to refresh the world-leading sensitivity to $0\nu\beta\beta$ in the coming decade [273].

This white paper outlines a hardware-AI codesign strategy that enables the deployment of ML algorithms directly on the FPGAs in the KamLAND2-Zen DAQ. By enabling real-time, position-aware adaptive triggering, this approach will enhance the experiment's sensitivity to $0\nu\beta\beta$ signatures by suppressing background events.

The main motivation for hardware-AI codesign in KamLAND2-Zen is to enhance the tagging of long-lived spallation isotopes, the major background in the $0\nu\beta\beta$ region of interest. The current KamLAND-Zen trigger system employs a simple, fixed-threshold hardware trigger. Therefore, many valuable low-energy events, particularly those relevant for long-lived spallation isotope tagging, are discarded before detailed analysis. To fully exploit the upgraded detector's capabilities, KamLAND2-Zen requires a shift from static, threshold-based triggers to intelligent, low-latency, physics-aware triggering.

As an initial proof of concept, PointNet was successfully deployed on RFSoC to reconstruct event position and energy using KamLAND-Zen Point Cloud data [271]. PointNet was chosen because of its distributed nature: It applies the same multilayer perceptron layers to each point individually before global aggregation. This design perfectly matches the KamLAND-Zen DAQ system, where 130 boards each can only access up to 16 points. The results show that PointNet can achieve 20 cm position resolution and a 0.06 MeV energy resolution [274]. Comparing PointNet to KamLAND-Zen's offline reconstruction method, while the position resolution is modestly degraded relative to offline reconstruction, the energy resolution is improved. The `cgra4ml` framework [47] was used to deploy the PointNet model onto an RFSoC heterogeneous computing system. As shown in Fig. 47, the model was able to reconstruct event position and energy with a 0.5 second latency. While this is still much longer than the desired latency for KamLAND2-Zen, it serves as the first successful demonstration of the feasibility of real-time event reconstruction on RFSoC.

Building on this foundation, the proposed KamLAND2-Zen trigger architecture centers on ultra-fast reconstruction using ridge regression implemented directly on front-end

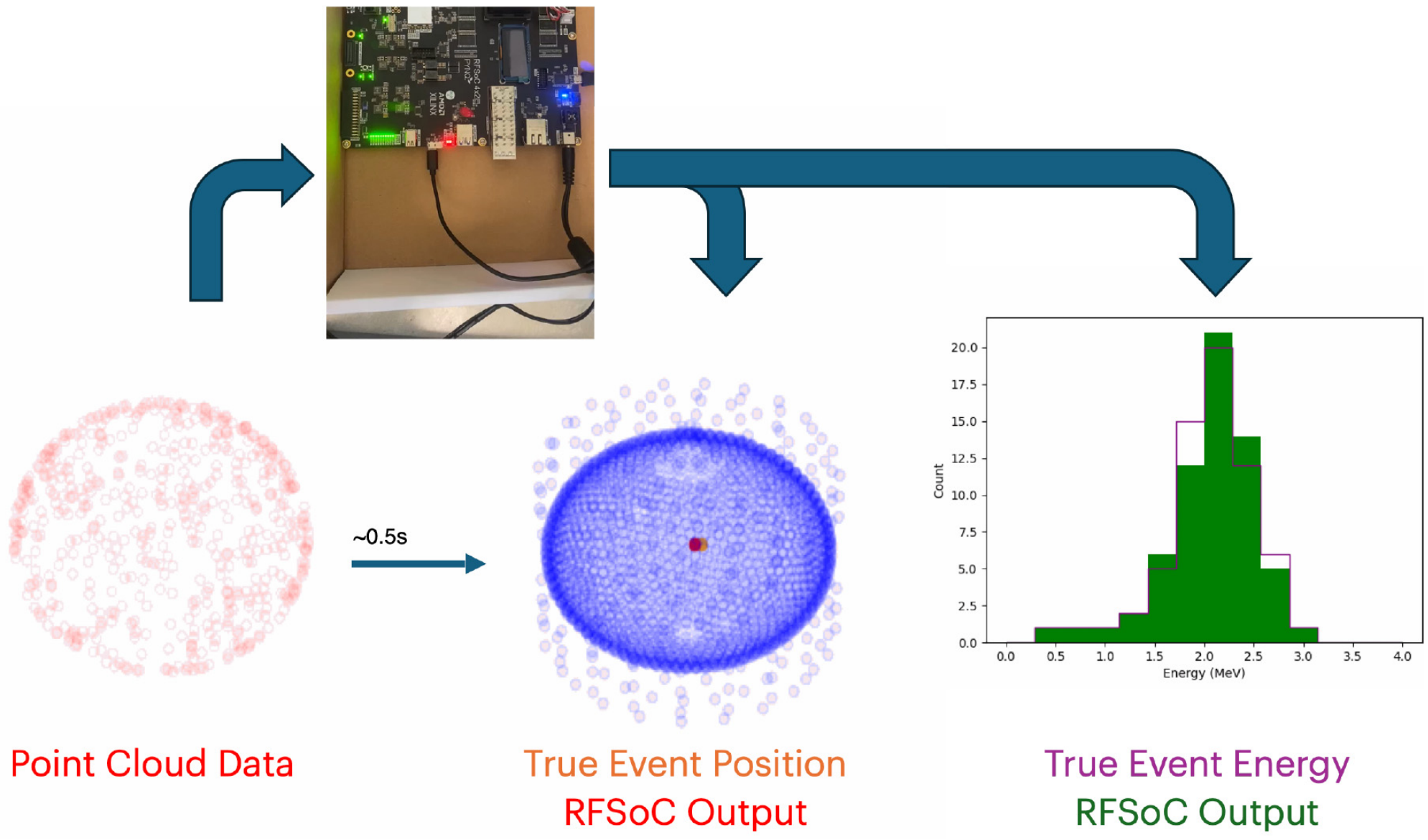

FIG. 47. Demonstration of event reconstruction with the PointNet model achieving $\mathcal{O}(s)$ latency on a RFSoC board.

FPGAs. Ridge regression has been demonstrated to accurately reproduce the KamLAND-Zen offline reconstruction [275] while requiring orders of magnitude fewer parameters than neural networks. This approach reproduces the behavior of established vertex fitters with orders-of-magnitude fewer parameters, making it well suited for low-latency hardware deployment.

The development program includes the following:

(1) *Front-end feature extraction*: Real-time charge integration to extract PMT charge and timing information on RFSoC boards.

(2) *Event-level inference*: A trained ridge regression model mapping PMT-level features to event energy and spatial coordinates $(x, y, z)$ within nanoseconds.

(3) *Validation and robustness studies*: Systematic evaluation of reconstruction accuracy as a function of energy and position, including tolerance to compromised PMT channels.

(4) *Interpretability analysis*: Leveraging model linearity to quantify the influence of individual PMTs and ensure physics transparency.

The key physics motivation of fast ML on RFSoC is long-lived spallation product rejection. The 2.2 MeV neutron capture events—the event when a neutron is captured by hydrogen and emits a 2.2 MeV photon, are one of the most important handles for rejecting the long-lived spallation backgrounds that dominate the $0\nu\beta\beta$ region of interest [276]. However, these events cannot be completely triggered by the fixed-threshold hardware trigger in current generation DAQ. To address this challenge, the collaboration plans to train and deploy a binary classification neural network that distinguishes neutron capture events from the abundant PMT afterpulses, and then perform software trigger based on the neutron capture score.

Moreover, the abovementioned real-time event reconstruction could provide additional handles in rejecting long-lived spallation product. The reconstructed position information allows us to trigger more frequently in the xenon-loaded fiducial volume while maintaining acceptable overall data rates; moreover, real-time energy and position information enables efficient triggering on other short-lived spallation products following cosmic muons, events that are largely missed under the current trigger scheme. Capturing these signals provides critical input for offline ML-based coincidence tagging, significantly improving suppression of long-lived spallation backgrounds.

### D. Accelerators

Particle accelerators are essential tools in modern science, and they have numerous applications in medicine and industry. Emerging applications require increasing degrees of precision and rapid on-demand customization of beams, sometimes with increasingly challenging-to-control characteristics (e.g., higher charge and energy, shorter duration). At present, the operation of these machines is handled by expert human operators and simple control algorithms (e.g., standard proportional–integral–derivative (PID) feedback, corrections based on linear physics models). There is substantial room for improvement in both the efficiency and quality of accelerator operations that can be provided by fast/edge ML systems, with early demonstrations across accelerators from LCLS to the Fermilab Booster [277].

#### 1. *Ultrafast x-ray diffraction*

State-of-the-art instruments such as SLAC's LCLS produce imaging data at extreme rates, enabled by megahertz-class detectors and coherent scattering geometries [278]. Techniques like x-ray photon correlation spectroscopy (XPCS) record speckle patterns that capture nanoscale, femtosecond-nanosecond material dynamics [279,280]. However, interpreting these patterns using conventional molecular-dynamics (MD) simulations requires modeling $10^{11}$ atoms, far beyond modern HPC limits [281]. This data-computation gap necessitates next-generation AI tools that combine physical modeling with scalable learning.

Recent work introduces a physics-informed AI framework capable of interpreting the massive, high-throughput imaging data produced by the fastest x-ray facilities in the United States. It comprises three parts as detailed below.

*Latent alignment of MD and experimental speckles.* Using self-DIstillation with NO labels version 2 (DINOv2) [282] and contrastive language-image pre-training (CLIP) [283], we construct a shared latent space aligning coarse MD simulations ($\sim 10^4$ atoms) with experimental speckle patterns. The resulting representation is designed to reveal physically interpretable structure—relaxation modes, spatial heterogeneity, and symmetry features—embedded within high-dimensional diffraction data.

*Diffusion-based speckle simulation with physics constraints.* Leveraging latent diffusion models [284] and flow-matching generators [285], we develop a simulation engine that produces physically realistic speckle patterns conditioned on MD embeddings and experiment metadata. Physics-based self-supervision losses [286] ensure that outputs remain consistent with known scattering laws, enabling high-fidelity simulation without the prohibitive cost of atomistic computation.

*Automated comparison, AD, and vision-language models-guided reasoning.* Multimodal AI systems such as Flamingo [287] and bootstrapping language-image pre-training (BLIP)-2 [288] automatically compare simulated and observed speckles, identify mismatches, and generate natural-language hypotheses regarding underlying molecular processes. This establishes an AI-augmented scientific pipeline for ultrafast imaging experiments, directly aligned with the goals of the Genesis Mission.

A key application domain is the glass transition, one of the most important unresolved problems in condensed matter physics [289,290]. Recent XPCS studies have revealed ultrafast relaxation pathways in glass-forming liquids [291], but extracting mechanistic insight requires a unifying computational framework across experiment and simulation—precisely the capability enabled by physics-informed AI. This integration of physics-grounded AI with world-leading x-ray infrastructure provides a prime example of the use of hardware-integrate AI/ML to accelelearte scientific discovery.

### 2. Beam optimization

The University of Hawai'i accelerator and free-electron laser (FEL) facility includes a microwave thermionic gun, a 2.856 GHz RF Linac, an infrared FEL oscillator, and an inverse Compton scattering (ICS) x-ray beamline. The beamline is compact and densely instrumented with steering magnets, quadrupoles, spectrometers, wire scanners, beam position monitors, and optical diagnostics. Previously demonstrated operating parameters include 50–60 pC micropulses, ~2 ps pulse duration, ~5 μs macropulse lengths, and beam energies near 40 MeV. The FEL produces ~3 μm IR radiation and the ICS chamber generates ~10 keV x-rays [292].

Start-to-end beam dynamics simulations using A Space charge TRacking Algorithm (ASTRA) [293], general particle tracer (GPT) [294], ELEctron Generation ANd Tracking (ELEGANT) [295], GINGER [296], and GENESIS [297] characterize transport from the cathode to the FEL and ICS interaction points. These simulations track transverse and longitudinal phase space, emittance growth, and timing sensitivity throughout the beamline. Even small variations in injector phase, magnet strengths, or cathode conditions lead to large downstream changes in beam quality, illustrating the highly coupled, high-dimensional, and nonlinear nature of the system. The machine therefore presents an ideal environment for data-driven optimization, where simulation-informed models can complement sparse experimental measurements.

Recent recommissioning work [298] has focused on establishing a stable operational baseline: repairing a major vacuum leak, installing a newly fabricated cathode, and retesting RF power delivery. Ongoing physics studies, such as FEL phase coherence and oscillator desynchronization [299], underscore the sensitivity of cavity performance to arrival-time jitter, energy chirp, and emittance, further motivating algorithmic control strategies capable of maintaining tight synchronization.

At this facility, opportunities for AI and machine learning span injector optimization, beam transport, lattice matching, undulator performance, and cavity tuning. Injector optimization involves adjusting heater power, RF phase, and solenoid fields to maximize current without inducing excessive beam loading or cathode back-heating. ML-based surrogate models or Bayesian optimization can map the high-dimensional response surface, enabling faster and safer tuning than grid scans or manual iteration.

Longitudinal phase-space control represents another major opportunity. Electron beam properties at the Linac exit determine FEL performance and ICS spectral quality; machine learning can help maintain minimal energy spread and compensate for drifts in RF amplitude or phase. Reinforcement learning or adaptive control algorithms could learn corrective actions based on diagnostics such as spectrometers.

Transverse matching to the undulator and ICS point is similarly suitable for ML-driven optimization. Quadrupole settings, chicane strengths, and steering corrections can be tuned using surrogate-assisted multi-objective methods that account for emittance, Twiss parameters, orbit stability, and radiation output simultaneously.

Finally, FEL performance can benefit from ML-guided optimization of cavity alignment, interferometer settings, and cavity length detuning. These adjustments traditionally rely on operator experience, but machine learning can search parameter space more systematically, maintain stability over long timescales, and adapt to thermal or mechanical drifts.

Overall, the UH accelerator provides a rich, controllable, and diagnostics-constrained environment ideally suited for AI/ML-enabled beam optimization and autonomous accelerator operation.

### 3. Beam tuning

In the past decade, the suite of AI/ML techniques that have successfully been applied to accelerators has grown dramatically (see Refs. [300–303] for overviews). At SLAC, substantial effort has been devoted to algorithm development for improved modeling and control of accelerators, open-source software development, and deployment of AI/ML tools into operations. A priority has been placed on developing tools that can aid fast beam customization for the many different operating modes of SLAC's scientific user facilities, such as facility for accelerator science and experimental tests (FACET)-II, megaelectronvolt ultrafast electron diffraction (MeV-UED), and the FEL at LCLS. To address the associated challenges, researchers at SLAC initially took two complementary paths: (1) they targeted ML-based optimization algorithms that could learn on-the-fly from limited measurements, to optimize a given configuration without the need for sophisticated system models or large amounts of previous data, and (2) they targeted methods to improve start-to-end modeling of accelerator systems, including hybrid physics and ML modeling approaches.

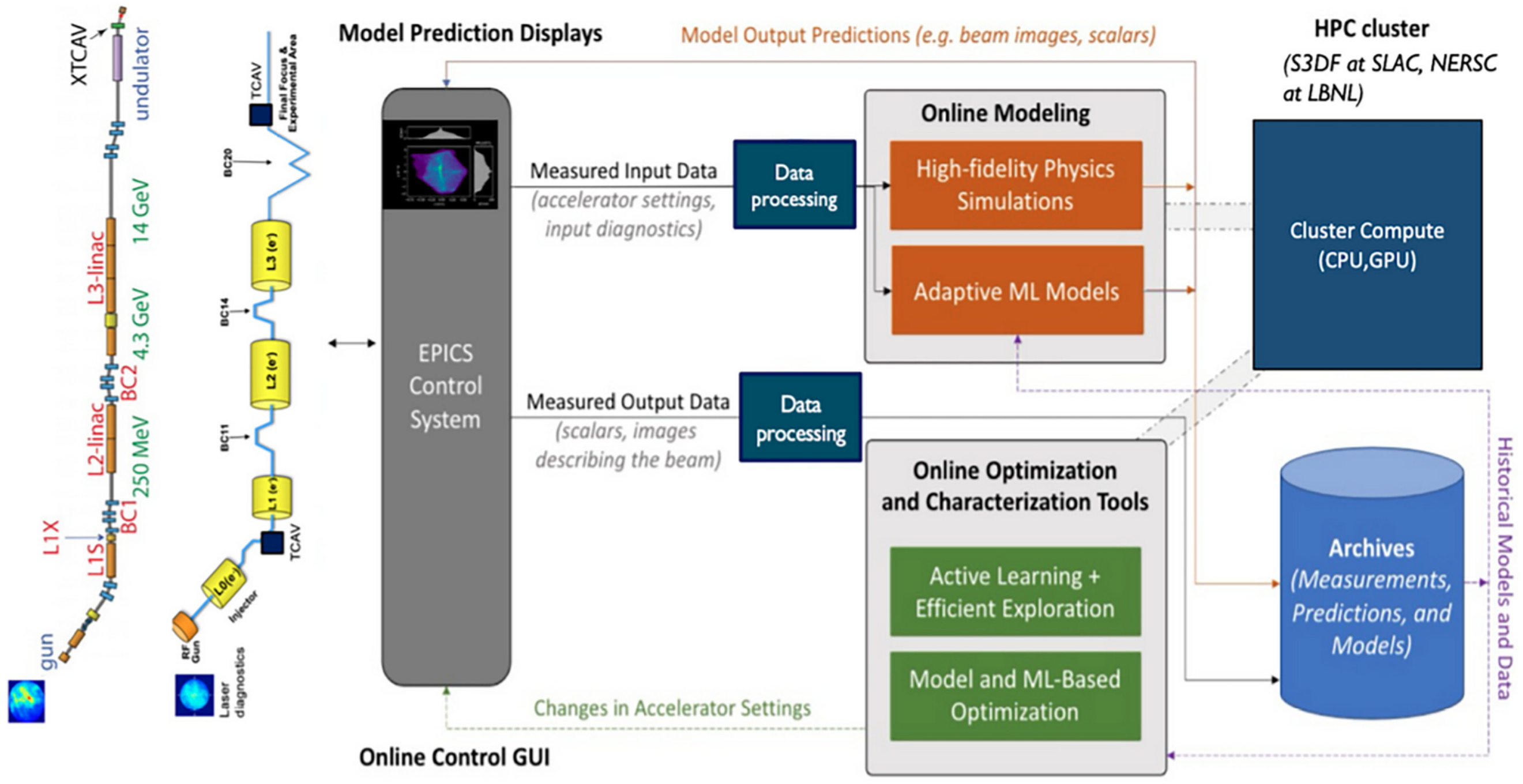

FIG. 48. Diagram of intelligent real-time control system for accelerator operations.

SLAC produced some of the first successful uses of Bayesian optimization (BO) for accelerators [304] and subsequently demonstrated many improvements, such as inclusion of learned output constraints (e.g., for avoiding beam losses), smoother local sampling (to avoid wasting time as variables settle), and trust region-based search for gradually exploring large parameter spaces (important, for example, in maintaining FEL pulse intensity while respecting constraints such as beam losses) [300,305]. SLAC researchers and collaborators have also demonstrated several approaches for using physics and ML-based system models to provide a jump start to optimizers (warm starts, physics-model informed kernels, and system models as priors) [306–308]. SLAC has also pioneered the use of methods for efficient exploration of parameter spaces in accelerators, such as Bayesian exploration [305]. In an example at FACET-II, SLAC researchers were able to get an 8× more efficient characterization of the injector system by using Bayesian exploration instead of standard N-D parameter scans. This data was then ready and well distributed for putting into an ML-based system model of the injector. SLAC has also developed approaches that combine on-the-fly system modeling and information-based sampling for multipoint optimization problems, such as emittance tuning; the team found that this resulted in a nearly 20 times speedup in emittance tuning [309].

Turning to digital twins and online modeling (Fig. 48), SLAC was an early leader in the development of ML-based surrogate models of accelerator systems based on both physics simulations and measurements [310–312]. The eventual aim is to produce digital twins of the accelerator that can aid operators in the control room by providing more insight into machine behavior, provide models to be used to jump start tuning, and provide models that can be used for experiment planning prior to beam times. SLAC produced several early demonstrations of physics and ML based digital twins for subsections of LCLS, including linking live-updating particle-in-cell physics simulations to control room operation [312]. For example, SLAC physicists used a physics simulation for the LCLS-II injector that was running live on the SLAC HPC cluster to provide insight into accelerator behavior that normally is not measurable. This helped the physicists to identify different settings of the injector to try to get to higher beam quality. They then fine-tuned the solution with BO, resulting in the best emittance that had been seen at that time during the LCLS-II injector commissioning, despite extensive previous tuning by hand. SLAC is continuing to expand this work toward a full digital twin of its facilities, both in terms of associated ML techniques and overarching software workflows. There is substantial ongoing collaboration between SLAC, LBNL, and other DOE partners in this space.

Alongside R&D, SLAC has also been developing software tools that are open source and can be used at a wide variety of particle accelerator facilities. For example, Badger and Xopt have been used in the control room of many facilities for online optimization (e.g., see Ref. [313]), and the lume software ecosystem for start-to-end physics simulations and ML-enhanced digital twins has been used worldwide [314]. These tools have been used on various problems with demonstrated improvements; just a few examples include sextupole tuning at FACET-II resulting in a 2× improvement in acceleration efficiency, emittance tuning with a 20× improvement in tuning speed, a storage lifetime record at European Synchrotron Radiation Facility (ESRF) [315], and improvement in FEL pulse intensity at LCLS (percent improvement is still highly variable, but 15%–20% improvement has been observed). Current work focuses on making these software tools and algorithms more robust and user-friendly, and relies heavily on feedback from the accelerator community.

We envision having accurate, fast-executing digital twins that adapt to the machine over time, provide initial settings, and can be used in model-based control. When exploring beam configurations for which there is little data, BO can provide an advantage in the initial search without the need for previous data. As data are collected, the digital twin can be updated to accurately model the configuration. Finally, RL and other data-intensive control techniques can then be brought into play for faster/continuous control for configurations where there is more data and accurate models are available. We envision these techniques will all be used in tandem during different types of operational states. Eventually, we also envision the accelerator and photon beamlines for LCLS being more closely linked in the modeling and tuning process, enabling the whole facility to be adjusted on-the-fly to better meet demands of scientific users and to take advantage of more subtle correlations across the whole facility [316].

## IV. COMMUNITY AND EDUCATION

### A. Related organizations and initiatives

The advent of ML-HEQUPP into the sciences has taken off across all subfields on an international scale; for example, the 2026 International Conference on High Energy Physics is the first in the series to have a dedicated track on “artificial intelligence, machine learning, and quantum computing in HEP” [317]. As a result, organic collaborations have emerged that aim to fully harness and democratize the use of these technologies in modern particle physics experiments. In the U.S., preexisting organizations such as the Division of Particles and Fields of the American Physical Society, particularly the CPAD [318], can act as facilitators. Significant overlaps also exist with physics organizations dedicated to specific current or future facilities [319], as these are the ultimate clients of ML-HEQUPP research. Details on several specific organizations follow below.

#### *1. Accelerated AI algorithms for data-driven discovery*

The Accelerated Artificial Intelligence Algorithms for Data-Driven Discovery (A3D3) Institute represents a large-scale, multi-institutional effort to enable real-time scientific discovery through the integration of domain science, advanced artificial intelligence algorithms, and specialized computing hardware (Fig. 49). As scientific experiments increasingly confront extreme data volumes and stringent latency requirements—exemplified by the LHC, where streaming rates reach terabytes per second and raw data bandwidths extend to the petabyte-per-second regime—the need for tightly codesigned AI and hardware solutions has become imperative. A3D3 addresses this challenge by uniting 21 partner institutions and more than 170 members across high-energy physics, multimessenger astronomy, neuroscience, and computer science and engineering. This interdisciplinary structure provides a collaborative environment for developing next-generation algorithms and acceleration strategies tailored to the unique constraints of real-time data-intensive applications.

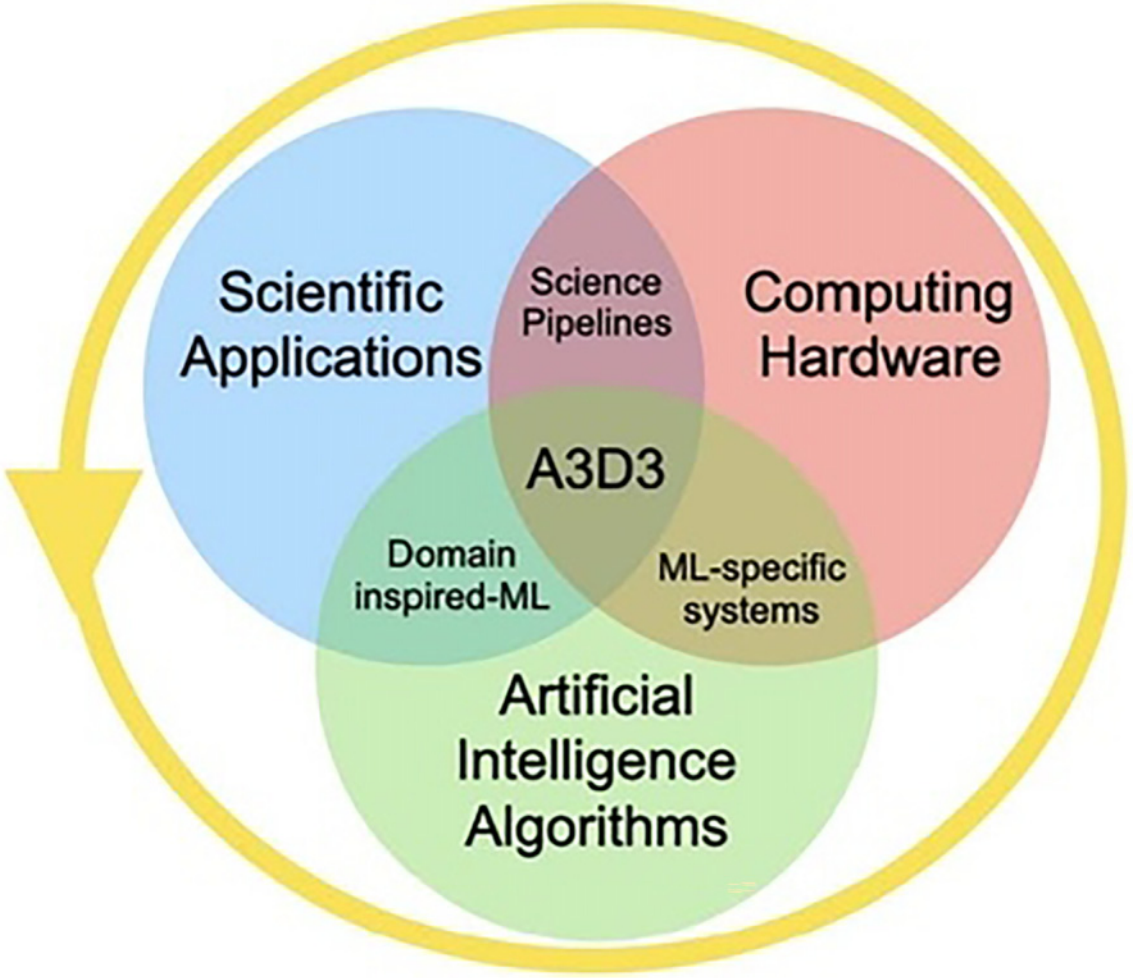

FIG. 49. A3D3: An National Science Foundation (NSF) harnessing the data revolution (HDR) Institute to enable real-time artificial intelligence at scale to advance scientific knowledge and accelerate discovery.

Here, we share an experience of research collaboration between NYCU (National Yang Ming Chiao Tung University, Taiwan) and the A3D3 ecosystem. We focus on FPGA-based acceleration of GNNs, neural decoding models, and high-throughput particle tracking pipelines. The collaborative workflow emphasizes iterative engagement with domain experts, including problem definition, algorithmic evaluation, software-hardware architectural design, prototyping, performance analysis, and model refinement. This approach has facilitated the creation of several end-to-end hardware-accelerated systems that bridge scientific needs with efficient ML-driven computation.

Beyond high-energy physics, the collaboration also developed an FPGA-accelerated deployment of the latent factor analysis via dynamical systems model for real-time neuroscience experiments. By extending the `hls4ml` framework, applying quantization-aware training, and designing a multimodel dataflow architecture, the system achieved submillisecond latency and preserved the result quality. These results demonstrate the generality of A3D3 methodology for applications requiring both low latency and high computational efficiency.

FIG. 50. NGT: A cross experiment CERN-based initiative aiming for enhancing the HL-LHC program.

FIG. 51. Logo for the Fast Machine Learning Community.

In addition to research, A3D3 promotes community engagement through workshops, conferences, summer schools, and ML hardware design competitions, strengthening cross-institutional training and collaboration. The recent admission of NYCU as a Technical Associate Institute of ATLAS further highlights the growing international role of Taiwan's scientific computing efforts.

Overall, the work presented here illustrates the transformative potential of coordinated AI-hardware codesign in addressing the demands of real-time scientific data processing. Through deep integration of algorithms, domain expertise, and specialized accelerators, A3D3 provides a blueprint for future breakthroughs in data-driven discovery.

#### 2. *Next generation triggers*

The Next Generation Triggers (NGT) project [320] at CERN is an on-going project that officially has started in January 2024 (Fig. 50). It is a cross-experimental community engulfing most CERN departments (experimental physics, theory, information technology), the two main experiments (ATLAS and CMS) and external collaborators. The main objective of the project focuses on extracting more physics information from the upcoming HL-LHC upgrade by developing neural network-based solutions. Due to the unprecedented conditions of HL-LHC such algorithms target different deployment algorithms from ASICs to FPGAs all the way to quantum-Computing-based solutions. The project is funded for 5 years from the Eric and Wendy Schmitt foundation.

FIG. 52. AIM (AI, Integrated Detector Concepts, Microelectronics) initiative of the U.S. HFCC organization.

As part of the mandate the project also evaluates technologies from the industry that could provide important performance upgrades for hardware-based trigger system. With that technologies like the AMD AI-engines are being explored in different ATLAS and CMS subsystems aiming to enhance the processing power of custom FPGA-based hardware boards. In addition NGT provides a platform for ML model R&D for the challenging conditions of HL-LHC. Industry-based developed models and techniques are tested in edge conditions allowing for a thorough survey of cutting-edge ML technologies.

#### 3. *FastML community*

The Fast Machine Learning (FastML) community [9] is a research collective of physicists, engineers, and computer scientists focused on deploying machine learning algorithms for challenging scientific applications requiring low latency and high throughput (Fig. 51). The community's projects span from real-time, on-detector inference to high-throughput heterogeneous computing. A broad range of scientific domains are covered, ranging from particle physics to include multimessenger astrophysics, quantum computing, astrophysics, neuroscience, and many more. The key driver within FastML is need to for hardware optimized algorithms to ensure either low latency, low power, or high throughput. ML with its intrinsic parallelizability, resulting from matrix multiply, when combined with heterogeneous computing, ends up often being the core element of optimization within FastML.

Central to the FastML ecosystem for latency and power optimized deployment of ML models is the `hls4ml` library [28,34,146], an open-source Python package that translates trained machine learning models into FPGA firmware using HLS. The library enables scientists to deploy neural networks, boosted decision trees, and other ML architectures with nanosecond-scale latency while fitting within the strict resource constraints of detector front ends. The `hls4ml` framework supports multiple HLS back ends (including AMD/Xilinx Vitis, Intel Quartus, Catapult HLS, and others) and has been extended to support quantization-aware training through integration with QKeras and QONNIX, enabling systematic optimization of bit-width configurations for hardware deployment.

For high-throughput systems, the FastML community focuses on the integration of inference-as-a-service as a paradigm for high-throughput computation. Much of this work centers around the SuperSONIC framework, which provides an optimized tool kit for efficient resource and scaling of heterogeneous computing targeting large-scale high-throughput scientific workflows. These scientific workflows are often found in later stages of the data acquisition chain of large-scale experiments, and typically exploit GPU, and CPU computation.

The FastML community organizes annual workshops that bring together domain scientists with ML and hardware experts, the first of the annual workshops happening in 2019 [321]. Recent editions include the 2024 workshop hosted by Purdue University and the 2025 conference at ETH Zurich, featuring presentations on topics ranging from graph neural networks for tracking to online deployment of real-time systems including AD at the LHC, and multimessenger alerts at LIGO, to include architectures like Kolmogorov-Arnold networks on FPGAs [322–324]. These workshops serve as forums for sharing best practices in codesign methodologies and fostering collaborations between experiments. Furthermore, these meetings aim to broaden collaboration with computing communities allowing for connections, and extended features in core software packages.

Beyond workshops, the community maintains extensive educational resources including tutorial notebooks, summer schools, and documentation. The `hls4ml` Summer School, supported by the NSF POSE program, provides hands-on training for students in hardware-accelerated ML, addressing the critical workforce development need for specialists who can bridge ML algorithms with FPGA implementation. Additionally, the community maintains a Slack space to enable collaboration. With over 2000 members on the Slack, the FastML community's open-source tools and collaborative culture have made it a cornerstone of the broader effort to embed intelligent processing at the edge of scientific experiments.

#### 4. U.S. Higgs factory circular collider organization

Guided by P5 recommendations, the U.S. DOE established the Higgs Factory Circular Collider (HFCC) organization to provide a strategic U.S. role in the planning and research for a future Higgs factory, specifically the FCC-ee (electron-positron Future Circular Collider) under study at CERN. This organization brings together national laboratories and universities to coordinate U.S. contributions in physics studies, accelerator design, detector R&D, software, computing, and collaboration planning with international partners. Its mission is to help shape the technical and scientific groundwork needed for meaningful U.S. participation in the FCC program, from preproject R&D to long-term experimental contributions, and to ensure alignment with the broader global particle physics community.

Within the HFCC structure, the AIM initiative focuses on three cross-cutting areas crucial for next-generation collider detectors: AI/ML, integrated detector concepts, and microelectronics development (Fig. 52). This group evolved from earlier detector electronics and ASIC efforts and now coordinates R&D that spans from smart readout to ML-optimized detector design and open-source tools. Its goals include developing AI/ML approaches suitable for complex detector systems (such as on-detector data processing), optimizing whole-detector concepts through simulation and design tools, and driving innovations in microelectronics (e.g., high-performance ASICs, integrated sensors, silicon photonics, and advanced integration techniques) that are essential to meet the extreme performance requirements of future FCC-ee experiments. The introduction of AIM to the U.S. detector R&D planning process highlights the importance of these critical emerging technologies and a holistic full-detector perspective in delivering on physics goals.

#### 5. Quantum computing for HEP

The intersection of quantum computing and high-energy physics represents an emerging frontier with potential implications for both theoretical simulations and experimental data analysis. Several community initiatives are coordinating efforts in this space, complementing the classical ML and FPGA-focused work described elsewhere in this white paper.

*QC4HEP working group.* The Quantum Computing for High-Energy Physics (QC4HEP) Working Group [74], organized by CERN, DESY, and IBM Quantum, brings together experts from academic and research institutions across four continents to explore quantum computing's potential as a transformative technology for the field. The working group's 2024 road map paper identified two core areas where quantum computing may impact HEP: algorithms for modeling high-energy physics problems (including lattice gauge theory simulations) and numerical methods for analyzing experimental results (including quantum machine learning approaches for event classification and reconstruction). The group holds annual meetings and actively develops benchmark applications targeting near-term quantum hardware.

*CERN quantum technology initiative (QTI).* CERN's Quantum Technology Initiative [325] provides a platform for collaboration between the HEP and quantum technology communities. Phase II of CERN QTI (launched January 2024) focuses on identifying applications of quantum computing and quantum sensing to deliver "quantum advantage" within CERN's physics program. The initiative offers extensive training resources, including online courses on quantum computing, the QTI Journal Club, and schools and workshops that have trained thousands of participants from the HEP community.

*Relevance to front-end ML.* While quantum computing remains in early stages for practical HEP deployment, several connections to front-end ML merit attention: Quantum machine learning algorithms are being explored for pattern recognition tasks such as tracking; hybrid quantum-classical workflows may eventually complement edge computing architectures; and quantum sensing technologies may inform future detector designs. As quantum hardware matures, the codesign principles developed for FPGA-based ML—including quantization, resource optimization, and latency-aware design—may prove valuable for quantum-classical hybrid systems.

### B. Workforce development

The demand for researchers who can develop low-latency, hardware-aware ML has grown rapidly across both industry and experimental physics. Even as the general AI/ML workforce expands, there remains a critical shortage of people who can work at the intersection of ML algorithms, FPGAs/ASICs, firmware, and front-end electronics. Real-time triggering, intelligent data reduction, and high-throughput readout are now central to modern experiments, and they require a workforce that understands both algorithmic design and hardware constraints. Developing this expertise is therefore an urgent community priority.

A major challenge is that most physics students begin without prior training in digital electronics or FPGA programming. In practice, it takes focused instruction, covering topics such as combinational and sequential logic, HDL coding, and vendor toolchains (e.g., AMD/Xilinx Vivado or Vitis, many of whose advanced synthesis and IP-core features require commercial licenses), as well as hands-on work with FPGA, before students can meaningfully contribute to front-end R&D. Model development itself is relatively accessible thanks to open-source ML templates in Keras, PyTorch, scikit-learn, XGBoost, and others; however, understanding how to convert these models into low-latency hardware implementations requires additional, specialized guidance.

A practical way to address these needs is to integrate electronics, ML, and hardware-software codesign directly into training programs. For example, the modern electronics course at the University of Hawai'i at Mānoa (the site of the original ML4FE workshop) provides hands-on instruction in digital circuit design with FPGAs and in deploying AI/ML models on hardware using HLS tools such as `hls4ml`. Through this type of curriculum, students learn to build low-latency inference pipelines under real detector constraints, including resource usage, timing closure, synchronization, and power. After such training, students can immediately apply these skills in collider experiments, for example in beam AD, waveform feature extraction, and other intelligent front-end tasks.

Similar efforts at KEK, particularly within the E-Sys group, involve students directly in firmware development, real-time ML for front-end systems, and mixed-signal design for instrumentation. Recent programs also incorporate AMD versal architecture, giving trainees exposure to advanced computational engines such as the AI engine and deep-learning processing unit, which are increasingly relevant for ML inference in next-generation readout systems. These initiatives demonstrate that effective education must be closely coupled to ongoing detector R&D, ensuring that trainees work with actual hardware, realistic latency constraints, and modern heterogeneous computing resources.

Beyond individual institutions, community-level initiatives play a critical role. The HEP community is uniquely positioned to serve as a hub for broader workforce development. By coordinating training materials, sharing curricula, and creating cross-institutional research opportunities, the community can accelerate the cultivation of specialists capable of advancing low-latency ML and intelligent front-end systems for future experiments while also contributing to the national need for a technically skilled workforce.

*Emerging training needs: quantum and analog computing.* As the field moves toward heterogeneous computing architectures that may eventually include quantum processors and analog accelerators, workforce development must anticipate these emerging technologies. The CERN Quantum Technology Initiative offers introductory courses on quantum computing that have reached thousands of participants, providing a model for community-wide training. Similarly, the QC4HEP Working Group's activities help build familiarity with quantum algorithms among HEP researchers.

For analog computing approaches discussed in Sec. II D 2 b, specialized training is needed to bridge the gap between traditional digital design and emerging paradigms such as in-memory computing, optical neural networks, and mixed-signal processing. While dedicated curricula for analog ML in HEP do not yet exist, the foundational skills developed through FPGA and ASIC training—including understanding of noise, precision trade-offs, and hardware-software codesign—provide a starting point. As analog computing matures for scientific applications, the community should consider developing targeted educational materials that address the unique challenges of deploying ML on these platforms.

### C. Interdisciplinary impacts

The body of ML-HEQUPP research in particle physics established FPGAs as a mature and reliable substrate for deterministic, high-frequency ML inference [48,146,326,327]—the same class of requirements encountered in biology-guided radiotherapy (BgRT)'s positron emission tomography (PET)-driven control cycle. In this way, the impact of these HEP-driven developments extends far beyond particle physics, offering interdisciplinary benefits in fields where high throughput, tight latency bounds, and limited power are equally critical. Medical imaging is one of the most direct beneficiaries. Here, low-latency edge inference enables closed-loop applications such as intraoperative imaging [328], adaptive filtering, and frame-rate enhancement in ultrasound/computed tomography [329], as well as on-device prescreening when data transmission is impractical. FPGAs have also been shown to support full image-reconstruction pipelines directly on hardware. For example, gamma-ray imaging systems have demonstrated filtered back projection and maximum-likelihood expectation maximization implemented and optimized on low-cost FPGA platforms using HLS, achieving multifold speedups while preserving image contrast despite reduced numerical precision. Such results highlight the feasibility of real-time or near-real-time reconstruction and localization on embedded FPGA hardware, reinforcing their suitability for advanced radiation-detection and imaging systems [330].

#### *1. Medical applications*

In the realm of radiation detectors, the main motivation for the adoption of front-end AI/ML is the reduction of the data throughput by means of preprocessing close to the signal source, i.e., in the detector. In-sensor processing allows for reduction of the amount of raw data to be transmitted, offering significant savings in terms of signal lines, bandwidth, power, data storage, and digital processing (mostly based on FPGA) requirements. Medical imaging is a key regime in which techniques often leveraged for data challenges in HEP can be highly useful.

The application that provides the context for the electronics described here is that of nuclear medical imaging for diagnostics, in particular, of tomographic techniques based on the detection of gamma rays such as PET and single-photon emission computed tomography. The field of view of scanners is getting longer, targeting total-body imaging, thus increasing the number of detection modules from tens to hundreds, with thousands of channels and thus making more and more crucial streamlining the dataflow by means of in-sensor preprocessing. ANNs have emerged as excellent tools for image processing, suitable to address the challenges of medical imaging, at multiple levels. Considering a gamma-ray detector consisting of scintillators readout by arrays of photodetectors, usually SiPM, ANNs are employed to estimate the scintillation position within the crystal.

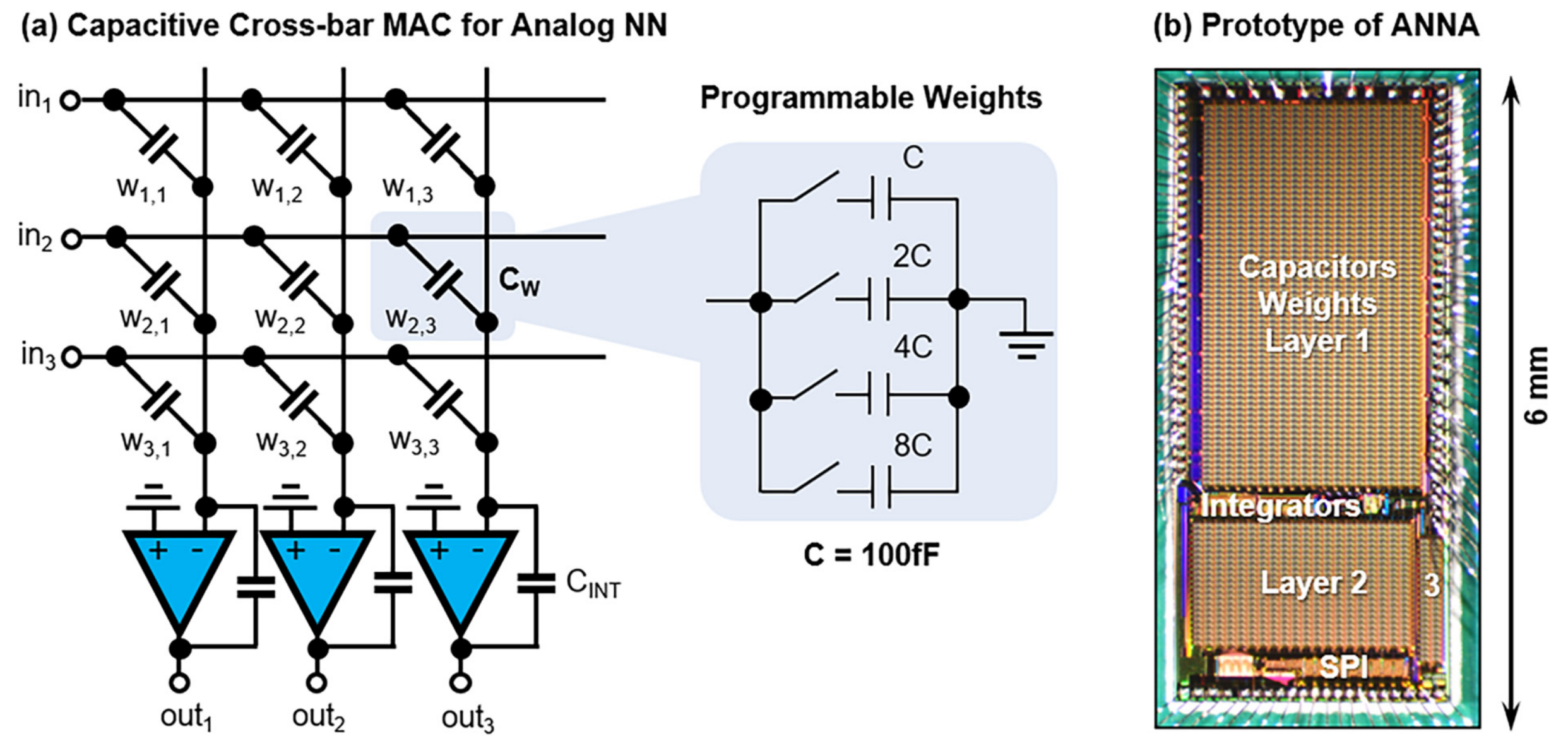

FIG. 53. (a) Charge-mode implementation of MAC operation for neurons with capacitive weights and its realization (b) in a legacy 0.35 μm technology share by SiPM front ends.

The core operation of each neuron in the simplest ANN topology (a feedforward perceptron) is the weighted sum of the inputs (e.g., multiply and accumulate or MAC) with a bias, followed by a nonlinear activation function. An analog implementation of MAC can be performed by manipulating electric charge. The input voltages are converted into charge by means of capacitors connected to the virtual ground of an integrator summing all input signals and converting the sum back into a voltage across the feedback capacitor. The weights are encoded in the values of capacitance, which, for instance, can be made programmable through a set of switched capacitors in parallel (Fig. 53). The value of weights, discretized in the first implementation in 4 bits + sign) is stored in local latches (inside the perceptron), adopting the advantageous in-memory computing paradigm. The activation function can be realized by means of the integrator saturating response (thus realizing a ReLU) or by adding nonlinear stages (with sigmoidal response) in cascade.

A first prototype of a fully connected three-layer ANN was designed [331] and fully characterized [332]. It features 64 inputs (meant to read a standard 8 × 8 arrays of SiPMs), 2 hidden layers with 20 neurons each and 2 outputs, which in this case represent the $x$ and $y$ coordinates of the scintillation point in the crystal. The nonideal behavior of the analog components, including operational amplifiers and switches, has to be carefully assessed and embedded in the training model to achieve better accuracy [333]. This first implementation was realized in the legacy node CMOS 0.35 μm, still used for analog front end thanks to good noise performance. Thus, the potential for a monolithic integration of the ANN with front-end electronics for SiPM [334,335] is demonstrated. Here, the least significant bit of weights is 100 fF. With a 10 MHz clock, the total inference time is 4.6 μs and the power consumption if 16.8 mW, corresponding to an energy cost of 76 nJ per inference and an efficiency of ∼50 GOPS/W. Such an efficiency places this result an order of magnitude below what is offered by FPGA-based, but the energy cost of analog-digital conversion of 64 signals, avoided by the ASIC, is not included in the comparison. Moreover, an improved topology has been designed in a more scaled node (110 nm), with a different recursive topology and improved integrator, based on an CMOS inverter allowing lower power consumption and smoother nonlinear response [336].

Numerous studies show that FPGA-accelerated image analysis pipelines achieve significant improvements in throughput and energy efficiency compared to CPU/GPU approaches, making FPGA ML an attractive technology for embedded clinical systems [328]. Modern FPGA-based PET electronics now integrate real-time signal processing, high-throughput data acquisition, and ML-enhanced reconstruction within tight resource budgets. In particular, systems combining custom front-end designs (e.g., delta-sigma ADC-based SiPM readouts) with deep neural networks synthesized via the `hls4ml` library demonstrate how pruning, quantization, and codesign can deliver low-latency, resource-efficient performance suitable for real-time PET imaging [337,338]. These trends align closely with the requirements of image-guided radiotherapy, where real-time reconstruction and decision-making must be both fast and predictable [339,340]. Tracking the tumor in real time [341,342] is crucial in radiotherapy, where external radiation beams must remain precisely aligned with a moving biological target. BgRT [343] represents a major advance toward this goal by integrating a Linear accelerator with PET [344,345], enabling radiation delivery guided directly by emissions originating inside the patient. For BgRT to function safely and effectively, a demanding control loop must be executed every 100 ms [346]. Extreme temporal constraints mirror the real-time challenges seen in HEP, where detectors must extract salient information from massive data rates on submicrosecond timescales.

#### 2. *Industrial applications*

Beyond medical imaging, related advances in aerospace [347], industrial sensing [348], and hyperspectral imaging [349] further demonstrate the versatility of FPGA-based AD. Implementations combining autoencoders with sequence models have enabled real-time AD across thousands of flight-test channels with orders-of-magnitude improvements in latency and power [350]. Closely related work in radiation detection shows that artificial neural networks implemented on FPGAs can classify high-rate detector pulses in real time, even in the presence of severe pileup. By recovering and classifying individual neutron and gamma events at megahertz-scale rates with microsecond-level latency, these systems illustrate how FPGA-based ML can operate reliably in intense radiation environments, a regime that shares important characteristics with nuclear imaging and radiotherapy instrumentation [351].

Similar results exist for RF and hyperspectral pipelines, where continuous-stream detection at high sample rates supports safety-critical operations such as structural health monitoring and fault prediction [352]. These examples reinforce a broader pattern: When data rates are extreme and latency budgets unforgiving, FPGAs provide a uniquely capable platform for real-time ML. Cybersecurity and network monitoring [353,354] also exploit this paradigm. High-throughput FPGA architectures hosting neural-network-based anomaly detectors [326,355,356] achieve line-rate processing of network traffic, enabling ML-driven intrusion detection that complements traditional signature-based systems. The deterministic timing, field-updatability, and multimodel flexibility of FPGAs make them ideal for deploying adaptive AD pipelines in rapidly evolving threat landscapes [357]. These characteristics parallel medical and HEP needs, where algorithms must be both robust and reconfigurable as detectors or clinical protocols evolve.

Real-time or near-real-time ML applied to postsilicon bug diagnosis offers a powerful way to enhance complex hardware systems by identifying anomalous behavior and localizing faults during validation [358–360]. By combining lightweight on-chip observability with FPGA-based analysis pipelines, these approaches accelerate failure localization and reliability validation in modern SoCs [360–362]. This convergence between microarchitectural debugging, HEP trigger systems, and medical-imaging pipelines illustrates the increasingly universal role of hardware-accelerated ML: It enables domain-specific intelligence to move closer to the data source, making real-time decision-making feasible in settings where software-only solutions cannot meet the timing constraints.

## V. KEY R&D TOPICS

Fully developing aforementioned technologies and realizing their potential to enhance physics experiments requires considerable R&D in both existing and fundamentally new directions. A set of prioritized R&D topics drafted at the conclusion of the ML4FE workshop have been expanded and refined following considerable scope expansion and input during the white paper process. These topics follow below, with emphasis both on the core technology and the follow-on implications for physics experiments and related infrastructure.

(1) Facilitating codesign through continued cross-frontier opportunities for collaboration between scientists, engineers, and technicians.

(2) Development of open-source interdisciplinary benchmarking data/simulation sets for electronics design;
e.g., "AI-ready data," including digitization-level detector simulations, readout demonstrators, end-to-end test benches, etc.

(3) Identify areas of greatest impact for hardware-based ML via classical versus ML benchmarking;
e.g., comparative studies of performance, resource efficiency, and scalability with and without AI/ML-based workflows, in order to guide targeted optimization efforts.

(4) Resource planning for U.S. analysis facilities, namely, considering GPUs and other accelerators, FPGA development environments, inference-as-a-service infrastructure, automation/validation tools, and quantum processors.

(5) End-to-end differentiable DAQ/heterogenous pipeline open-source optimization tools;
addressing need for multistage instrument design and ML operations considering data rates, costs, and system boundaries.

(6) Design and fabrication of smart ASICs and/or embedded FPGAs;
including 3D integration of advanced sensors + embedded intelligence, toward self-driving/real-time-capable instruments and/or real-time training.

(7) Expand open-source tools to synthesize advanced ML for hardware implementation;
e.g., streaming, structured pruning, high-granularity quantization, quantum algorithms for complex inference at trigger/DAQ level.

(8) Intelligence in cold electronics;
for real-time applications with LAr time projection chambers/calorimeters, quantum devices/sensors.

(9) Analog compute;
toward the frontier of fast and low-power computation needs for advanced detectors (4D tracking, five-dimensional high-granularity calorimetry, streaming).

(10) Tracking and incorporating advances in commercial technology;
e.g., performance comparisons on new industrial hardware for modern physics applications and challenges.

(11) Filling essential workforce development gaps with key emerging technologies at the university and postgraduate levels.

## VI. CONCLUSIONS

The onset of the AI/ML era has set fundamental scientific research at an inflection point. The introduction of cutting-edge technologies related to AI/ML, microelectronics, high-performance computing, and quantum processors is essential to deliver an ever more ambitious physics program intent on discovering physics and measuring the standard model to extreme precision. However, the novelty of these technologies means that such an introduction must be done carefully and intentionally to avoid compromising the quality and integrity of experimental and theoretical scientific pursuits. Continued community-centered conversations across science, engineering, government, and industry, supported by a robust and purposeful basic research funding program in the U.S., is the key to success in this process. With such a system in place, the U.S. scientific community can look forward to yet unimaginable advances in enabling technologies and scientific efforts, on a global scale defined by U.S. leadership, with an era-changing long-term outlook for our understanding of the universe.

Project administration: Julia Gonski, Jenni Ott. Writing review & editing: Julia Gonski, Jenni Ott, Shiva Abbaszadeh, Sagar Addepalli, Matteo Cremonesi, Jennet Dickinson, Giuseppe Di Guglielmo, Erdem Yigit Ertorer, Lindsey Gray, Ryan Herbst, Christian Herwig, Tae Min Hong, Benedikt Maier, Maryam Bayat Makou, David W. Miller, Mark S. Neubauer, Cristián Peña, Dylan Rankin, Seon-Hee Seo, Giordon Stark, Alexander Tapper, Audrey Corbeil Therrien, Ioannis Xiotidis, Keisuke Yoshihara. Writing - original draft: Julia Gonski, Jenni Ott, Shiva Abbaszadeh, Sagar Addepalli, Matteo Cremonesi, Jennet Dickinson, Giuseppe Di Guglielmo, Erdem Yigit Ertorer, Lindsey Gray, Ryan Herbst, Christian Herwig, Tae Min Hong, Benedikt Maier, Maryam Bayat Makou, David W. Miller, Mark S. Neubauer, Cristián Peña, Dylan Rankin, Seon-Hee Seo, Giordon Stark, Alexander Tapper, Audrey Corbeil Therrien, Ioannis Xiotidis, Keisuke Yoshihara, Nural Akchurin, Carlos Argüelles, Saptaparna Bhattacharya, Lorenzo Borella, Christian Boutan, Tom Braine, James Brau, Antonio Chahine, Talal Ahmed Chowdhury, Yuan-Tang Chou, Seokju Chung, Marco Carminati, Ryan Coffee, Alberto Coppi, Mariarosaria D'Alfonso, Abhilasha Dave, Chance Desmet, Angela Di Fulvio, Karri DiPetrillo, Javier Duarte, Auralee Edelen, Jan Eysermans, Yongbin Feng, Emmett Forrestel, Loredana Gastaldo, Julián Garca Pardiñas, Lino Gerlach, Abarajithan Gnaneswaran, Loukas Gouskos, Katya Govorkova, Carl Grace, Christopher Grant, Philip Harris, Ciaran Hasnip, Timon Heim, Abraham Holtermann, Tae Min Hong, Gian Michele Innocenti, Koji Ishidoshiro, Miaochen Jin, Jyothisraj Johnson, Stephen Jones, Andreas Jung, Georgia Karagiorgi, Ryan Kastner, Nicholas Kamp, Doojin Kim, Kyoungchul Kong, Katie Kudela, Jelena Lalic, Bo-Cheng Lai, Yun-Tsung Lai, Tommy Lam, Jeffrey Lazar, Aobo Li, Zepeng Li, Haoyun Liu, Qibin Liu, Vladimir Lončar, Luca Macchiarulo, Christopher Madrid, Zhenghua Ma, Prashansa Mukim, Victoria Nguyen, Sungbin Oh, Isobel Ojalvo, Hideyoshi Ozaki, Simone Pagan Griso, Myeonghun Park, Christoph Paus, Santosh Parajuli, Benjamin Parpillon, Sara Pozzi, Ema Puljak, Benjamin Ramhorst, Amy Roberts, Larry Ruckman, Kate Scholberg, Sebastian Schmitt, Noah Singer, Eluned Anne Smith, Alexandre Sousa, Michael Spannowsky, Kaito Sugizaki, Sioni Summers, Yanwen Sun, Daniel Tapia Takaki, Antonino Tumeo, Caterina Vernieri, Belina von Krosigk, Yash Vora, Linyan Wan, Michael H.L. S. Wang, Amanda Weinstein, Andy White, Simon Williams, Liangyu Wu, Felix Yu.

## DATA AVAILABILITY

There are no publicly available research data or software supporting this manuscript. Requests for further information or data should be sent to the authors.